%% file: Main.tex
\documentclass[journal]{IEEEtran}
\usepackage{url}
\usepackage{graphicx}
\usepackage{color}
\usepackage{cite}
\usepackage{subfigure}
\usepackage{verbatim}
\usepackage{amsmath}
\usepackage{amssymb}
\usepackage{multirow}
\usepackage{adjustbox}
\usepackage{textcomp}

\usepackage{algorithmic}
\usepackage[lined,boxed,ruled,commentsnumbered]{algorithm2e}
\usepackage[table,xcdraw]{xcolor}
\usepackage{tabularx,booktabs}
\usepackage[flushleft]{threeparttable}
\usepackage{longtable}
\usepackage{subfiles}

\makeatletter
\newcommand*\bigcdot{\mathpalette\bigcdot@{1.5}}
\newcommand*\bigcdot@[2]{\mathbin{\vcenter{\hbox{\scalebox{#2}{$\m@th#1\bullet$}}}}}
\makeatother

\usepackage{acro}
\subfile{Glossary}

\begin{document}
	

	
	\title{Security and privacy for 6G: A survey on prospective technologies and challenges}

	\author{\IEEEauthorblockN{Van-Linh Nguyen, \textit{Member, IEEE}, Po-Ching Lin, \textit{Member, IEEE}, Bo-Chao Cheng, \\ Ren-Hung Hwang, \textit{Senior Member, IEEE},  Ying-Dar Lin, \textit{Fellow, IEEE}}  \thanks{Van-Linh Nguyen, Po-Ching Lin, Ren-Hung Hwang  are with the Department of Computer Science and Information Engineering (\textit{e-mail: \{nvlinh, pclin, rhhwang\}@cs.ccu.edu.tw}), Bo-Chao Cheng is with the Department of Communications Engineering (\textit{e-mail: bcheng@ccu.edu.tw}), National Chung Cheng University, Chiayi, Taiwan.}
	\thanks{Van-Linh Nguyen is also with the Department of Information Technology, Thai Nguyen University of Information and Communication Technology, Thai Nguyen, Vietnam (\textit{e-mail: nvlinh@ictu.edu.vn}).} \thanks{Ren-Hung Hwang is also with Advanced Institute of Manufacturing with High-tech Innovations, National Chung Cheng University and a Jointly Appointed Professor of AI college, National Yang Ming Chiao Tung University, Taiwan.} \thanks{Ying-Dar Lin is with the Department of Computer Science, National Yang Ming Chiao Tung University, Taiwan (\textit{e-mail: ydlin@cs.nctu.edu.tw}).}}
	
	\markboth{ }%
    {\MakeLowercase{\textit{ Nguyen et al.}}: Security and privacy for 6G: A survey on prospective technologies and challenges}
	
	\maketitle

	\begin{abstract}
	
		Sixth-generation (6G) mobile networks will have to cope with diverse threats on a space-air-ground integrated network environment, novel technologies, and an accessible user information explosion. However, for now, security and privacy issues for 6G remain largely in concept. This survey provides a systematic overview of security and privacy issues based on prospective technologies for 6G in the physical, connection, and service layers, as well as through lessons learned from the failures of existing security architectures and state-of-the-art defenses. Two key lessons learned are as follows. First, other than inheriting vulnerabilities from the previous generations, 6G has new threat vectors from new radio technologies, such as the exposed location of radio stripes in ultra-massive MIMO systems at Terahertz bands and attacks against pervasive intelligence. Second, physical layer protection, deep network slicing, quantum-safe communications, artificial intelligence (AI) security, platform-agnostic security, real-time adaptive security, and novel data protection mechanisms such as distributed ledgers and differential privacy are the top promising techniques to mitigate the attack magnitude and personal data breaches substantially.
     
	\end{abstract}
	
	
	
\section{Introduction}
\label{sec:introduction}

\subfile{Section_I_Introduction}

\section{Related work and our survey position}
\label{sec:related-work}

\subfile{Section_II_Related_Work}

 \section{Security issues and the evolution of security architecture in legacy mobile networks}
 \label{sec:overview-core-security-architecture}

\subfile{Section_III_Security_legacy}

\section{6G network vision and potential changes of its security architecture}
 \label{sec:6G-security-architecture}

\subfile{Section_IV_6G_Core_security}

\section{Security in the physical layer}
\label{sec:security-physical-layer}

\subfile{Section_V_PHY}

\section{Security in the connection layer} 
\label{sec:security-connection-layer}

\subfile{Section_VI_Connection}

\section{Security in the service layer} 
\label{sec:security-service-layer}

\subfile{Section_VII_Service}

\section{Artificial Intelligence's impact on 6G security}
\label{sec:security-ai}

\subfile{Section_VIII_Sec_AI}

\section{Privacy challenges and preservation approaches}
\label{sec:privacy-in-6G}

\subfile{Section_IX_Privacy}

\section{Discussion on future research directions and open issues}
\label{sec:discussion-future-research}

\subfile{Section_X_Discussion}

\section{Conclusion}
\label{sec:conclusion}

Security and privacy have been the pillars of the success of mobile networks. In 6G, when the right to Internet access is guaranteed to everyone, the networks will become a gigantic connected world, with heterogeneous domains of enterprise and telecom networks, virtual and physical, satellites, terrestrial nodes, and so on. The more complicated the networks are, the more risks we face. Other than traditional security concerns such as virus/malware/DDoS attacks/deepfake, learning-empowered attacks and massive data breaches may occur more commonly in 6G because of the increase in connected devices and novel technologies. This work has provided an overview of security and privacy issues of prospective technologies for the physical, connection, service layers of 6G. Based on the lessons learned from the survey, we have outlined an assessment of the prospective technologies for 6G security and privacy issues, such as physical layer security, QKD, deep slicing, and distributed ledgers. However, satisfying real-time protection requirements and energy efficiency are still major challenges for such technologies. Without these features, many 6G security services likely fall short of their own goals. Finally, we believe that supply chain security -- while not a technical issue -- will play a central role in keeping the development of 6G security on the right track. The success of several initiatives, such as open RAN and open-source security, will be an essential part of improving supply chain security.

\section*{Acknowledgment}

This work was supported in part by the Ministry of Education and Training (MOET) of Vietnam and Thai Nguyen University under Grant No B2021-TNA-02.

\bibliographystyle{ieeetr}
\bibliography{References}

\end{document}

%% file: Glossary.tex
\DeclareAcronym{IoE}{short = IoE,long=Internet of Everything}
\DeclareAcronym{KPIs}{short = KPIs,long=Key Performance Indicators}
\DeclareAcronym{ITU}{short = ITU,long=International Telecommunication Union}
\DeclareAcronym{AI}{short = AI,long=Artificial Intelligence }
\DeclareAcronym{SMS}{short =SMS,long=Short Message Service}
\DeclareAcronym{3GPP}{short =3GPP,long=3rd Generation Partnership Project}
\DeclareAcronym{V2X}{short =V2X,long=Vehicle-to-everything}
\DeclareAcronym{V2V}{short =V2V,long=Vehicle-to-Vehicle}
\DeclareAcronym{V2I}{short =V2I,long=Vehicle-to-Infrastructure}
\DeclareAcronym{DDoS}{short =DDoS,long=Distributed Denial-of-Service}
\DeclareAcronym{DoS}{short =DoS,long=Denial-of-Service}
\DeclareAcronym{DNS}{short =DNS,long=Domain Name System}
\DeclareAcronym{CCTV}{short =CCTV,long=Closed Circuit Television}
\DeclareAcronym{IoT}{short =IoT,long=Internet of Things}
\DeclareAcronym{MEC}{short =MEC,long=Multi-access Edge Computing}
\DeclareAcronym{5G}{short =5G,long= fifth-generation}
\DeclareAcronym{SDN}{short =SDN,long=Software defined Networking}
\DeclareAcronym{NFV}{short =NFV,long=Network Function Virtualization}
\DeclareAcronym{SDA}{short =SDA,long=Service-defined Architecture}
\DeclareAcronym{RAN}{short =RAN,long=Radio Access Network}
\DeclareAcronym{RSU}{short =RSU,long=Road-side Unit}
\DeclareAcronym{eMBB}{short =eMBB,long=enhanced Mobile Broad Band}
\DeclareAcronym{uRLLC}{short =uRLLC,long=Ultra-Reliable Low-Latency Communication}
\DeclareAcronym{mMTC}{short =mMTC,long=massive Machine Type Communications}
\DeclareAcronym{LPW}{short =LPW,long=Low-power Wireless}
\DeclareAcronym{LPAN}{short =LPAN,long=Low-power Personal Network}
\DeclareAcronym{LPWAN}{short =LPWAN,long=Low-power Wide Area Network}
\DeclareAcronym{EDA}{short =EDA,long=Energy depletion attacks}
\DeclareAcronym{PCF}{short =PCF,long=Policy Control Function}
\DeclareAcronym{NRF}{short =NRF,long=Network Repository Function}
\DeclareAcronym{NSSF}{short =NSSF,long=Network Slice Selection Function}
\DeclareAcronym{UDM}{short =UDM,long=Unified Data Management}
\DeclareAcronym{UPF}{short =UPF,long=User Plane Function}
\DeclareAcronym{AMF}{short =AMF,long=Access and Mobility Function}
\DeclareAcronym{SMF}{short =SMF,long=Session Management Function}
\DeclareAcronym{LADN}{short =LADN,long=Local Area Data Network }
\DeclareAcronym{MIMO}{short =MIMO,long=Multiple-input and Multiple-output}
\DeclareAcronym{SFC}{short =SFC,long=Service function chaining}
\DeclareAcronym{SECaaS}{short =SECaaS,long=Security as a service}
\DeclareAcronym{PISA}{short =PISA,long=Protocol-Independent Switch Architecture}
\DeclareAcronym{ONOS}{short =ONOS,long=Open Network Operating System}
\DeclareAcronym{IDS}{short =IDS,long=Intrusion Detection System}
\DeclareAcronym{IPS}{short =IPS,long=Intrusion Prevention System}
\DeclareAcronym{PKI}{short =PKI,long=Public-key Infrastructure}
\DeclareAcronym{UE}{short =UE,long=User Equippment}
\DeclareAcronym{NTP}{short =NTP,long=Network Transfer Protocol}
\DeclareAcronym{UDP}{short =UDP,long=User Diagram Protocol}
\DeclareAcronym{ICMP}{short =ICMP,long=Internet Control Message Protocol}
\DeclareAcronym{LTE}{short =LTE,long=Long-term Evolution}
\DeclareAcronym{TEID}{short =TEID,long=Tunnel Endpoint Identifier}
\DeclareAcronym{GTP}{short =GTP,long=General Packet Radio Service tunneling protocol}
\DeclareAcronym{GTP-U}{short =GTP-U,long=GPRS Tunnelling Protocol – User}
\DeclareAcronym{GTP-C}{short =GTP-C,long=GPRS Tunnelling Protocol – Control}
\DeclareAcronym{SGW}{short =SGW,long=Serving Gateway}
\DeclareAcronym{PGW}{short =PGW,long=Packet Gateway}
\DeclareAcronym{PDP}{short =PDP,long=Packet Data Protocol}
\DeclareAcronym{MSISDN}{short =MSISDN,long=Mobile Station International Subscriber Directory Number}
\DeclareAcronym{IMSI}{short =IMSI,long=International Mobile Subscriber Identity}
\DeclareAcronym{MME}{short =MME,long=Mobility Management Entity}
\DeclareAcronym{WSMP}{short =WSMP,long=Wave Short Message Protocol}
\DeclareAcronym{ADAS}{short =ADAS,long=Advanced Driver-Assistance Systems}
\DeclareAcronym{CACC}{short =CACC,long=Cooperative Adaptive Cruise Control}
\DeclareAcronym{VANET}{short =VANET,long=Vehicular Ad-hoc Networks}
\DeclareAcronym{RSSI}{short =RSSI,long=Radio Signal Strength Indicator}
\DeclareAcronym{AoA}{short =AoA,long=Angle of radio Arrival}
\DeclareAcronym{TDOA}{short =TDOA,long=Time Difference of Arrival}
\DeclareAcronym{LOS}{short =LOS,long=Light-of-Sight}
\DeclareAcronym{non-LOS}{short =non-LOS,long=non-Light-of-Sight}
\DeclareAcronym{NLOS}{short =NLOS,long=non-Light-of-Sight}
\DeclareAcronym{C-ITS}{short =C-ITS,long=Cooperative Intelligent Transportation Systems}
\DeclareAcronym{LIDAR}{short =LIDAR,long=LIght Detection and Ranging}
\DeclareAcronym{VLC}{short =VLC,long=visible light communication}
\DeclareAcronym{BSM}{short =BSM,long=Basic Safety Message}
\DeclareAcronym{CAM}{short =CAM,long=Cooperative Awareness Message}
\DeclareAcronym{OFDMA}{short =OFDMA,long=Orthogonal Frequency Division Multiple Access}
\DeclareAcronym{ULA}{short =ULA,long=Uniform Linear Array}
\DeclareAcronym{PEB}{short =PEB,long=Position Error Bound}
\DeclareAcronym{UKF}{short =UKF,long=Unscented Kalman Filter}
\DeclareAcronym{HD}{short =HD,long=High-resolution Dynamic}
\DeclareAcronym{CV}{short =CV,long=Constant Velocity}
\DeclareAcronym{CA}{short =CA,long=Constant Acceleration}
\DeclareAcronym{CTRA}{short =CTRA,long=Constant Turn Rate and Acceleration}
\DeclareAcronym{CSAV}{short =CSAV,long=Constant Steering Angle and Velocity}
\DeclareAcronym{CCA}{short =CCA,long=Constant Curvature and Acceleration}
\DeclareAcronym{IMM}{short =IMM,long=Interacting Multiple Model}
\DeclareAcronym{NIES}{short =NIES,long=Normalised Innovation Error Squared}
\DeclareAcronym{LIS}{short =LIS,long=Large Intelligent Surface}
\DeclareAcronym{FOV}{short =FOV,long=Field of View}
\DeclareAcronym{6G}{short =6G,long=sixth-generation}
\DeclareAcronym{4G}{short =4G,long=fourth-generation}
\DeclareAcronym{AIML}{short =AI/ML,long=Artificial Intelligence and Machine Learning}
\DeclareAcronym{GSM}{short =GSM,long=Global System for Mobile}
\DeclareAcronym{UMTS}{short =UMTS,long=Universal Mobile Telecommunications System}
\DeclareAcronym{SIM}{short =SIM,long=Subscriber Identity Module}
\DeclareAcronym{USIM}{short =USIM,long=Universal Subscriber Identity Module}
\DeclareAcronym{TMSI}{short =TMSI,long=Temporary Mobile Subscriber Identity}
\DeclareAcronym{HSPA}{short =HSPA,long=High-Speed Packet Access}
\DeclareAcronym{AKA}{short =AKA,long=Authentication and Key Agreement}
\DeclareAcronym{EPS}{short=EPS,long=Evolved Packet System}
\DeclareAcronym{E-UTRAN}{short=E-UTRAN,long=Evolved-Universal Terrestrial Radio Access Network}
\DeclareAcronym{ESN}{short =ESN,long=Electronic Serial Numbers}
\DeclareAcronym{MDN}{short =MDN,long=Mobile Directory Numbers}
\DeclareAcronym{MSC}{short =MSC,long=Mobile Switching Center}
\DeclareAcronym{AuC}{short =AuC,long=Authentication Center}
\DeclareAcronym{HLR}{short =HLR,long=Home Location Register}
\DeclareAcronym{VLR}{short =VLR,long=Visitor Location Register}
\DeclareAcronym{M2M}{short =M2M,long=Machine-to-Machine}
\DeclareAcronym{OFDM}{short =OFDM,long=Orthogonal Frequency Division Multiplexing}
\DeclareAcronym{eMBMS}{short =eMBMS,long=enhanced Multimedia Broadcast Multicast Service}
\DeclareAcronym{UMTS-FDD}{short =UMTS-FDD,long= UMTS–frequency-division duplexing}
\DeclareAcronym{AN}{short =AN,long= Access Network}
\DeclareAcronym{SN}{short =SN,long= Serving Network}
\DeclareAcronym{HN}{short =HN,long= Home Network}
\DeclareAcronym{EPC}{short =EPC,long= Evolved Packet Core}
\DeclareAcronym{ePDG}{short =ePDG,long= Evolved Packet Data Gateway}
\DeclareAcronym{HSS}{short =HSS,long=Home Subscriber Server}
\DeclareAcronym{SEAF}{short =SEAF,long=Security Anchor Function}
\DeclareAcronym{AUSF}{short =AUSF,long=Authentication Server Function}
\DeclareAcronym{ARPF}{short =ARPF,long=Authentication Credential Repository and Processing Function}
\DeclareAcronym{SUCI}{short =SUCI,long=Subscription Concealed Identifier}
\DeclareAcronym{SUPI}{short =SUPI,long=Subscription Permanent Identifier}
\DeclareAcronym{LI}{short =LI,long=Lawful Interception}
\DeclareAcronym{SIDF}{short =SIDF,long=Subscription Identifier De-concealing Function}
\DeclareAcronym{GDPR}{short =GDPR,long=General Data Protection Regulation}
\DeclareAcronym{BTS}{short =BTS,long=Base Stations}
\DeclareAcronym{CU}{short =CU,long=Centralized Unit}
\DeclareAcronym{DU}{short =DU,long=Distributed Unit}
\DeclareAcronym{BSC}{short =BSC,long=Base Station Controller}
\DeclareAcronym{SGSN}{short =SGSN,long=Serving GPRS Support Node}
\DeclareAcronym{GMSC}{short =GMSC,long=Gateway Mobile Switching Centre}
\DeclareAcronym{GGSN}{short =GGSN,long=Gateway GPRS Support Node}
\DeclareAcronym{RNC}{short =RNC,long=Radio Network Controller}
\DeclareAcronym{PCRF}{short =PCRF,long=Policy and Charging Rules Function}

\DeclareAcronym{NaaS}{short =NaaS,long=Network as a service}
\DeclareAcronym{CP}{short =CP,long=Control Plane}
\DeclareAcronym{vCU}{short =vCU,long=virtualized Central Unit}
\DeclareAcronym{LLF}{short =LLF,long=Lower-Level Function}
\DeclareAcronym{QKD}{short =QKD,long=Quantum Key Distribution}
\DeclareAcronym{GUTI}{short =GUTI,long=Global Unique Temporary Identifier}
\DeclareAcronym{NOMA}{short =NOMA,long=Non-Orthogonal Multiple Access}
\DeclareAcronym{IRS}{short =IRS,long=Intelligent Reflecting Surfaces}
\DeclareAcronym{MTD}{short =MTD,long=Moving Target Defense}
\DeclareAcronym{LDPC}{short =LDPC,long=Low-Density Parity-check Code}
\DeclareAcronym{PET}{short =PET,long=Privacy Enhancing Technologies}
\DeclareAcronym{SD-WAN}{short =SD-WAN,long=Software-Defined Wide Area Network}
\DeclareAcronym{SD-LAN}{short =SD-LAN,long=Software-Defined Local Area Network}
\DeclareAcronym{CSI}{short =CSI,long=Channel State Information}
\DeclareAcronym{SA}{short =SA,long=Standalone}
\DeclareAcronym{NSA}{short =NSA,long=Non-Standalone}
\DeclareAcronym{VNF}{short =VNF,long=Virtualized Network Function}
\DeclareAcronym{VM}{short =VM,long=Virtual Machine}
\DeclareAcronym{vRAN}{short =vRAN,long=Virtualized Radio Access Network}
\DeclareAcronym{C-RAN}{short =C-RAN,long=Cloud RAN}
\DeclareAcronym{TLS}{short =TLS,long=Transport Layer Security}
\DeclareAcronym{EAP}{short =EAP,long=Extensible Authentication Protocol}

%% file: Section_I_Introduction.tex
6G is the sixth generation standard for cellular communications that are currently under development to succeed 5G. 6G offers an ambitious vision of truly autonomous networks that will be commercially deployed someday in the 2030s \cite{DOCOMO2020}. 6G will be able to support speeds of over 1Tbps, 50 times faster than 5G, while latency is projected at 10-100$\mu$s \cite{DOCOMO2020}. Researchers expect that this standard will expand connectivity for both conventional coverage areas in 5G and space-air-ground-sea applications. The coverage and network capability will enable a wide range of digital services such as wearable displays, implantable devices, telepresence applications (rendering of 3D holographic representation of each participant in a meeting), mixed reality, tactile Internet \cite{Saad2020,5GIA2020}, and autonomous driving\cite{Zhang2020,Tang2020}.  With the substantial increase of coverage and network heterogeneity, there are severe concerns that 6G security and privacy can be worse than the previous generations. For example, the involvement of connected devices in every aspect of humans (e.g., implants/cyborgs) poses serious concerns of potential leaks of personal information (e.g., health records). Potential loss from security attacks could be irrecoverable, not only about finance or personal reputation as currently but also about life (e.g., fatal crash because of attacks into autonomous driving). Further, the achievements of artificial intelligence can be abused for massive online surveillance\cite{Sun2020}. By contrast, novel technologies such as quantum-safe communications and distributed ledgers promise to significantly improve 6G security and privacy. Many believe that robust security and enhanced privacy technologies will be key provisions to the success of 6G.

\begin{figure}[!ht]
   \centering
   \begin{center}
			\includegraphics[width=1\linewidth]{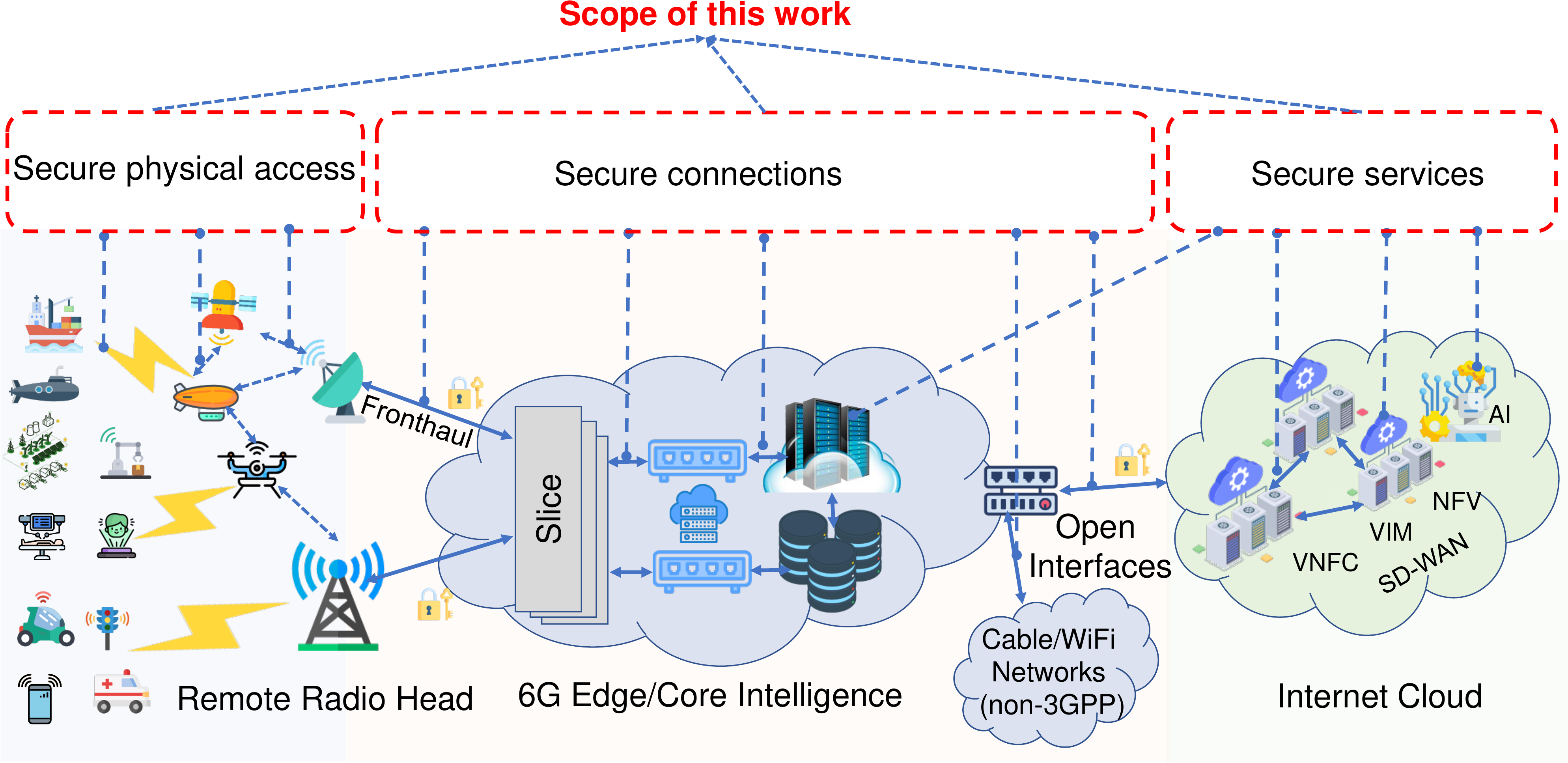}
	\end{center}
	
    \caption{A concept of generic mobile networks for 6G and the scope of this work. Security and privacy preservation technologies in the physical layer, connection layer, and service layer are pillars of 6G networks.}
    \label{fig:intro-security-story}
\end{figure}

\begin{figure*}[!ht]
   \centering
   \begin{center}
			\includegraphics[width=1\linewidth]{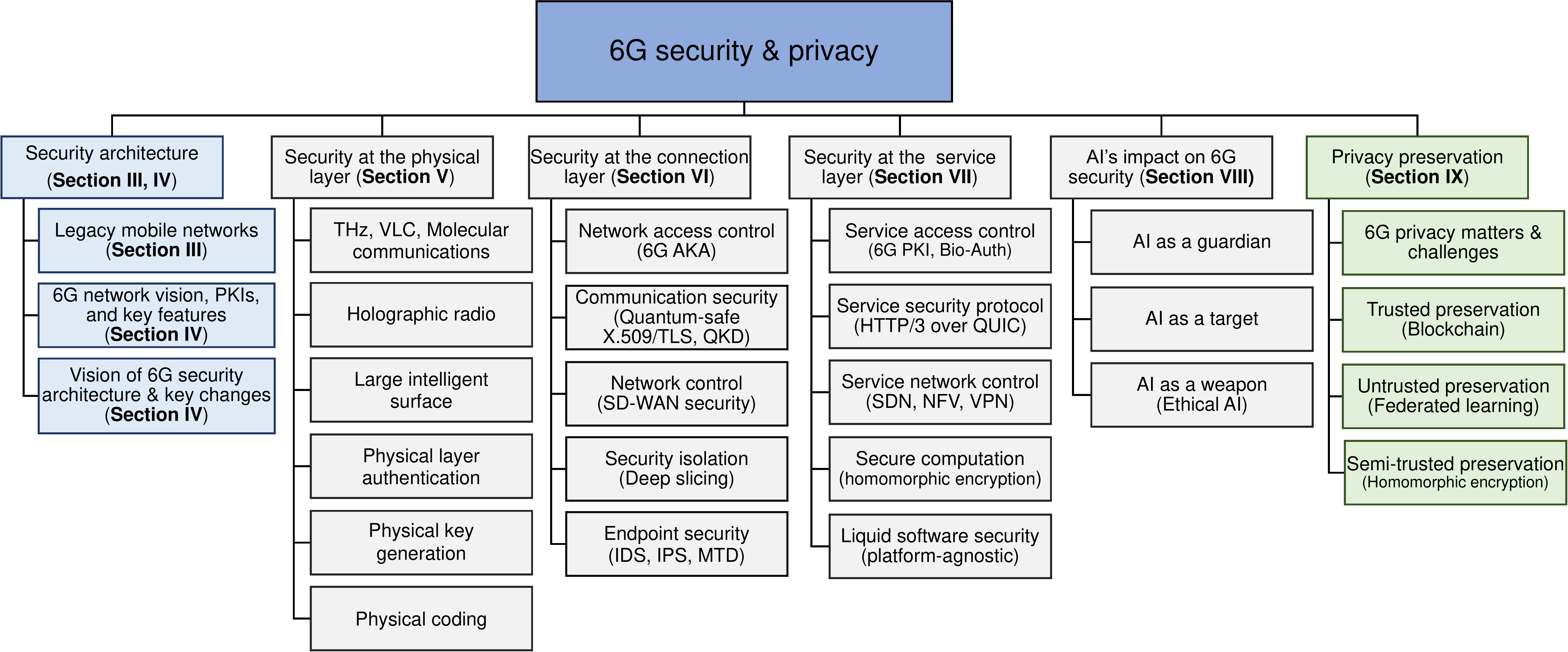}
	\end{center}
	
    \caption{A taxonomy of key points of our survey on security \& privacy for 6G: security architecture, specific security technologies in the three layers, privacy enhancing models, and AI's impact on those fields.}
    \label{fig:survey-structure}
\end{figure*}

However, research on security and privacy issues for 6G networks is still at an early stage. Existing studies on such issues are mostly for the \ac{IoT} networks \cite{Ferrag2019}, 5G networks \cite{3GPP33501,NIST2020,Ahmad19}, quick excursions into specific technologies, such as \ac{AIML} \cite{Sun2020,Gui2020}, or fragmented ideas in generic surveys about 6G concepts \cite{Yang2019}. Lack of related work is because many fundamental components of 6G networks remain largely undefined. However, the evolution of prior networks (from 4G to 5G) indicates that every network generation tends to stay in use for years. For capital expenditure and payback optimization, many network providers favor rolling out a new network generation while still enhancing the technologies on existing infrastructure \cite{DSS}. Moreover, backward compatibility is necessary to maintain connectivity for deployed devices. As the basis of inheritance, 6G security and privacy will presumably adopt many features from 5G.

Inspired by security evolution in prior generations, this work provides a systematic review of existing research efforts on security and privacy for 6G networks. The survey reviews the issues of 6G enabling technologies and state-of-the-art defense methods. Also, we adopt the three-layer architecture in 6G-enabled IoT networks \cite{DaiNing19,GuoXian21}, including physical (perception) layer, connection (network) layer, and service (application) layer, to classify the attacks and corresponding solutions. This layer-based review approach can particularly benefit the operators and developers of interest in preventing the specific attacks on the fundamental protocols that impact many 6G applications. By conducting the problems in each technology, our goal is to provide a holistic view of the evolution of core security and privacy issues, along with the remaining challenges for further enhancements. Figure~\ref{fig:intro-security-story} illustrates the scope of this work. The lessons learned from the survey indicate that 6G networks will need significant upgrades of security protection and data privacy preservation. To this end, the study aims to answer the fundamental question: What are the major potential changes of 6G security infrastructure from the prior generations? What are new challenges and prospective approaches for privacy preservation in 6G to satisfy the requirements in laws, such as \ac{GDPR}?


\subsection{Review methodology}
   
   Rather than using all brand-new technologies, 6G security and privacy will continue the trajectory of many enabling technologies from prior generations because of costly deployment. Inspired by this inheritance, we outline potential changes of 6G security and privacy, seen through the mirrors of the lessons learned from two aspects. The first is to fix remaining security issues and enhance privacy preservation in security architecture and several legacy technologies, such as \ac{SDN}. The other is to address potential security and privacy issues in futuristic technologies which will highly impact 6G, such as THz communications. As illustrated in Figure~\ref{fig:survey-structure}, this work covers discussions on every aspect, from security architecture to specific technologies in each layer, or factors that impact 6G security and privacy.  

   Security attacks are often the main motivations that drive how security systems in the next generation should be changed to counter the exploit of old vulnerabilities. Such attacks often reveal weaknesses of system or protocol flaws that the creators may never have thought of at the design stage. Learning from such attacks is a straightforward approach to get to know how different the security enhancements of mobile networks are -- sometimes for fixing known vulnerabilities exploited by attacks in prior generations. For example, authentication protocol flaws revealed by security attacks often require upgrades of core network functions to fix. However, most vendors are opposed to such a replacement because of the high cost. These not-yet-been-fixed issues thus become potential targets for upgrades in 6G, or are a basis for future research. 
   

\subsection{Contributions}

In summary, the main contributions of this article are as follows. First, the work provides a systematic overview of the evolution of security architecture and vulnerabilities in legacy networks. By investigating the shortcomings of the standards and technical insights of protocol flaws in such networks, required enhancements to 6G security and privacy are highlighted. Second, our survey provides a holistic view of security and privacy issues and how the existing solutions must be changed to satisfy the new demands in 6G. Since 6G will continue on the techno-economic trajectory of 5G, a systematic review on transition and possible changes of 6G security and privacy can shed light on the best plan for the operators/developers to upgrade the security infrastructure/defense systems at the right time. Finally, our discussions about lessons learned from the shortcomings of existing security architecture and remaining technical challenges may help researchers/developers quickly identify relevant issues and starting points for further works. To the best of our knowledge, this survey is the first attempt to provide a thorough review of security and privacy for 6G from security architecture, specific technologies in the layers, to AI's impact on those fields.  

\subsection{Structure of the paper}

The rest of this paper is organized as follows. Section~\ref{sec:related-work} presents related work and point of departure of our survey. Section~\ref{sec:overview-core-security-architecture} reviews lessons learned from security issues and the evolution of security architecture in legacy networks (from 1G to 5G). Section~\ref{sec:6G-security-architecture} describes 6G networks and our vision about potential changes in its security architecture. The problems and prospective technologies of 6G security in the \textit{physical layer}, \textit{connection layer}, and \textit{service layer} are then discussed in Section~\ref{sec:security-physical-layer}, Section~\ref{sec:security-connection-layer}, and Section~\ref{sec:security-service-layer}, respectively. The impact of \textit{artificial intelligence} on 6G security is detailed in Section~\ref{sec:security-ai}. Privacy concerns and potential solutions in 6G are presented in Section~\ref{sec:privacy-in-6G}. Section~\ref{sec:discussion-future-research} discusses several other aspects of 6G security\& privacy in the coming years. Section~\ref{sec:conclusion} concludes this paper. The key points of our survey are shown in Figure~\ref{fig:survey-structure}. The acronyms used in this work are listed as follows.

\renewcommand{\IEEEiedlistdecl}{\IEEEsetlabelwidth{SONET}}
\printacronyms[sort=true]
\renewcommand{\IEEEiedlistdecl}{\relax}

%% file: Section_II_Related_Work.tex
Security and privacy for 6G are still in the infancy of their development cycle, as are related surveys. Existing studies on security and privacy issues \cite{Ali2020,Ahmad19,Ferrag2019,Farris19} can be categorized into three groups: (1) security and privacy preservation for IoT networks and their subsidiaries, e.g., wireless sensor networks, vehicular networks, (2) security and privacy issues for existing 4G/5G cellular networks, and (3) security and privacy in 6G by analyzing issues around specific key technologies, such as machine learning. 

IoT security and privacy is not a new topic. Several technologies of IoT security and privacy may partially contribute to the 6G field. For example, the authors of \cite{Neshenko19} present security-related challenges and sources of threats in IoT, highlighting various prospective technologies for 6G, such as blockchain. A narrower approach is to address security in specific network types, which may become a reality in 6G, such as vehicular networks \cite{Hasan20} and 6G-enabled IoT networks \cite{GuoXian21,Promwongsa21}. However, discussions of specific attacks in 6G networks in the mirror of IoT security are somewhat limited. Moreover, given little vision of 6G concepts at the time of the studies of \cite{Cara20,Cao20}, many predictions on the security issues for next-generation networks favor referring to conventional threats such as side-channel attacks and \ac{DDoS} attacks. Several field surveys such as security in 6G heterogeneous vehicular networks \cite{HuiLong21}, tactile Internet \cite{Promwongsa21}, security in edge-computing-assisted IoT networks \cite{Alwarafy21} indicate that ultra-high reliability and real-time protection will be key requirements of many applications in future networks. However, 6G communication models and security architecture are not explicitly included in the discussions of the surveys.

\begin{table*}[ht]
\caption{Several related state-of-the-art studies and our survey position}
\label{tab:related-work}
\begin{adjustbox}{width=1\textwidth}
\small
\begin{tabular}{lllllllllll}
\hline

\multirow{2}{*}{\textbf{The paper}} & \multirow{2}{*}{\textbf{Year}} & \multirow{2}{*}{\textbf{Paper type}} & \multirow{2}{*}{\textbf{\begin{tabular}[c]{@{}l@{}}Mobile\\ network \\ generation\end{tabular}}} & \multicolumn{6}{c}{\textbf{Content coverage}} & \textbf{} \\ \cline{5-10}
 &  &   &  &  \textbf{\begin{tabular}[c]{@{}l@{}}Core architecture\end{tabular}} &  \textbf{\begin{tabular}[c]{@{}l@{}}Physical layer\\ security\end{tabular}} &  \textbf{\begin{tabular}[c]{@{}l@{}}Connection layer\\ security\end{tabular}} &  \textbf{\begin{tabular}[c]{@{}l@{}}Service layer\\ security\end{tabular}} &  \textbf{\begin{tabular}[c]{@{}l@{}}AI\\ security\end{tabular}} &  \textbf{\begin{tabular}[c]{@{}l@{}}Privacy\\ preservation\end{tabular}} &  \textbf{Contributions} \\ \hline
Dai et al. \cite{Dai19} & 2019 & Short paper & 5G/B5G & &  &  &  &  $\bigcdot$ & $\bigcdot$ & \begin{tabular}[c]{@{}l@{}} $\blacktriangleright$ Blockchain and deep learning\\ for 5G networks and beyond\end{tabular}  \\
Ahmad et al. \cite{Ahmad19} \textcolor{red}{*} & 2019 & Comprehensive survey & 5G/XG & $\bigcdot$ & & $\bigcdot$ & $\bigcdot$ &  & $\bigcdot$ & \begin{tabular}[c]{@{}l@{}} $\blacktriangleright$Security and privacy in 5G and\\ overview of the changes of\\ security and privacy in the next-\\ generation network (not specify 6G)\end{tabular} \\
Farris et al. \cite{Farris19} & 2019 & Comprehensive survey & 5G/B5G & & & $\bigcdot$ & $\bigcdot$ &  &  & $\blacktriangleright$ SDN/NFV security survey \\
Chica et al. \cite{Chica2020} & 2020 & Comprehensive survey & 5G/B5G & &  &  & $\bigcdot$ &  &  & $\blacktriangleright$ SDN security survey \\
Ranaweera et al. \cite{Ranaweera21} & 2021 & Comprehensive survey & 5G/6G &  &   &  & $\bigcdot$  & &  & \begin{tabular}[c]{@{}l@{}}$\blacktriangleright$Multi-Access Edge security\end{tabular} \\
Wang et al. \cite{Wang2020} & 2020 & Short survey & 6G &  &  & & $\bigcdot$ &  & $\bigcdot$ & \begin{tabular}[c]{@{}l@{}}$\blacktriangleright$An overview of security and \\ privacy in key applications/\\ technologies of 6G networks\end{tabular} \\
Ylianttila et al. \cite{Ylianttila2020} \textcolor{red}{*} & 2020 & White paper & 6G & & $\bigcdot$ & $\bigcdot$ &  &  & $\bigcdot$ & \begin{tabular}[c]{@{}l@{}}$\blacktriangleright$6G research challenges for  trust, \\ security and privacy\end{tabular} \\
Kantola \cite{Kantola2020TrustNF} & 2020 & Short survey & 5G/6G &  & & $\bigcdot$ &  &  & $\bigcdot$ & $\blacktriangleright$Trust networking \\
Sheth et al. \cite{Sheth2020} & 2020 & Comprehensive survey & 6G &  & & &  & $\bigcdot$ &  & \begin{tabular}[c]{@{}l@{}}$\blacktriangleright$A taxonomy of AI techniques \\ for 6G networks\end{tabular} \\
Sun et al. \cite{Sun2020} & 2020 & Comprehensive survey & 6G &  & & &  & $\bigcdot$ & $\bigcdot$ & \begin{tabular}[c]{@{}l@{}}$\blacktriangleright$Machine learning for 6G privacy\end{tabular} \\
Mahmoud et al. \cite{Mahmoud21} & 2021 & Comprehensive survey & 6G & & $\bigcdot$ &  &  &  & $\bigcdot$ & \begin{tabular}[c]{@{}l@{}}$\blacktriangleright$A generic survey about 6G\\ technologies, applications, \\ challenges\end{tabular} \\
Xie et al. \cite{Xie21} & 2021 & Comprehensive survey & 5G/6G  &  & $\bigcdot$ & &  &  &  & $\blacktriangleright$Physical layer authentication \\
Huang et al. \cite{Huang21} & 2021 & Short survey & 5G/6G &  &  & & & $\bigcdot$ &  & \begin{tabular}[c]{@{}l@{}} $\blacktriangleright$Intelligent Reflecting Surface \\ aided pilot contamination attack\\ \& defense\end{tabular} \\
Pirandola et al. \cite{Pirandola20} & 2021 & Comprehensive survey & 6G & & $\bigcdot$ & $\bigcdot$ &  &  &  & \begin{tabular}[c]{@{}l@{}}$\blacktriangleright$Advances in quantum cryptography\end{tabular} \\
Xu et al. \cite{Xu21} & 2021 & Short brief & 5G/6G &  & $\bigcdot$  &  &  & &  & \begin{tabular}[c]{@{}l@{}}$\blacktriangleright$Pilot spoofing attack in \\ Massive MIMO systems\\  \& countermeasures\end{tabular} \\
Porambage et al. \cite{Porambage21} \textcolor{red}{*} & 2021 & Comprehensive survey & 6G &  &   &  & $\bigcdot$  & &  $\bigcdot$& \begin{tabular}[c]{@{}l@{}}$\blacktriangleright$Security and privacy briefs for \\ 6G applications\end{tabular} \\
\textbf{Our survey} & 2021 & Comprehensive survey & B5G/6G & $\bigcdot$ & $\bigcdot$ & $\bigcdot$ & $\bigcdot$& $\bigcdot$ & $\bigcdot$ & \begin{tabular}[c]{@{}l@{}}$\blacktriangleright$A comprehensive security survey \\ on the prospective technologies \\ and challenges for 6G on all layers \\  and core architecture \end{tabular} \\ \hline
\end{tabular}
\end{adjustbox}
\begin{tablenotes}
	\item \textcolor{red}{*} The studies are closely related to our work.
\end{tablenotes}
\end{table*}

Security and privacy topics for cellular networks have attracted much attention because of the popularity of mobile devices in our daily lives. Most current security and privacy surveys of cellular networks are for 5G. For example, the authors of \cite{Ahmad19,Ferrag2019} present various aspects of 5G security architecture standards \cite{3GPP33501,NIST2020,Ericsson2020,Huawei2019,Rupprecht18} relevant to security attacks and privacy problems in multiple layers, such as core/backhaul networks. The limitations and security challenges of specific technologies such as \ac{SDN} \cite{Chica2020}, programmable networking, and \ac{NFV} \cite{Farris19,ETSI013}, particularly for \ac{RAN} \cite{Parvez18} and \ac{MEC} \cite{Alwarafy20,Ranaweera21} -- which promise to play the key roles in 6G -- are hot topics and have been partially addressed. The open discussions and technical materials from these 5G security and privacy surveys \cite{Ahmad19,Farris19,Chica2020} and specific technologies \cite{Hamamreh19,Dai19} are important sources for our review of the legacy technologies that impact 6G security and privacy. The overview of several state-of-the-art studies in this field is listed as the top five entries in Table~\ref{tab:related-work}. 

Exploring new technologies for 6G security and privacy started to gain traction recently. The overviews of the studies in this direction are listed in the last twelve entries of Table~\ref{tab:related-work}. Closely related to our work, Wang et al. \cite{Wang2020} sketch a quick overview of the vision of 6G security and privacy through the mirror of emerging applications, e.g., wireless brain-computer interactions and multi-sensory XR applications. However, that study lacks a relevant technical discussion on how such security architecture will evolve or any detail of AI's progress in enhancing security. Ylianttila et al. \cite{Ylianttila2020} highlight several key technologies of 6G security and privacy along with their challenges that remain, but no discussion on connection and service layer security. On the other hand, many authors have carried out surveys on narrower issues, such as physical layer security \cite{Xie21,Huang21,Xu21}, quantum-safe security technologies \cite{Pirandola20,Wright21}, AI-driven security \cite{Sun2020,Sheth2020}, trusted networks \cite{Kantola2020TrustNF}, which are supposed to be the top priorities in 6G. Recently, the authors in \cite{Porambage21} highlight possible security and privacy challenges in different 6G technologies and applications. However, there is no comprehensive survey to provide a holistic view of 6G security and privacy issues in the context of overall security architecture, protection solutions in the fundamental communication technologies, or how they evolve from the legacy networks to satisfy the new demands in 6G applications. The knowledge on security transition and the feasibility of prospective technologies can shed light on the best plan for the operators/service providers/developers to upgrade their security infrastructure/defense systems at the right time.

%% file: Section_III_Security_legacy.tex
A preferable way to comprehensively understand 6G security and privacy is to look at lessons learned from would-be failures of the current security architectures and legacy technologies if applied to satisfy 6G requirements. This section presents our assessment of security attacks and privacy issues for various generations of cellular networks. We then highlight lessons learned from the prior security transitions to envisage potential changes and enhancements for security and privacy preservation features in 6G.

\subsection{1G,2G,3G - Security issues of phased-out and phasing-out networks}

There are many memorable milestones with the development of the first three generations of mobile networks, particularly in security. The first generation (1G) of mobile networks provided neither security nor privacy. Launched in the 1980s and 1990s, respectively, the second generation (2G) and the third-generation (3G) played a critical role in completely transforming the era of analog phone services (1G) to IP-based networks (3G). Although many operators around the world, particularly in developing countries, still offer 2G and 3G services, both networks are scheduled to be entirely switched off in the next five years \cite{2GPhaseOut}. 2G and 3G gave many valuable lessons of how security issues can be exploited by attackers. For example, the most infamous attack on 2G and 3G was \ac{IMSI}-catcher\cite{IMSICatcher,Dabrowski14}, where the attacker exploited unencrypted identity information during authentication and paging procedures to track mobile subscribers. Many law enforcement and intelligence agencies in some countries still use IMSI-based tracking to follow crimes\cite{IMSICatcher}. On the other hand, absence of end-to-end encryption in communications was the root cause of many eavesdropping attacks such as man-in-the-middle, phone fraud, and SMS interception \cite{Cattaneo13}. Downgrade attacks \cite{Broek15,Dabrowski14} were also a headache in the dual-network infrastructure of 2G and 3G. In such an attack, an attacker would force a victim to connect to 2G networks which do not require mutual authentication. After downgrading successfully, the attacker could launch another man-in-the-middle (MITM) attack \cite{Meyer2004} and freely collect the IMSI of UE for further location tracking. Many security attacks, which appeared for the first time in 2G and 3G, such as signalling DoS attacks \cite{Mavoungou16} and energy depletion attacks \cite{Nguyen2019}, have not yet been resolved.

\subsection{4G, 5G - Security issues and enhancement of operating networks in the next five years}

 Launched from 2009, the \ac{LTE}-Advanced standard (official technology of 4G in ITU requirements \cite{3GPPRel10}) has evolved strongly over the years, resulting in the most widely deployed network \cite{LTE5GDeployment}. Compared to 3G, 4G security has been enhanced significantly. For example, 4G/LTE \cite{TS33.401} Evolved Packet System-Authentication and Key Agreement (EPS-AKA) included a series of security enhancements for interconnection between 3GPP and non-3GPP networks. A new cipher and integrity checking mechanism \cite{TS33.401} was also introduced to protect the signalling data between UEs and the core network. With EPS-related cryptographic key involvement, a UE can verify the \ac{SN}'s identity. Tunnel encryption was first proposed with the support of IPSec. Despite many upgrades of security features, the 4G/LTE security architecture has several weaknesses. First, 4G is not immune to DDoS attacks, which can be launched from malicious mobile applications to overload the Home Subscriber Server (HSS) and Mobility Management Entity (MME) servers with multiple authentication requests. Such overloading potentially blocks the access of legitimate subscribers to the network \cite{Paolini12}. Although IMSI/\ac{GUTI}, a temporary identifier, is used to hide a subscriber's long-term identity, researchers found that TMSI/GUTI allocation frequency is predictable \cite{Hong18} and can be used for tracking the location of any subscriber. 

 The authentication decision model of a home network to consult a serving network during UE authentication in 4G also has many security flaws. Because the decision is made solely by the serving network, a well-organized attacker can create fake serving networks to track subscribers \cite{Hussain18,HussainNDSS19}. Another big vulnerability lies in the Voice over LTE (VoLTE) service \cite{LiVoLTE15,Kim15}, which uses packet-based LTE networks and IP protocol to establish voice and media calls. The authors of \cite{Rupprecht20} found the problem of keystream reuse in VoLTE -- the packets of the first call are encrypted with the same keystream as those of the second call -- being exploited to decrypt and access the contents of a recorded target call. Criminals use VoLTE to spoof a caller, launch denial-of-service attacks, subdue voice calls, and strip the victim's mobile account of money \cite{Kim15}.

\begin{table*}[ht]
\caption{Security issues of the networks from 1G to 6G}
\label{tab:security-architecture-vulnerabilities}
\begin{adjustbox}{width=1\textwidth}
\begin{tabular}{clllllllllllll}
\hline
\rowcolor[HTML]{EFEFEF} 
\cellcolor[HTML]{EFEFEF} & \cellcolor[HTML]{EFEFEF} & \cellcolor[HTML]{EFEFEF} & \cellcolor[HTML]{EFEFEF} & \multicolumn{6}{c}{\cellcolor[HTML]{EFEFEF}\textbf{Affected Networks}} & \cellcolor[HTML]{EFEFEF} & \cellcolor[HTML]{EFEFEF} & \cellcolor[HTML]{EFEFEF} & \cellcolor[HTML]{EFEFEF} \\
\rowcolor[HTML]{EFEFEF} 
\multirow{-2}{*}{\cellcolor[HTML]{EFEFEF}\textbf{Target}} & \multirow{-2}{*}{\cellcolor[HTML]{EFEFEF}\textbf{\begin{tabular}[c]{@{}l@{}}Attack name\end{tabular}}} & \multirow{-2}{*}{\cellcolor[HTML]{EFEFEF}\textbf{\begin{tabular}[c]{@{}l@{}}Time exposed\\  /authors\end{tabular}}} & \multirow{-2}{*}{\cellcolor[HTML]{EFEFEF}\textbf{\begin{tabular}[c]{@{}l@{}}Method to \\ uncover\end{tabular}}} & 1G & 2G & 3G & 4G & 5G &6G & \multirow{-2}{*}{\cellcolor[HTML]{EFEFEF}\textbf{\begin{tabular}[c]{@{}l@{}}Protocol \\ flaws, bugs\end{tabular}}} & \multirow{-2}{*}{\cellcolor[HTML]{EFEFEF}\textbf{\begin{tabular}[c]{@{}l@{}}Risk \\ level\end{tabular}}} & \multirow{-2}{*}{\cellcolor[HTML]{EFEFEF}\textbf{Consequence}} & \multirow{-2}{*}{\cellcolor[HTML]{EFEFEF}\textbf{\begin{tabular}[c]{@{}l@{}}Status of \\ attack fix \textcolor{red}{*}\end{tabular}}} \\ \hline

  & \begin{tabular}[c]{@{}l@{}}Signalling DoS \\ attacks\end{tabular} & \begin{tabular}[c]{@{}l@{}} $\blacktriangleright$ 2007: \cite{Lee007}\\ 2012: \cite{Paolini12}\\ 2016: \cite{Mavoungou16}\end{tabular} & Signalling storm &  & $\bigcdot$ & $\bigcdot$ & $\bigcdot$ & $\bigcdot$ & $\bigcdot$ & \begin{tabular}[c]{@{}l@{}}Limited radio \\ resources\end{tabular} & High & Block user access & No \\

  & \begin{tabular}[c]{@{}l@{}}Paging DoS \\ attacks\end{tabular} & $\blacktriangleright$ 2013: \cite{Golde13} & \begin{tabular}[c]{@{}l@{}}Hijack paging \\ procedure\end{tabular} &  & $\bigcdot$ &  &  &  &  & \begin{tabular}[c]{@{}l@{}}Paging procedure \\
bugs\end{tabular}  & High & Block user access & \begin{tabular}[c]{@{}l@{}}Partially\\ fixed\end{tabular} \\

  & \begin{tabular}[c]{@{}l@{}}DDoS\\ Authentication \\ server\end{tabular} & $\blacktriangleright$ 2009: \cite{Traynor009} & \begin{tabular}[c]{@{}l@{}}Use phone \\ botnets\end{tabular} &  & $\bigcdot$ & $\bigcdot$ &  &  & $\bigcdot$ &  \begin{tabular}[c]{@{}l@{}}Core networks\\ connect IP\\ services\end{tabular} & High & Block user access & No \\

  & \begin{tabular}[c]{@{}l@{}}SMS saturation \\ attacks\end{tabular} & $\blacktriangleright$ 2009: \cite{Traynor09} & \begin{tabular}[c]{@{}l@{}}Send massive SMS\\ to block voice\\ calls\end{tabular} &  & $\bigcdot$ & $\bigcdot$ &  & & & \begin{tabular}[c]{@{}l@{}}Text messages use\\ the same control \\ channels as voice \\ calls\end{tabular} & High & \begin{tabular}[c]{@{}l@{}}Block voice \\ communications\end{tabular} & \begin{tabular}[c]{@{}l@{}}Partially\\ fixed\end{tabular} \\
 
\multirow{-5}{*}[45pt]{ \begin{tabular}[c]{@{}l@{}}Availability\end{tabular}} & \begin{tabular}[c]{@{}l@{}}Energy depletion \\ attacks\end{tabular} & $\blacktriangleright$ 2019: \cite{Nguyen2019} & \begin{tabular}[c]{@{}l@{}}Send random false \\ authentication\\ messages,\end{tabular} &  &  & $\bigcdot$ & $\bigcdot$ & $\bigcdot$ & $\bigcdot$ & \begin{tabular}[c]{@{}l@{}}Weak\\ authentication\end{tabular} & High & \begin{tabular}[c]{@{}l@{}}Disable IoT devices,\\ shutdown low-power\\ sensor networks\end{tabular} & No \\ \hline

  & Cloning attacks & $\blacktriangleright$ 1995: \cite{Chandra08} & \begin{tabular}[c]{@{}l@{}}Clone the \\ victim's ESN, \\ MDN\end{tabular} & $\bigcdot$ &  &  &  & &  & \begin{tabular}[c]{@{}l@{}}No protection\\ for identity\end{tabular} & High & Phone fraud & Yes \\

  & \begin{tabular}[c]{@{}l@{}}SIM card \\ rooting\end{tabular} & $\blacktriangleright$ 2013: \cite{Nohl13} & Exploit sloppy encryption & & $\bigcdot$ & $\bigcdot$ & $\bigcdot$ &  & & \begin{tabular}[c]{@{}l@{}} Implementation \\ flaws\end{tabular} & High & Clone SIM card & \begin{tabular}[c]{@{}l@{}}Partially \\ fixed\end{tabular} \\

 & Partitioning attacks & $\blacktriangleright$ 2002: \cite{Rao2002} & \begin{tabular}[c]{@{}l@{}} Use side-channel \\ attacks\end{tabular} &  & $\bigcdot$ & $\bigcdot$ &  & &  & \begin{tabular}[c]{@{}l@{}}Insecure wireless \\ channels\end{tabular} & High & Clone SIM card & Yes \\

  & Impersonation attacks & $\blacktriangleright$ 2020: \cite{RupprechtNDSS2020} & \begin{tabular}[c]{@{}l@{}} Exploit no integrity \\ protection of \\ the user plane\end{tabular} &  &  & $\bigcdot$ & $\bigcdot$  & $\bigcdot$ & $\bigcdot$ & \begin{tabular}[c]{@{}l@{}}Authentication\\ protocol bugs\end{tabular}  & High & Impersonate a user & No \\

\multirow{-5}{*}[25pt]{ \begin{tabular}[c]{@{}l@{}} Integrity\end{tabular}} & Voice IP attacks & \begin{tabular}[c]{@{}l@{}} $\blacktriangleright$ 2015: \cite{Kim15}\\ 2020: \cite{Rupprecht20}\end{tabular} & \begin{tabular}[c]{@{}l@{}} Exploit SIP leaked \\ session info,  hidden\\ data channels, \\ keystream reuse \\ in subsequent calls\end{tabular} &  &  &  & $\bigcdot$ & $\bigcdot$ & $\bigcdot$ & \begin{tabular}[c]{@{}l@{}}IP-based service\\ vulnerability\end{tabular} & High & \begin{tabular}[c]{@{}l@{}}Caller spoofing,\\ service interruption,\\ subdue voice calls, \\ over-billing\end{tabular} & \begin{tabular}[c]{@{}l@{}}Partially\\ fixed\end{tabular} \\ \hline

\multicolumn{1}{l}{ } & SMS interception & $\blacktriangleright$ 2013: \cite{Cattaneo13} & Reverse engineering &  & $\bigcdot$ &  &  &  & & \begin{tabular}[c]{@{}l@{}}Authentication\\ protocol bugs\end{tabular} & Medium & User tracking & Yes \\

\multicolumn{1}{l}{ } & IMSI-catcher & \begin{tabular}[c]{@{}l@{}} 2013: \cite{Cattaneo13}\\ 2014:\cite{Dabrowski14}\\ 2015: \cite{ShaikBANS15}\end{tabular} & \begin{tabular}[c]{@{}l@{}}Use downgrade attacks, \\ unencrypted\\ paging information\end{tabular} &  & $\bigcdot$ & $\bigcdot$ &  & & - & \begin{tabular}[c]{@{}l@{}}Authentication\\ protocol bugs\end{tabular} & High & User tracking & \begin{tabular}[c]{@{}l@{}}Partially\\ fixed\end{tabular} \\

\multicolumn{1}{c}{\multirow{-3}{*}[25pt]{ \begin{tabular}[c]{@{}l@{}}Confidentiality\end{tabular}}} & Traceability attack & \begin{tabular}[c]{@{}l@{}} $\blacktriangleright$2015: \cite{ShaikBANS15,  Mavoungou16}\\ 2018: \cite{Hussain18}\\ 2019: \cite{HussainNDSS19}\end{tabular} & \begin{tabular}[c]{@{}l@{}}Exploit information from \\ failure messages,\\ paging errors\end{tabular} &  &  & $\bigcdot$ & $\bigcdot$ & $\bigcdot$ & $\bigcdot$ & \begin{tabular}[c]{@{}l@{}}Authentication\\ protocol bugs\end{tabular} & High & User tracking & No \\ \hline

\end{tabular}
\end{adjustbox}
\begin{tablenotes}
	\item \textcolor{red}{*}If the status of attack fix is \textit{yes}, it means the attacker cannot use that approach to launch a similar attack on the latest standard.
\end{tablenotes}
\end{table*}

 \textit{Featured upgrades in 5G security}
 
 5G has been upgraded significantly in terms of both security architecture and authentication protocols to satisfy a service-oriented network model as well as fixing many vulnerabilities in 4G. According to the latest specifications \cite{3GPP33501}, other than the five security features in 4G networks, 5G adds a new domain: service-based architecture (SBA) security. 
 5G is also the first standard to have its authentication architecture as a unified framework. This platform supports both 3GGP and non-3GPP access networks, e.g., Wi-Fi and cable networks. With a unified platform, 5G enables \textit{one authentication execution}, in which a UE can be authenticated in a 3GPP access network, and then move to other non-3GPP network without the need for reauthentication \cite{Huawei2019}. 
 
 5G protects UE identities better than 4G. 5G specifically designs \ac{SUCI}, an encrypted form of the \ac{SUPI}, to conceal the subscriber's real information in the authentication stage \cite{Verizon5GPrivacy}. With the enhancement, a UE's permanent identifier, e.g., the IMSI, will not be sent over 5G networks in plaintext. This feature is a major security update over earlier network generations. 5G also made its first attempt to support lawful interceptions. For example, in some special cases, e.g., the court issues subpoena for investigating a crime, the operators can provide lawful interception services to the authorized law enforcement agents. 5G enables two new additional authentication methods: EAP-AKA' and EAP-TLS. EAP-AKA' has the same mission and security capabilities as 5G-AKA but for the difference in the message format and the role of entities \cite{3GPP33501}. EAP-TLS, as defined in RFC 5216, is designed for subscriber authentication in IoT or private networks. Many potential non-USIM devices, such as laptop or IoT devices, now are able to subscribe and access to the 5G core by using EAP-TLS, which was impossible in earlier generations.

  \textit{Security vulnerabilities of 5G}  

   Because of its complexity, 5G has security weaknesses. First, the 5G-AKA protocol fails to meet several goals that it is expected to have. For example, the agreement between subscribers and serving networks is weak \cite{Basin18} because of the lack of a binding assumption on the channel between the serving network and the home network. This vulnerability could allow an attacker to transfer the network bill to someone else for his access on a serving network. Although 5G-AKA can defeat IMSI-catcher attacks \cite{ShaikBANS15,IMSICatcher}, researchers of \cite{Basin18} found that user tracking is still possible in 5G by observing synchronization failure messages over time. In another work \cite{HussainNDSS19}, the authors propose that a seemly harmless service, like paging, can be exploited to locate a user with fewer than 10 calls. Finally, the issue of using a rogue base station to fool a UE into disclosing its SUPI, e.g., by leveraging a spoofed pre-authentication message, has not yet been fixed in 5G\cite{Jover19}.

\subsection{Network deployment strategies' impact on security architecture and security transition}
\label{sub:5G-sa-nsa-strategy}

  Because of cost, operators may select either of two deployment strategies: \ac{SA} vs. \ac{NSA}. NSA provides control signalling of a new standard to the base stations of older standards, whereas in SA, the base stations of a new standard are directly connected to the core network without an intermediate carry of the old infrastructure. Deployment in an NSA strategy has two benefits: (1) much lower cost than SA and (2) reusing existing facilities. By contrast, deployment of the SA strategy requires high CAPEX but can provide services with the full capacity of the new standard. Most operators may prioritize the deployment of NSA to rapidly bring new technology to the market and monetize their investment gradually rather than move all at once.

  From a security perspective, the selection of either deployment strategy has particular impacts. Unlike SA-deployed networks with the full benefits of native security in a new standard, operators have to support a transition procedure which can be a potential security risk. To support NSA in 5G, operators must deploy EUTRA-NR Dual Connectivity, where 4G-LTE is the master radio access technology and 5G-NR serves as secondary radio access technology with UEs connected to both radios \cite{NIST2020}.  However, the use of confidentiality protection is optional \cite{3GPP33501} in the dual authentication and may open the door for potential exploitation if not set up correctly.

\subsection{Lessons learned from the security issues and enhancement from 1G to 5G for 6G security}

    Every network generation has its shortcomings. Although many methods have been developed to mitigate exploitation, several vulnerabilities remain as a result of the complexity of replacing core protocols. Table~\ref{tab:security-architecture-vulnerabilities} summarizes several \textit{known} security attacks and privacy violations against the security architectures and core authentication protocols across the prior generations, in conjunction with their statuses of being fixed. The attacks are categorized by the security principles (confidentiality, integrity, availability). Following the summary, signalling DoS, DDoS against authentication servers, energy depletion attacks, and user tracking are four of the many attacks that will continue to be a headache with 6G security architecture and applications. This stems from the fact that the flaws in the underlying protocol designs (e.g., weak authentication) and the limitations of radio resources are generic issues of all network generations, and to fix them perfectly is a challenge. In summary, four key lessons learned from the security issues and enhancement of legacy networks are as follows.

    \begin{enumerate}
       
        \item \textit{New applications are often sources of security threats which in turn call for security enhancement}. New applications offer brand features that highlight the enhancements of new network standards compared to those of earlier generations. However, they potentially result in new vulnerabilities. For example, the VoLTE protocol contains a security problem of keystream reuse in two subsequent calls. Exploiting this vulnerability, an attacker can decrypt the contents of an encrypted VoLTE call and eavesdrop on phone calls \cite{Rupprecht20,LiVoLTE15}. Many studies \cite{Ylianttila2020,Saad2020,6GSumit2019,5GIA2020} forecast that 6G will host a wave of new applications such as mixed reality and autonomous driving, which will probably also be susceptible to impersonation and DoS attacks. Enhancing the security capabilities of the technologies, before they go into operation, is thus an important issue.
        
        \item \textit{Supporting a legacy protocol in a new protocol deployment could expose old vulnerabilities}. The root causes are the challenge of protecting devices on old network parts with weak security capabilities and the incompatibility of core security functions of two different network standards. To address the incompatibility, the new standard must often switch to ask the old architecture to authenticate old devices. This access control model potentially exposes old vulnerabilities in an earlier standard. For example, an attacker can launch a downgrade attack \cite{Broek15,Dabrowski14} to force 4G-LTE devices to connect to 2G/3G networks. Then, by exploiting the vulnerabilities of no mutual verification between the UE and authentication servers in the 2G/3G standards, the attacker can freely collect the IMSI of UE and track UE location. If 6G supports legacy 4G/5G devices, security issues to guarantee the compatibility for dual network access authentication and identity management should be considered seriously.
        
        \item \textit{Fewer changes on protocol designs but more changes on protocol implementations are good to introduce fewer new vulnerabilities, but fixing more existing vulnerabilities faster}. This starts from the fact that fixing security architecture and protocol flaws, such as in AKA and subscriber identity management, often requires large-scale core equipment upgrades, even to end-user devices. The change can lead to a burden in finance, which many operators and subscribers may not be ready to accept. A new architecture and protocol design also needs substantial time to have its security capabilities verified, so as to avoid new vulnerabilities introduced in a real environment. A workable schedule is to prioritize security patches for protocol implementation or updates for intrusion prevention systems at the endpoints that can mitigate the impact of existing vulnerabilities. However, in the long term, a design upgrade that thoroughly eliminates the flaws and weaknesses of the old architecture is still critical. 
        
        \item \textit{Mutual authentication and end-to-end encryption are still a challenge and the subject to demand a breakthrough}. Absence of these two features is a major source of many notorious attacks such as fake operators, eavesdropping, and traceability attacks. Even 5G is likely to fail to meet these security goals, since implementing these two features faces challenges of high computation and communication overload. Without a breakthrough in processing capacity and management models, a mandate of strong end-to-end encryption and mutual authentication in 6G can impact on many latency-sensitive services. Any delay to have such features in 6G can practically sink the hope of thoroughly fixing the existing security issues.   
  
   \end{enumerate}

%% file: Section_IV_6G_Core_security.tex
Each new generation of cellular networks always attempts to define or upgrade at least one of the security architecture components, such as new authentication and key management, to address challenges from new applications and business models. This section gives an overview of 6G roadmap and new changes of 6G enabling technologies in three layers (physical layer, connection/network layer, service/application layer) based on current studies and recent 6G white papers \cite{Mahmood2020,Ziegler2020,Rajatheva2020,Taleb2020}. Through changes of those components, we outline security requirements and potential solutions, particularly in security architecture.

\subsection{6G applications, network vision, and potential security threats that impact on 6G security and privacy}
\label{sub:network-design}

\begin{figure*}[ht]
    \centering
    \begin{center}
			\includegraphics[width=1\linewidth]{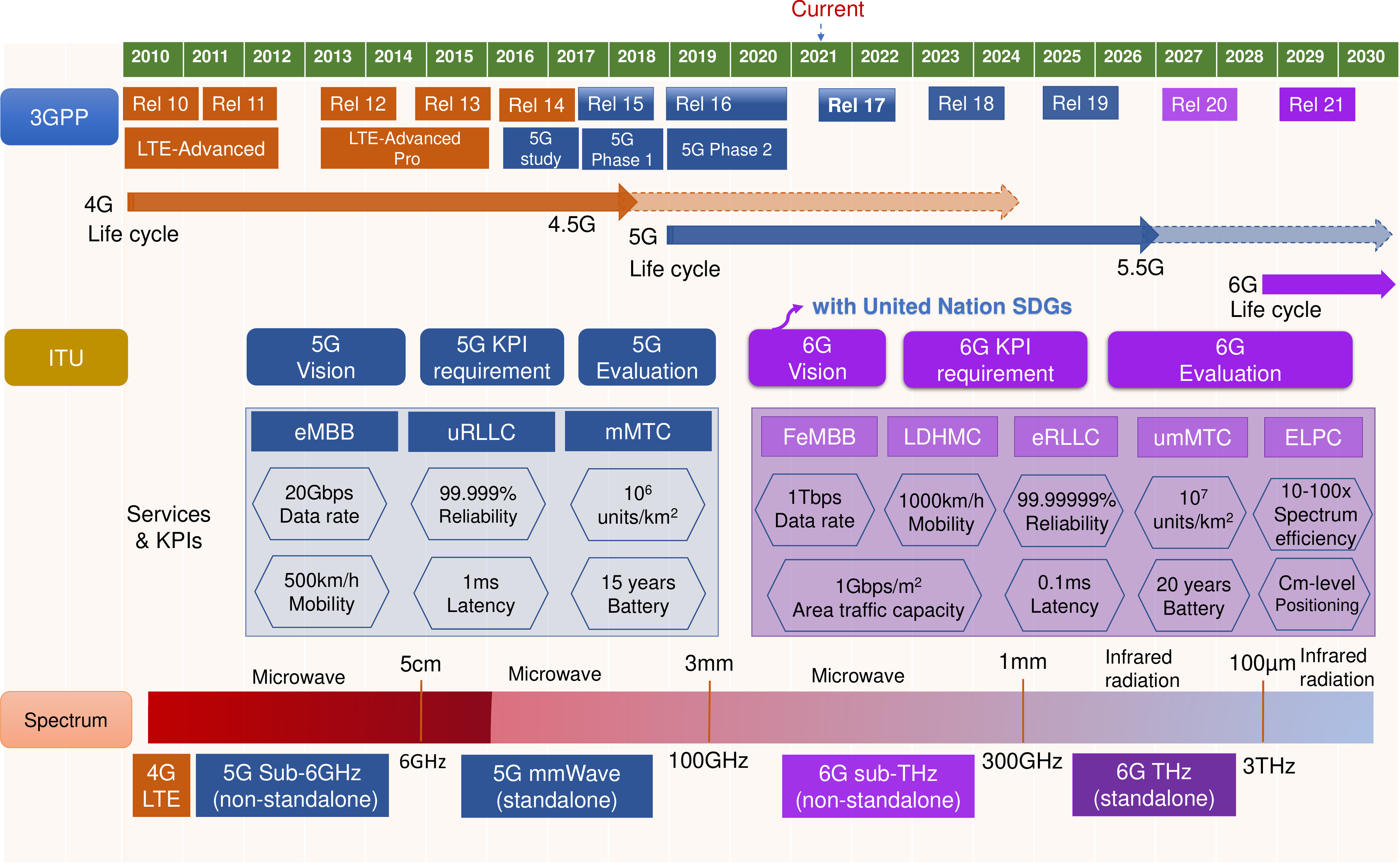}
	\end{center}
	
    \caption{The roadmap of 6G in comparison with 5G life cycle. KPI: Key Performance Indicator; eMBB: enhanced mobile broadband; uRLLC: ultra-reliable and low-latency communications; mMTC: ultra-massive machine-type communications; FeMBB: further-enhanced mobile broadband; ERLLC: extremely reliable and low-latency communications; LDHMC: long-distance and high-mobility communications; ELPC: extremely low-power communications; umMTC: ultra-massive machine-type communications; THz: Terahertz.}
    \label{fig:6G-networks-roadmap}
\end{figure*}

6G starts the wheel of its typical 10-year evolution cycle. Based on the studies in the literature (e.g.,\cite{Saad2020,DOCOMO2020,Gui2020,Rappaport19}), we summarize a roadmap of envisioned 6G development, \ac{KPIs}, and spectrum usage in comparison to those of prior generations in Figure~\ref{fig:6G-networks-roadmap}. Notably, 6G will enable network speeds over 1Tbps, latency under 1ms, and energy efficiency 10-100 times better than 5G. 6G is also set with the long-term goals in mind for a sustainable and carbon-neutral world in United Nations Sustainable Development Goals (UN SDGs) at 2030s or the Internet of Senses \cite{Prytz2020}. To achieve the requirements of \ac{KPIs}, 6G is targeted at three significant changes in the air-interface level: (1) pushing communications to higher frequency bands, (2) creating smart radio environments through reconfigurable surfaces, and (3) removing the conventional cell structures, such as cell-free \ac{MIMO} \cite{Matthaiou21}). Given the spectrum at lower frequencies in earlier generations of networks no longer applied/used, besides using compatible frequencies (6G mmWave), 6G will explore new high-frequency spectrum and photonic signals (300GHz-3THz) \cite{Rappaport19,Gui2020,Saad2020} for native 6G communications (6G THz). Also, multi-user MIMO, holographic radio beamforming, orbital angular momentum, and VLC are important technologies to enhance 6G communications \cite{Rajatheva2020,Taleb2020}. A brief selection of key enabling technologies for 6G physical layer is listed in the first row of Table~\ref{tab:6G-enabling-technologies}.

At the connection (network) level, 6G network architecture will have several significant changes from 5G. First, 6G may achieve the concept of \ac{NaaS} and network automation. NaaS allows users/enterprises to personalize networks based on their needs. This per-user basis model requires a new network design. Intent-based networking, end-to-end softwarization, cloudization, and deep slicing/function virtualization are key technologies to achieve the target. Second, the rapid shift of 5G infrastructure deployments towards cloud-based networks and open source, particularly for core/RAN network components, signals the ``full openness'' era of 6G. The development model can unlock the potential of 6G \cite{Tataria2020}, as features and improvement can be contributed at scale. This trend is primarily driven by top vendors (e.g., AT\&T) to avoid depending entirely on specific equipment makers. 6G can be the first truly AI-empowered wireless cellular network. This vision will transfer the concept of ``\textit{connected things}'' in 5G to ``\textit{connected intelligence}'' in 6G, where most network functions and nodes will likely be controlled by AI \cite{Letaief2019}. According to \cite{Taleb2020}, Deterministic networking (DetNet) or Time-Sensitive Networking (TSN), which aims to provide guaranteed latency and zero data loss, can be an important upgrade in 6G to satisfy the requirement of 6G ultra-reliable and low-latency communications (uRLLC).

\begin{table*}[ht]
\caption{Comparison of key enabling technologies for 5G and 6G }
\label{tab:6G-enabling-technologies}
\begin{adjustbox}{width=1\textwidth}
\begin{tabular}{lllll}
\hline
\rowcolor[HTML]{EFEFEF} 
\multicolumn{2}{l}{\cellcolor[HTML]{EFEFEF}\textbf{Layer}} &
  \textbf{5G} &
  \textbf{6G} &
  Challenges \\ \hline
 &
  \begin{tabular}[c]{@{}l@{}}Spectrum \&\\ communication\end{tabular} &
  \begin{tabular}[c]{@{}l@{}}$\bigcdot$ Sub-6GHz\\ $\bigcdot$ mmWave communication\end{tabular} &
  \begin{tabular}[c]{@{}l@{}}$\bigcdot$ mmWave communications \\ $\bigcdot$ Terahertz communications\\ $\bigcdot$ Visible Light Communication (VLC)\\ $\bigcdot$ Molecular communication\end{tabular} &
  \begin{tabular}[c]{@{}l@{}}$\blacktriangleright$ High-frequency processing in THz consumes much \\energy, no commercial THz chip \\ $\blacktriangleright$ Molecular communications are still under development\end{tabular} \\
 &
  \begin{tabular}[c]{@{}l@{}}Antenna \\ modulation\end{tabular} &
  \begin{tabular}[c]{@{}l@{}}$\bigcdot$ Massive MIMO, NOMA\\ $\bigcdot$ Beamforming\\ $\bigcdot$ Intelligent Reflecting Surfaces\end{tabular} &
  \begin{tabular}[c]{@{}l@{}} $\bigcdot$ Ultra-massive MIMO, Cell-free MIMO\\ $\bigcdot$ NOMA \\ $\bigcdot$ Holographic radio\\ $\bigcdot$ Large intelligent surfaces (LIS)\end{tabular} &
  \begin{tabular}[c]{@{}l@{}}$\blacktriangleright$ Most technologies are still at the early stage of research \\ and development \\ $\blacktriangleright$ Heterogeneity of  6G devices and services\end{tabular} \\
\multirow{-3}{*}[20pt]{Physical layer} &
  Coding &
  $\bigcdot$ LDPC, Polar code &
  $\bigcdot$ Multiuser LDPC, space-time coding &
  \begin{tabular}[c]{@{}l@{}}$\blacktriangleright$ High energy consumption, challenge to process terabits \\ per second\end{tabular} \\
\rowcolor[HTML]{EFEFEF} 
\begin{tabular}[c]{@{}l@{}}Connection layer\\ (Network layer)\end{tabular} &
  \begin{tabular}[c]{@{}l@{}}Networking\\ \& features\end{tabular} &
  \begin{tabular}[c]{@{}l@{}} $\bigcdot$ SDN\\ $\bigcdot$ NFV\\ $\bigcdot$ Network slicing\\ $\bigcdot$ Blockchain\\ $\bigcdot$ UAV networks\\ $\bigcdot$ Low-power networks\end{tabular} &
  \begin{tabular}[c]{@{}l@{}}$\bigcdot$ SD-WAN\\ $\bigcdot$ NFV \\ $\bigcdot$ Deep slicing\\ $\bigcdot$ Blockchain, distributed ledgers\\$\bigcdot$ Deterministic networking \\ $\bigcdot$ Quantum communications\\ $\bigcdot$ Space-air-ground-sea integrated networks\end{tabular} &
  \begin{tabular}[c]{@{}l@{}}$\blacktriangleright$ Challenge to transform the current Internet into the fully \\ software-defined networks\\ $\blacktriangleright$ High cost to deploy deep slicing\\ $\blacktriangleright$ Quantum computers have not yet been realized\\ $\blacktriangleright$ The complexity to manage space-air-ground-sea \\ networks over the worldwide area \\ $\blacktriangleright$ Few commercial blockchains/distributed ledgers\end{tabular} \\
Service layer &
  \begin{tabular}[c]{@{}l@{}}Edge/Cloud\\ features\end{tabular} &
  \begin{tabular}[c]{@{}l@{}}$\bigcdot$ Cloud services\\ $\bigcdot$ Container-based virtualization\\ $\bigcdot$ Edge computing\\ $\bigcdot$ Massive IoT services\end{tabular} &
  \begin{tabular}[c]{@{}l@{}}$\bigcdot$ Services everywhere\\ $\bigcdot$ Container-based virtualization \\ $\bigcdot$ Zero-touch service orchestration\\ $\bigcdot$ Distributed/autonomous computing\\ $\bigcdot$ Quantum computing\end{tabular} &
  \begin{tabular}[c]{@{}l@{}}$\blacktriangleright$ Quantum computers have not yet been realized\\ $\blacktriangleright$ Edge servers have not yet fully deployed\\ $\blacktriangleright$ High energy consumption\end{tabular} \\
\rowcolor[HTML]{EFEFEF} 
AI in use&
  \begin{tabular}[c]{@{}l@{}}AI model\\ \& capability\end{tabular} &
  \begin{tabular}[c]{@{}l@{}}$\bigcdot$ Machine learning\\ $\bigcdot$ Deep learning\end{tabular} &
  \begin{tabular}[c]{@{}l@{}}$\bigcdot$ Trustable AI\\$\bigcdot$ Explainable AI\\$\bigcdot$ Superintelligent AI\end{tabular} &
  \begin{tabular}[c]{@{}l@{}}$\blacktriangleright$ Current AI has no creativity \\ $\blacktriangleright$ Design of low-complexity AI solutions\end{tabular} \\ \hline
\end{tabular}
\end{adjustbox}
\end{table*}

With the new network capabilities, 6G extends the capacity of the three key services of 5G (uRRLC, eMBB and mMTC) while also enables new features such as long-distance and high-mobility communications (LDHMC) and extremely-low power communications (ELPC) -- as illustrated in Figure~\ref{fig:6G-networks-roadmap}. Notably, new applications are supposed to appear in 6G, including extended reality/digital twin, tactile/haptic Internet, smart medical nano-robot, fully automated driving, and holographic telepresence \cite{Saad2020,5GIA2020}. Extended reality (XR) refers to the combination of Virtual Reality (VR), Augmented Reality (AR), and Mixed Reality (MR) to enable real-and-virtual combined environments and human-machine interactions in real-time. In 6G, XR will be a key technology for many futuristic industries such as manufacturing, education, and training, including practicing for hazardous environments that are too risky or expensive to carry out on the field. Similarly, the digital twin extends the combination of virtual reality technologies (VR/AR/MR) to simulate the real-world counterpart (copy of physical objects and their behavior in real-time), such as modeling urban environments for planning practice. Tactile/haptic Internet denotes an internet network of ultra-low latency, extremely high availability, reliability, and security that can enable humans and machines to interact with each other in real-time. Finally, autonomous driving and holographic telepresence are the prospective technologies in 6G that also require ultra-low latency and extremely high reliability. To support these time-sensitive applications, more computing power will not only be placed at the edge but at every hop of 6G communications. AI technologies are expected to play a key role in boosting intelligence and autonomous capability in such computing nodes. Following \cite{5GIA2020,Zhang2019}, other enabling technologies for 6G service layer and AI are summarized in the third and fourth rows of Table~\ref{tab:6G-enabling-technologies}.

\subsection{Space-air-ground-sea integrated networks in 6G: the premise for convergence of heterogeneous networks}

 \begin{figure}[!ht]
    \centering
    \begin{center}
			\includegraphics[width=1\linewidth]{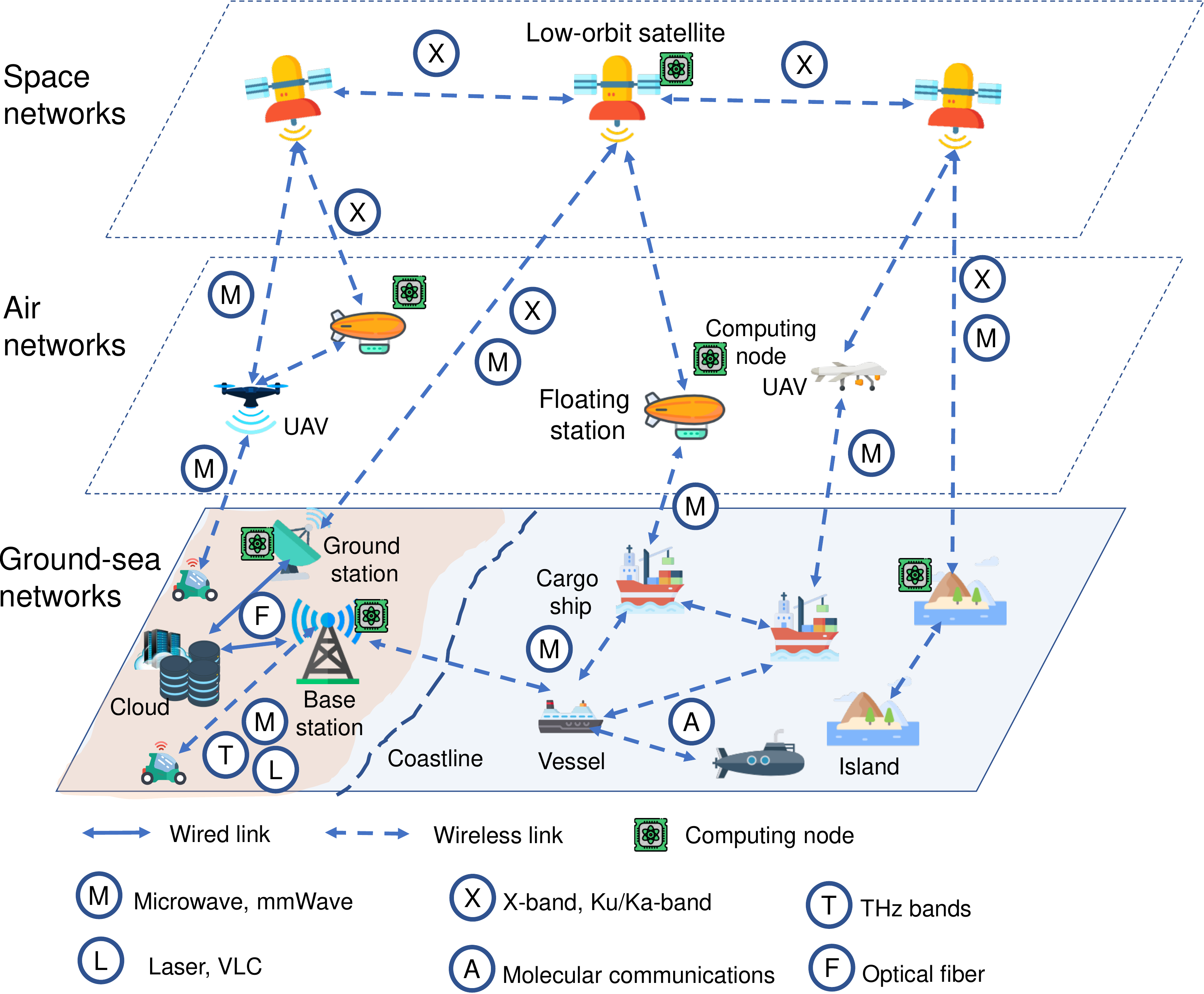}
	\end{center}
    \caption{An overview of 6G networks with the integration of non-terrestrial networks and mobile terrestrial systems, including the communication technologies for the typical links (e.g., satellite-ground link, satellite-aerial link, inter-satellite link, aerial-sea link, aerial-ground link, point-to-point link (ground), ground-sea link, ship-to-ship link). }
    \label{fig:6G-space-air-sea-ground-networks}
\end{figure}

3GPP has initiated the first discussions on the integration of non-terrestrial networks with mobile terrestrial systems that will be likely ready to use in the early 2030s \cite{saarnisaari20206g}, the era of 6G. In essence, the space-air-ground-sea integrated networks model can enable high-rate, reliable transmission and extremely broader coverage than four separate network segments. Figure~\ref{fig:6G-space-air-sea-ground-networks} illustrates the shape of 6G space-air-ground-sea integrated networks. The networks will use high-altitude platforms (e.g., floating stations), low-earth-orbit satellites, terrestrial base stations, and on-board ships to relay signals among the end-devices in coverage. With the expansion of coverage, 6G will bring up broadband transmission for maritime communications that will significantly benefit users in offshore ships and islands \cite{Wei21,WangMin21}. The benefits may be beyond civil applications of infotainment or remote learning, e.g., enhancing efforts to rescue persons and ships in distress. However, many challenges still should be overcome, such as protocol optimization designs to reduce the propagation delay, operate different \ac{RAN} capabilities, and keep service continuity (between mobile terrestrial systems and non-terrestrial networks) \cite{saarnisaari20206g}. Cooperation for infrastructure sharing among different operators is also critical since the satellite-based access systems and ground stations may be located in different regions.

\begin{figure}[ht]
    \centering
    \begin{center}
			\includegraphics[width=1\linewidth]{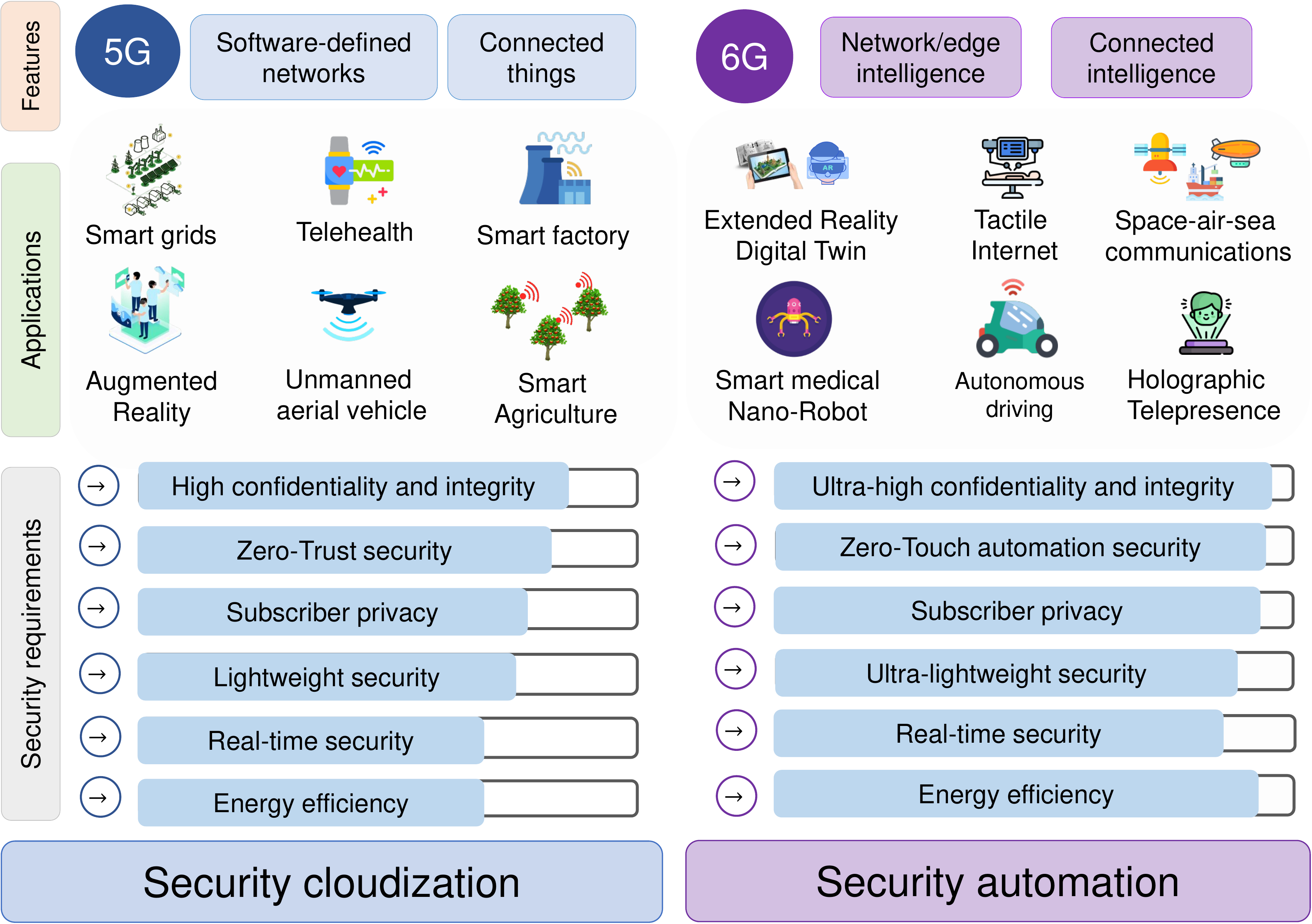}
	\end{center}
	
    \caption{An overview of typical 6G vs 5G applications and security requirements in comparison. 6G security will upgrade 5G security with new capability in terms of intelligence, automation, and energy efficiency.}
    \label{fig:6G-5G-security-requirements}
\end{figure}

 From security perspective, non-terrestrial networks are also vulnerable to wireless-targeted security attacks such as jamming and signalling DoS attacks \cite{Porambage21}. Degradation of signals because of such attacks can interrupt communications and delay delivery of messages in critical applications. Moreover, with the important role of high-altitude platforms, masquerading base stations becomes a real threat in hostile areas. With the high heterogeneity of network structure, the appearance of new 6G applications and enabling technologies creates new challenges for security and privacy preservation. Figure~\ref{fig:6G-5G-security-requirements} highlights six typical applications and six key security requirements for 5G and 6G networks, summarized from the literature \cite{Porambage21,Wang2020,Happa19,Ren20,Wei21,Szymanski16,Promwongsa21}. Notably, besides enhancing most 5G security principles, real-time security protection and full security automation will be the two key features to appear in 6G, given the rise of many time-sensitive applications and the complexity of integrated networks. Many believe that advanced AI and high-performance computing will be key factors to realize the goals. In this vision, 6G will continue the trajectory of migrating enterprise security services into the edge nodes and clouds in 5G while advancing pervasive data intelligence usage to enhance the protection capability. Finally, Zero Trust and Zero Touch architecture, where the protection system is supposed not to trust anything and keep verifying every entity before granting them access or the hassles of passwords are expected to be eliminated, is likely the mainstream of 6G security systems to stop data breaches. 
 
\begin{table*}[ht]
\caption{Security and privacy issues of several 6G applications}
\label{tab:6g-application-security}
\begin{adjustbox}{width=1\textwidth}
\begin{tabular}{llllllllllll}
\hline
\rowcolor[HTML]{EFEFEF} 
\cellcolor[HTML]{EFEFEF} &
 \cellcolor[HTML]{EFEFEF} &
  \cellcolor[HTML]{EFEFEF} &
  \cellcolor[HTML]{EFEFEF} &
  \multicolumn{6}{c}{\cellcolor[HTML]{EFEFEF}\textbf{Security requirement}} &
  \cellcolor[HTML]{EFEFEF} &
  \cellcolor[HTML]{EFEFEF} \\ \cline{5-10}
\rowcolor[HTML]{EFEFEF} 
\multirow{-2}{*}{\cellcolor[HTML]{EFEFEF}\textbf{6G applications}} &
\multirow{-2}{*}{\cellcolor[HTML]{EFEFEF}\textbf{Reference}} &
  \multirow{-2}{*}{\cellcolor[HTML]{EFEFEF}\textbf{Potential security issues}} &
  \multirow{-2}{*}{\cellcolor[HTML]{EFEFEF}\textbf{Potential privacy issues}} &
  \begin{tabular}[c]{@{}l@{}}High \\ confidentiality\\ \& integrity\end{tabular} &
  \begin{tabular}[c]{@{}l@{}}Zero \\ Touch\end{tabular} &
  \begin{tabular}[c]{@{}l@{}}Subscriber\\ privacy\end{tabular} &
  \begin{tabular}[c]{@{}l@{}}Ultra-\\lightweight \\ security\end{tabular} &
  \begin{tabular}[c]{@{}l@{}}Real-time\\ security\end{tabular} &
  \begin{tabular}[c]{@{}l@{}}Energy \\ efficiency\end{tabular} &
  \multirow{-2}{*}{\cellcolor[HTML]{EFEFEF}\textbf{Key solutions}} &
  \multirow{-2}{*}{\cellcolor[HTML]{EFEFEF}\textbf{Open challenges}} \\ \hline
\begin{tabular}[c]{@{}l@{}}Extended Reality\\ Digital Twin\end{tabular} & \cite{Happa19} &
  \begin{tabular}[c]{@{}l@{}}Embed malicious content \\ into XR applications to \\ attract click, deepfake XR \\ services, malware injection,\\ DoS against XR services, \\ physical damage\end{tabular} &
  \begin{tabular}[c]{@{}l@{}}Expose biometric data such \\ as iris or retina scans, \\ fingerprints and handprints, \\ face geometry, and \\ voiceprints\end{tabular} &
  M &
  H &
  H &
  H &
  M &
  M &
  \begin{tabular}[c]{@{}l@{}}Security edge\\ protection, \\ differential \\ privacy, IDS/MTD\end{tabular} &
  \begin{tabular}[c]{@{}l@{}}Practical implementation\\ of real-time security\end{tabular}  \\
Tactile interaction & \cite{Promwongsa21,Szymanski16} &
  \begin{tabular}[c]{@{}l@{}}DoS against tactile services, \\ Man-in-the-middle attacks\end{tabular} &
  \begin{tabular}[c]{@{}l@{}}Expose biometric data \\ such as fingerprints\end{tabular} &
  M &
  H &
  H &
  M &
  H &
  L &
  \begin{tabular}[c]{@{}l@{}}Physical layer security\\ quantum-safe \\ communications \\ IDS/MTD \end{tabular}  &
  \begin{tabular}[c]{@{}l@{}}Practical implementation\\ of real-time security\end{tabular} \\
\begin{tabular}[c]{@{}l@{}}Space-air-sea\\ communications\end{tabular} & \cite{Wei21,Wang2020} &
  \begin{tabular}[c]{@{}l@{}}Jamming, DoS attacks, \\ eavesdropping, API \\ vulnerabilities\end{tabular} &
  \begin{tabular}[c]{@{}l@{}} Signalling-based location \\ tracking, expose identity\end{tabular} &
  H &
  H &
  H &
  M &
  M &
  M &
  \begin{tabular}[c]{@{}l@{}}End-to-end security, \\ non-ID, blockchain,\\ distributed ledgers, \\ quantum communications, \\ firewall/IDS/MTD\end{tabular} &
  \begin{tabular}[c]{@{}l@{}}Practical implementation\\ of blockchain/distributed \\ ledgers, quantum \\ communications, end-to\\ end security\end{tabular}  \\
\begin{tabular}[c]{@{}l@{}}Smart medical\\ Nano-Robot\end{tabular} & \cite{DRESSLER2012} &
  \begin{tabular}[c]{@{}l@{}}Inject malware to create \\ malfunction device cycles \\ and cause physical damage\end{tabular} &
  \begin{tabular}[c]{@{}l@{}}Expose body health \\ information such as heat\\ rate, blood pressure,\\ pathological behavior..\end{tabular} &
  H &
  H &
  H &
  H &
  L &
  H &
  \begin{tabular}[c]{@{}l@{}}Physical layer \\ security, IDS/MTD\end{tabular}  &
  \begin{tabular}[c]{@{}l@{}}High-performance \\ edge security, efficient\\ lightweight security,\\ energy efficiency\end{tabular} \\
\begin{tabular}[c]{@{}l@{}}Autonomous \\ driving\end{tabular} & \cite{Ren20} &
  \begin{tabular}[c]{@{}l@{}}Jamming V2X\\ DoS attacks, eavesdropping, \\ Fake beacon messages\\ to create virtual traffic \\ jam, sudden crash...\end{tabular} &
  \begin{tabular}[c]{@{}l@{}}Location tracking, \\ compromised credentials \\ (pseudonyms)\end{tabular} &
  H &
  H &
  H &
  L &
  H &
  M &
  \begin{tabular}[c]{@{}l@{}} Blockchain, distributed\\ ledgers, misbehavior\\ detection, physical \\ security isolation,\\ IDS/MTD\end{tabular} &
  \begin{tabular}[c]{@{}l@{}}Practical implementation\\ of blockchain/distributed \\ ledgers, real-time edge \\ security\end{tabular}  \\
\begin{tabular}[c]{@{}l@{}}Holographic \\ telepresence\end{tabular} & \cite{CHuang20Holo,Promwongsa21} &
  \begin{tabular}[c]{@{}l@{}}DoS attacks, eavesdropping,\\ deepfake agent\end{tabular} &
  \begin{tabular}[c]{@{}l@{}}Expose personal behavior, \\ social habits, biometric \\ data\end{tabular} &
  M &
  H &
  H &
  M &
  H &
  L &
  \begin{tabular}[c]{@{}l@{}}Physical layer\\ security, IDS/MTD\end{tabular} &
  \begin{tabular}[c]{@{}l@{}}Ultra-lightweight\\ security, energy \\ efficiency\end{tabular}  \\ \hline
\end{tabular}
\end{adjustbox}
\begin{tablenotes}
	\item L: Low; M: Medium; H: High; DoS: Denial-of-Service, V2X: Vehicle-to-Everything; XR: eXtended Reality
	\item IDS: Intrusion Detection System; MTD: Moving Target Defense
\end{tablenotes}
\end{table*}

 \subsection{Overview of several security and privacy issues on typical 6G applications: finding the origin of the issues}
 
Table~\ref{tab:6g-application-security} summarizes security and privacy issues on the applications, security requirements, potential solutions, and remaining challenges. In general, different applications have different requirements in terms of security. Similarly, the attacks vary by application. However, several attacks can impact multiple applications. For example, an attacker can launch signalling DoS attacks to overload and degrade the performance of nearly all the services (e.g., XR, tactile Internet, autonomous driving). On the other hand, eavesdropping, jamming, or API vulnerabilities are common in autonomous driving and space-air-sea networks. Because of the impact of an attack on multiple applications, in the next sections, we adopt the three-layer architecture to classify attacks and corresponding defense solutions for 6G enabling technologies instead of focusing on several applications only. This layer-based review approach can significantly benefit the operators and developers of interest in preventing specific attacks that cause high risk to many applications while not limited to the six typical ones listed in this subsection. The flaws of related protocols and the remaining challenges of the technologies are then the starting points for further enhancements.

\subsection{6G for the expectation of the Internet of Everything and security evolution}

6G is expected to be the mobile generation network for heterogeneous technologies. According to \cite{Saad2020,Zong2019}, the human-to-machine and machine-to-machine interaction promise to become more common to support wireless brain-computer applications, smart implants, textile integrated fabrics, and autonomous health in upcoming decades. In this vision, 6G will enlarge the interconnectivity beyond the things for a world of the Internet of Everything (IoE) where many wireless technologies coexist. 6G IoE will expand the common uses of IoT applications and offer them for every industry. Due to the significant expansion of the network coverage, many self-organizing networks (SON) will be proposed and operated in practice, where AI-empowered automation control systems can assist in simplifying the planning, configuration, management, optimization, and healing of mobile radio access networks and beyond. In 6G, SON can be enhanced to self-sustainable networks (SSNs) \cite{Saad2020} to eliminate the need for separate charging of heterogeneous devices, e.g., by combining energy-efficient communication, energy harvesting/power transfer techniques, and offloading of energy-intensive processing from the devices to the computing nodes (servers located at the edge or floating stations as the illustration in Figure~\ref{fig:6G-space-air-sea-ground-networks}). 

With the transformation of the network technologies and applications towards autonomous functions and heterogeneous connectivity, 6G security and privacy issues will significantly differ from conventional IoT and computer security. First, contemporary IT security is not designed for serving a massive number of connected and high-mobility heterogeneous devices. These large-scale IoT devices, which are not equipped with strong cryptography schemes or mutual authentication as a result of their resource/energy constraints, could introduce enormous concerns of the weak links for intrusion. Second, security aspects of the new protocols in 6G IoE communications such as brain-computer interactions and molecular communications will require specific security protection requirements (e.g., trustable communications or real-time response), due to the direct risks to human life. However, 6G may still inherit several security threats from legacy networks and IoT/IT environments due to using the same fundamental protocols, e.g., TCP/IP. Some example attacks are DoS, eavesdropping, vulnerabilities exploitation, and spoofing, as shown in Table~\ref{tab:6g-application-security}. With heterogeneous networks' high complexity, 6G needs new security and privacy preservation approaches, e.g., more automation and interoperability support. In the following subsection, we envisage 6G overall security architecture and sketch out the potential changes, which are then detailed in the next sections.

\subsection{Vision of 6G security architecture and potential changes}
\label{subsec:vision-6G-security-architecture}

\begin{figure}[ht]
    \centering
    \begin{center}
			\includegraphics[width=1\linewidth]{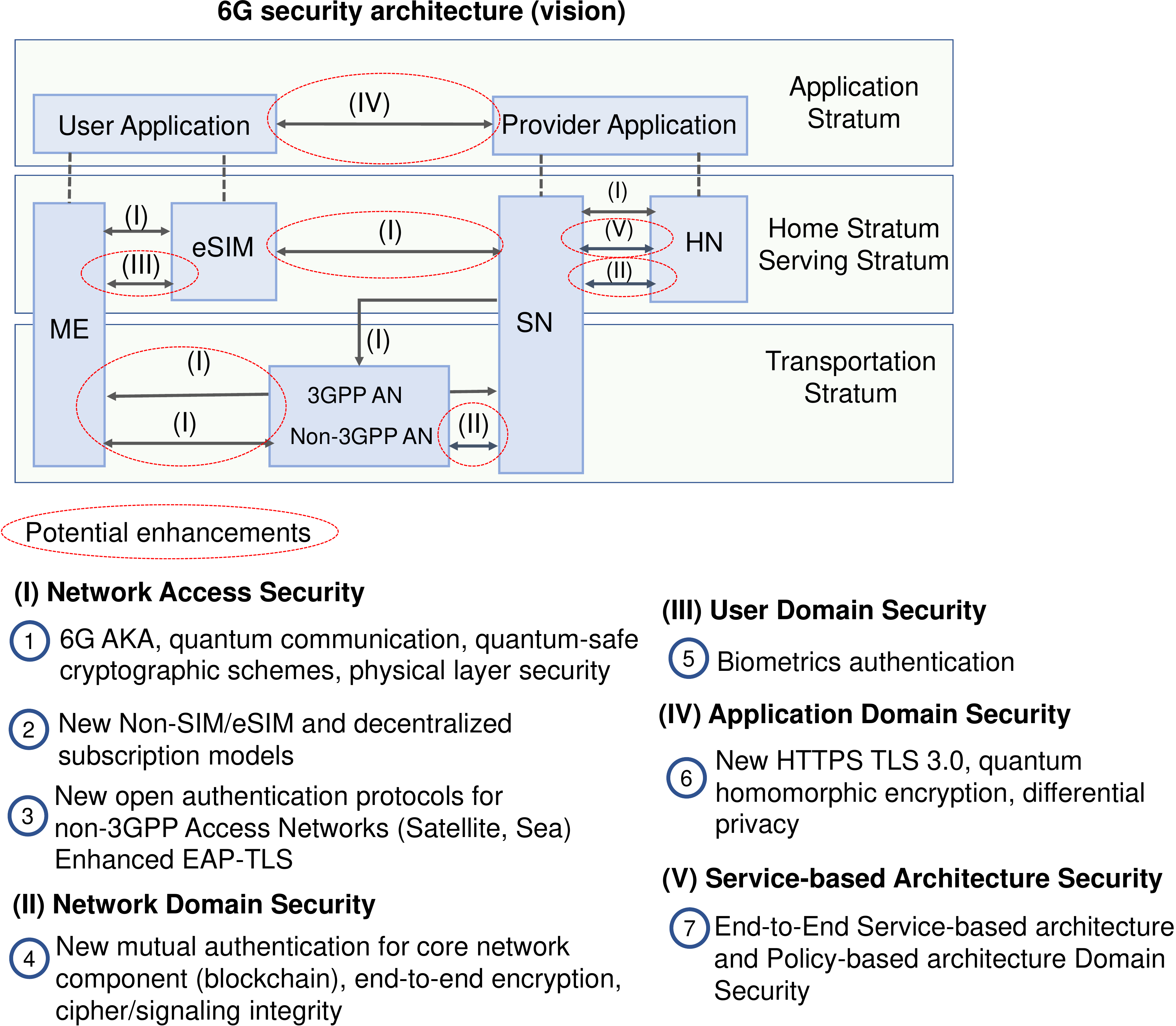}
	\end{center}
	
    \caption{Abstract 6G security architecture and potential changes for core components in 6G security architecture (marked by red circles). ME: Mobile Equipment, SN: Serving Network, HN: Home network, AN: Access network.}
    \label{fig:6G-security-architecture}
\end{figure}

\begin{figure}[ht]
    \centering
    \begin{center}
			\includegraphics[width=1\linewidth]{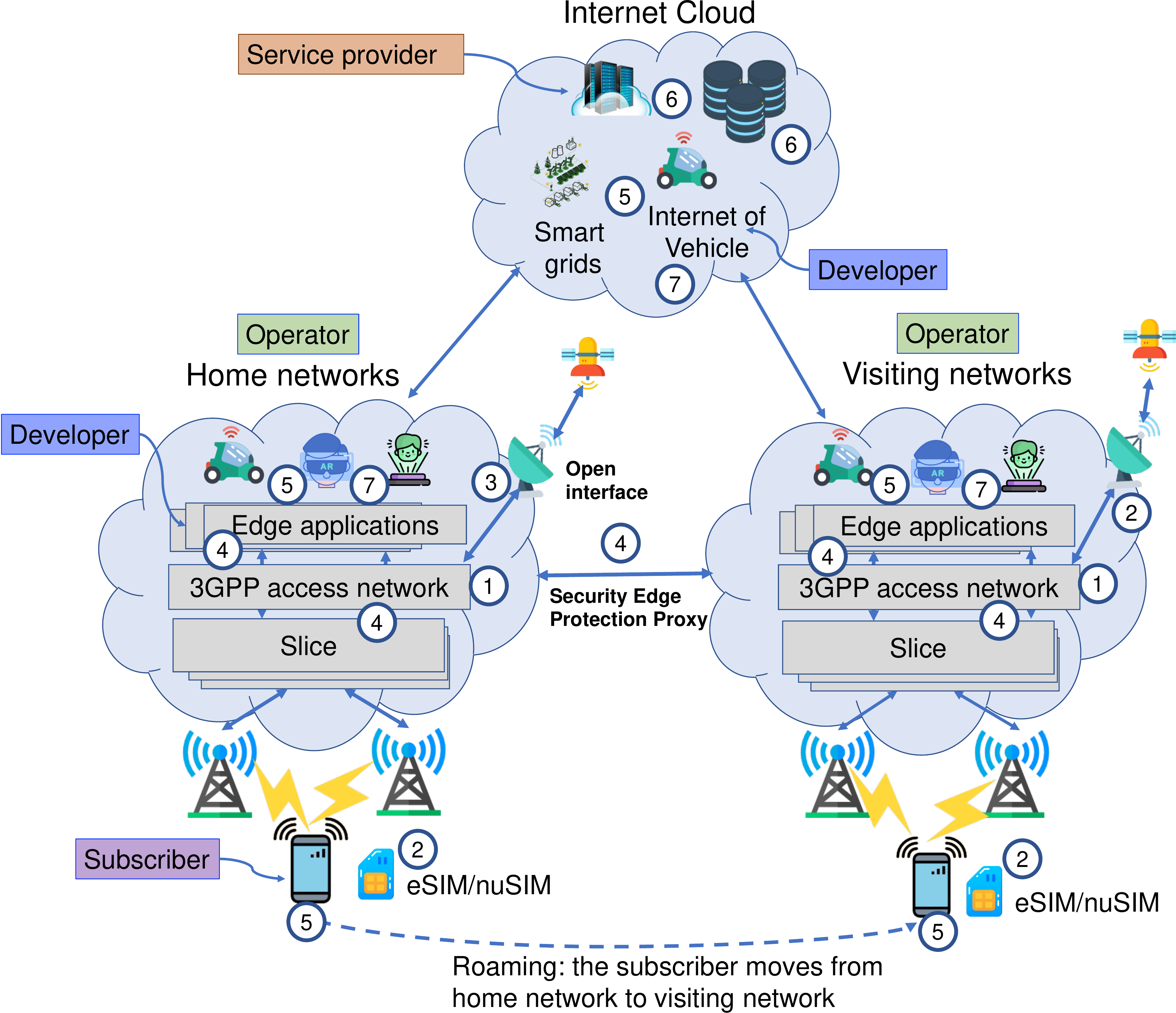}
	\end{center}
	
    \caption{The illustration of seven changes of 6G security components (\textcircled{\raisebox{-0.9pt}{1}}---\textcircled{\raisebox{-0.9pt}{7}} in Figure~\ref{fig:6G-security-architecture}) and the potential impact on the related stakeholders: operator, subscriber, and service provider/developer.}
   \label{fig:6G-security-architecture-visual}
\end{figure}

 6G will certainly require adjustment in its security architecture to satisfy new applications and the expansion of the space-air-ground-sea integrated network model (as presented above). Figure~\ref{fig:6G-security-architecture} envisions an abstract 6G security architecture with several key potential changes in core components of current 3GPP security architecture. Figure~\ref{fig:6G-security-architecture-visual} illustrates six new changes for core components of 6G security architecture and their potential impact on related stakeholders: network operators (\textcircled{\raisebox{-0.9pt}{1}}---\textcircled{\raisebox{-0.9pt}{4}}), subscribers (\textcircled{\raisebox{-0.9pt}{2}},\textcircled{\raisebox{-0.9pt}{5}}), and service providers/developers (\textcircled{\raisebox{-0.9pt}{5}},\textcircled{\raisebox{-0.9pt}{6}}, \textcircled{\raisebox{-0.9pt}{7}}). In the chain, network operators provide network connectivity for subscribers and probably the dedicated Internet infrastructure for service providers, tenants, and developers. Therefore, network operators will be the main stakeholders to upgrade the network access and network domain security infrastructure. The service providers offer value-added services for subscribers (infotainment, web) and probably platforms for developers/tenants (cloud storage, data analytic). At this aspect, the service providers will take into account the upgrade on the application domain and service-based architecture security. Since developers develop and maintain cloud/edge applications (XR/AR game), the potential upgrades of enhancing security for application development or supporting new security API (following the services provided by the third-party providers) will be their task. In 6G, network operators can play the role of a service provider, e.g., mobile storage. As a result, they may also involve in many parts of enhancing service security. Finally, subscribers may see little impact from the upgrades, e.g., by changing a new device or the SIM card registration.
 
\textcircled{\raisebox{-0.9pt}{1}} \textit{6G requires new authentication model and cryptographic schemes for communication security. 6G-AKA, quantum-safe cryptographic schemes, and physical layer security are the top candidates}. The changes in 6G network design towards cloud-based and open programmable networking platforms urge reform of authentication architecture. 6G will likely inherit some security models from 5G, such as a unified authentication platform for both open and access network agnostic. However, many new functions need to be added to complement these. For example, a 6G-AKA protocol needs to certify the clear roles of which component, \ac{AUSF} or \ac{SEAF}, will decide authentication in cross-slice communications. 6G-AKA needs to be able to verify the claimed identity of an endpoint in a deep-sliced and open programmable networking platform. Other than improving network access control, physical layer security can be a game changer to protect 6G IoT networks against conventional threats such as impersonation attacks. Details about physical layer security for 6G are presented in Section~\ref{sec:security-physical-layer}. Quantum-safe cryptographic schemes \cite{Hoffstein1998} are prospective technologies for secure communications in 6G, particularly for highly sensitive and strategic sectors such as banking or defense. These technologies are a remarkable upgrade in 6G. In turn, crypto-systems such as elliptic curve cryptography (ECC) may not disappear soon, at least not until their replacement by quantum technology is technically practicable. Details of quantum-safe cryptographic schemes communication security are presented in Section~\ref{subsec:communication-security}.

\begin{table*}[ht]
\caption{Security domains, stakeholders, and standardization bodies.}
\label{tab:6g-5g-security-changes}
\begin{adjustbox}{width=1\textwidth}
\small
\begin{tabular}{lllll}
\hline
\rowcolor[HTML]{EFEFEF} 
\textbf{Security domain} &
  \textbf{Features} &
  \textbf{Main stakeholders} &
  \textbf{Standardization bodies} &
  \textbf{Potential upgrades in 6G} \\ \hline
\begin{tabular}[c]{@{}l@{}}Network access security \end{tabular} &
  \begin{tabular}[c]{@{}l@{}}Enable a UE to authenticate and access\\ to access network securely (3GPP \\ access and Non-3GPP access), prevent\\ attacks on the radio interfaces\end{tabular} &
  \begin{tabular}[c]{@{}l@{}}$\bigcdot$ Network operator\end{tabular} &
  \begin{tabular}[c]{@{}l@{}} $\bigcdot$ 3GPP, ETSI, ITU-T, NGMN \\ $\bigcdot$ 5G PPP, IETF, NIST\end{tabular} &  \begin{tabular}[c]{@{}l@{}}$\blacktriangleright$ Physical layer security (Section$\sim$V) \\ $\blacktriangleright$ 6G authentication and key management, \\endpoint security (Section$\sim$VI) \\ $\blacktriangleright$ Trust networks (blockchain) (Section$\sim$VI)\end{tabular} 
   \\ \rowcolor[HTML]{EFEFEF} 
\begin{tabular}[c]{@{}l@{}}Network domain security\end{tabular} &
  \begin{tabular}[c]{@{}l@{}}Enable network nodes to securely exchange\\ signalling data and user plane data (e.g.,\\ serving  networks and home networks)\end{tabular} &
  \begin{tabular}[c]{@{}l@{}} $\bigcdot$ Network operator\end{tabular} &
  \begin{tabular}[c]{@{}l@{}} $\bigcdot$ 3GPP, ETSI, ITU-T, NGMN \\ $\bigcdot$ 5G PPP, IETF, NIST \end{tabular} & \begin{tabular}[c]{@{}l@{}}$\blacktriangleright$ Enhanced IPSec/TLS (Section$\sim$VI)\end{tabular} 
   \\
\begin{tabular}[c]{@{}l@{}}User domain security\end{tabular} &
  Secure the user access to mobile equipment &
 \begin{tabular}[c]{@{}l@{}} $\bigcdot$ Subscriber \\ $\bigcdot$ Network operator \\ $\bigcdot$ Device vendors \end{tabular}&
  \begin{tabular}[c]{@{}l@{}}$\bigcdot$ 3GPP, ETSI \end{tabular} & \begin{tabular}[c]{@{}l@{}} $\blacktriangleright$ Distributed subscription (Section$\sim$VI) \end{tabular}
   \\ \rowcolor[HTML]{EFEFEF} 
\begin{tabular}[c]{@{}l@{}}Application domain security\end{tabular} &
  \begin{tabular}[c]{@{}l@{}}Enable user applications and provider \\ applications to exchange data securely\end{tabular} &
  \begin{tabular}[c]{@{}l@{}}$\bigcdot$ Developer \\ $\bigcdot$  Service provider\end{tabular} &
  \begin{tabular}[c]{@{}l@{}} $\bigcdot$ 5G PPP, IETF \end{tabular} & \begin{tabular}[c]{@{}l@{}} \\ $\blacktriangleright$ Biometric authentication (Section$\sim$VII) \\ $\blacktriangleright$ Enhanced HTTPS (Section$\sim$VII) \\ $\blacktriangleright$ AI-empowered security (Section$\sim$VIII) \\ $\blacktriangleright$ Enhanced privacy (Section$\sim$IX)\end{tabular}
   \\
\begin{tabular}[c]{@{}l@{}}Service-based architecture\\ security\end{tabular} &
  \begin{tabular}[c]{@{}l@{}}Enable network functions to securely \\ communicate within the serving network\\ domain and the other domains, e.g., network\\ function registration, discovery,\end{tabular} &
  \begin{tabular}[c]{@{}l@{}}$\bigcdot$ Service provider \\ $\bigcdot$ Developer\end{tabular} & 
  \begin{tabular}[c]{@{}l@{}}$\bigcdot$ 5G PPP, IETF\end{tabular}& \begin{tabular}[c]{@{}l@{}}$\blacktriangleright$ End-to-end SECSaaS (Section$\sim$VII) \end{tabular} 
   \\ \hline
\end{tabular}
\end{adjustbox}
\begin{tablenotes}
	\item $\diamond$ 3rd Generation Partnership Project (3GPP)'s major security fields: Security architecture, authentication, RAN security, subscriber privacy, network slicing 
	\item $\diamond$ European Telecommunications Standards Institute (ETSI) 's major security fields: NFV, MEC security, security architecture, privacy 
	\item $\diamond$ Union International Telecommunications Standardization Sector (ITU-T)'s major security fields: Cybersecurity, trust networks, authentication
	\item $\diamond$ Next Generation Mobile Networks (NGMN)'s major security fields: Subscriber privacy, MEC security, network slicing 
     \item $\diamond$ 5G Infrastructure Public Private Partnership (5G PPP)'s major security fields: Subscriber privacy, security architecture, authentication
     \item $\diamond$ Internet Engineering Task Force (IETF)'s major security fields: Security protocol draft standards in RFC, public-key infrastructure 
     \item $\diamond$ National Institute of Standards and Technology (NIST)'s major security fields: Information security standards (ISO 27001, NIST SP 800-53,...)
\end{tablenotes}
\end{table*}

       \textcircled{\raisebox{-0.9pt}{2}} \textit{New user identity management model will be the major change of 6G subscriber management, compared with 5G}. Despite many enhancements in security and convenience, SIM cards and the identity management model have seen no significant change since 2G. The requirement of plugging it into devices practically limits many IoT applications. Using an eSIM (a SIM card embedded in a mobile device) or non-SIM model can remove the barriers to 6G implant devices. This shift will require a fundamental change in identity storage or release, e.g., a SIM could be part of a system-on-chip (nuSIM) \cite{Ylianttila2020}. A decentralized subscription model can be another significant upgrade from the centralized authentication and authorization paradigm of 5G. Currently, a visited network can neither authenticate the UE in 5G nor sell any service (e.g., VR/AR content) to subscribers \cite{Velde20}. A roaming agreement must be made to connect with the UE's home network, which charges all the services. Such a model will protect the revenue for home operators, which at least comes with a revenue-sharing agreement with the visited networks. Because of the hostility of operators, this roaming model has not as yet been changed from 2G. From a user's perspective, such roaming is quite inconvenient, particularly if the user leaves the home network's coverage. It will be a significant upgrade in 6G if the visited network can authenticate the UE for the services it provides to the UE, and as does the home network. However, it is unclear how one would balance the right of operators if using a decentralized subscription. Until there is a new business model to satisfy operators (e.g., monetizing from extra application services), or a mutual trust protocol among operators, a centralized subscription model and authentication for roaming connections between the home network and a serving network will need improvements in the coming years. Given the inevitable transition from traditional telephone services to VoIP and Internet services, the revenue drop of the roaming division may motivate operators to accept the sell-access-as-service model. In that way, the current authentication model may have more space to be simplified in 6G. 
       
        \textcircled{\raisebox{-0.9pt}{3}} \textit{New open authentication protocols are necessary because of the expansion of 6G to non-terrestrial networks, such as satellite and maritime communications, new open authentication protocols are necessary}. 6G can open more interfaces for heterogeneous applications such as space-air-ground-sea networks. Other than EAP-AKA and EAP-TLS, this means that more standards may be rolled out to support authentication in maritime communications. However, compared to traditional ground or satellite networks, the integrated network will be affected by limited and unbalanced network resources \cite{JLiu18}. Supporting open authentication protocols for the integrated network is highly recommended and a subject of many ongoing efforts, e.g., in \cite{Yao20}.  
             
        \textcircled{\raisebox{-0.9pt}{4}} \textit{Mutual authentication is critical for the goal of 6G trust networks}. In 5G, mutual authentication is still based on a conventional symmetric key model. However, Blockchain and Distributed Ledger Technologies (DLT) can be prospective solutions to change the way of protecting confidentiality and integrity in 6G \cite{Maksymyuka2020,Nguyen2020}. Blockchain and DLT guarantee mutual trust, high privacy preservation, and single-failure disruption prevention. They can also enhance the communication reliability of 6G key entities, such as authentication servers or between a Serving Network and a Home Network. However, blockchain and DLT are at a very early stage due to many of their fundamental components still under active development. Storing and processing large-scale records/blocks at the nodes in time need further breakthroughs in the coming years since operations of the state-of-the-art blockchain solutions require huge memory and resources (for mining power-of-work). Such a computation burden can significantly impact network performance and device energy, limiting many potential applications.

    \textcircled{\raisebox{-0.9pt}{5}} \textit{Biometric authentication or a passwordless service access control model is a long-awaited feature for security in 6G applications}. Password-based protection models have been essential for protecting many applications for decades. However, they have many shortcomings. Some are easy to be compromised, costly in storage, and hard to memorize. Future technology like brainwave/heartbeat-based authentication can provide more theft-resistant and enhance user experience: citizens can use their bio-identity to access the network and services anywhere without tracking many passwords. This \textit{passwordless} authentication will be a massive leap forward for a 6G security posture. More details about such authentication schemes are given in Section~\ref{subsec:biometric-authentication}.
    
    \textcircled{\raisebox{-0.9pt}{6}} \textit{Enhanced HTTPS TLS and homomorphic encryptions are the future technologies to enhance service and data security in 6G applications}. Notably, both enhanced HTTPS TLS and homomorphic encryption will be equipped with quantum-resistant algorithms to resist quantum attacks (e.g., AES-256 or ECC P-384). Besides, homomorphic encryption allows performing operations, such as search and query, on encrypted data directly without decryption. Therefore, subscribers can send their data to third parties (e.g., operators, cloud providers) for storage or processing. More details about enhanced HTTPS TLS and homomorphic encryption are given in  Section~\ref{subsec:homorphic-encryption}.

   \textcircled{\raisebox{-0.9pt}{7}} \textit{Service-based Security Architecture in 5G is upgraded into End-to-End service-based and Policy-based security architecture in 6G}. A Service-based Architecture Domain security is the pillar of 5G security architecture. 6G will take this feature to a new level, End-to-End Service-based Architecture or even Policy-based architecture domain security, to satisfy the personalization and micro-deployment flexibility. Furthermore, applications such as mixed reality may move closer to a UE, i.e., on edge nodes. For optimization purposes, the protection model for communication between a UE and such applications may work based on flows or flexible control by policies from the control plane unit of the serving network.

 Table~\ref{tab:6g-5g-security-changes} summarizes 6G security components, the related main stakeholders, corresponding standardization bodies and mentioned potential upgrades for each of them. More details of the changes are subsequently presented in the next sections.

\subsection{Summary of lessons learned from key possible changes of 6G security}

This section reviews the key changes of 6G in terms of enabling technologies, security requirements, and security architecture. In summary, three lessons learned from these potential changes for 6G security architecture are as follows.
 
 \begin{enumerate}
     
    \item \textit{Many 5G features will not fade away but be fully supported with further security enhancement for 6G usage}. United authentication framework and security isolation in 5G technologies will continue to play a central role in 6G to converge the authentication features for multi-access networks. However, given the expansion of network coverage to the space-air-ground-sea integrated environment, security capabilities of the 5G-AKA framework will be the target of further enhancements. 
     
    \item \textit{Deploying non-SIM-based identity management will be a huge step for 6G security}. This ambitious goal is to reform the current SIM-based identity management to using a non-SIM card and decentralized subscription model. Pursuing this goal will be an important step towards removing the barriers to implant devices and user experience in 6G. However, it is unclear whether operators will support such bold changes. Enhancing eSIM technologies can be a reasonable approach to prepare for a future leap of the transition. 
     
     \item \textit{A unified authentication framework, passwordless authentication, and open security model are the future of 6G security, but in relevant context}. A unified authentication framework and open interface model can enable access control and security architecture simplification for space-air-ground integrated networks -- a key goal of 6G. Meanwhile, by decreasing the complexity of memorizing and storing login information as in conventional password-based models, \textit{passwordless} authentication is a long-awaited feature that will enhance user experience and convergence of access control that many 6G services expect. However, building a comprehensive architecture to support the features for all 6G applications will require long-term effort. Implementing the features for the applications that need them first and then expanding the implementation to the whole network -- when the infrastructure is ready -- is likely the best approach.

 \end{enumerate}

The following sections cover security attacks and prospective defense technologies, which will powerfully dominate 6G physical layer, connection layer, and service layer, based on the summary in Table~\ref{tab:6g-5g-security-changes}.

%% file: Section_V_PHY.tex
Since the physical layer is the cornerstone of wireless communications, protecting physical-layer information can prevent many conventional attacks on radio signals such as eavesdropping and jamming that nearly impact on every 6G application. The premise of physical layer security is to exploit the characteristics of wireless channels (e.g., fading, noise) to enhance confidentiality and perform lightweight authentication. Low complexity of physical layer security will particularly benefit 6G low-cost IoT devices, which often lack energy and computation capacity to run advanced authentication mechanisms. Besides, relying on physical laws, physical layer security is robust against cryptanalysis, which has been the top concern of conventional cryptographic algorithms. Physical layer security can be implemented at the base stations/IoT gateways of the operators or in the signal modulation algorithms. The following subsections look at key security concerns and several prominent defense approaches for enabling technologies in 6G. 

\subsection{Security in 6G mmWave communications}

\begin{figure}
    \centering
    \begin{center}
			\includegraphics[width=1\linewidth]{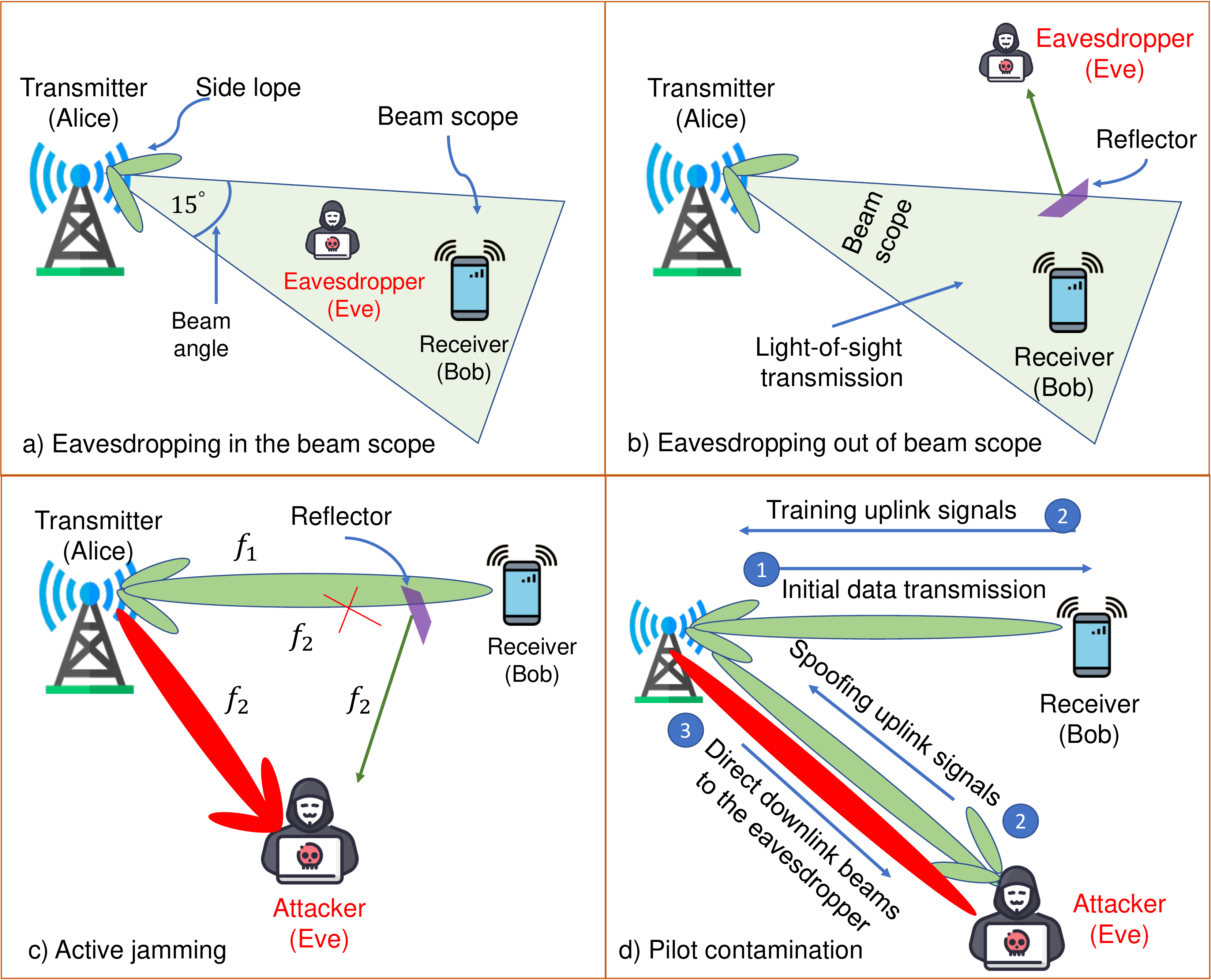}
	\end{center}
   \caption{Illustration of three common attacks against 5G physical layer technologies that can occur in future 6G compatible networks: a) Eavesdropper wiretaps beamforming signals in the beam scope of the transmitter (Alice); b) Eavesdropper wiretaps beamforming signals from the transmitter by placing a reflector in the beam scope of the transmitter (Alice); c) A jammer generates signals at the frequency $f_2$ (based on the information obtained from eavesdropping attack) that can partially degrade transmission between the Alice and Bob; d) Pilot contamination attacks.}
   \label{fig:5G-eavesdropping-attacks}
\end{figure}

As we noted in the previous section, when 6G networks start to roll out, a substantial number of 5G devices will still be on. Therefore, mmWave and massive MIMO are still crucial physical layer technologies in 6G compatible (non-standalone) networks. Figure~\ref{fig:5G-eavesdropping-attacks} illustrates three common attacks in mmWave MIMO networks: eavesdropping, jamming, pilot contamination attacks (PCA). Eavesdropping is carried out through inferring and wiretapping (sniffing) open (unsecured) wireless communications. In 6G mmWave MIMO networks, beamforming technology can benefit security. As illustrated in Fig.~\ref{fig:5G-eavesdropping-attacks}, an eavesdropper (Eve) must locate in the beam scope (Figure~\ref{fig:5G-eavesdropping-attacks}.a) or use a reflector (Figure~\ref{fig:5G-eavesdropping-attacks}.a) to wiretap the channel. The eavesdropper can be a legitimate person (internal) of the network (e.g., employee) or someone outside (external). Based on the information about the transmission signals (e.g., the frequency $f_2$) between the transmitter (Alice) and the receiver (Bob), the eavesdropper can carry out two other attacks. One is jamming attack (Figure~\ref{fig:5G-eavesdropping-attacks}.c), where the jammer can inject radio signals (same frequency as Bob's $f_2$), so as to occupy a shared wireless channel. Such aggressive injection prevents legitimate users (e.g., Bob) from using a wireless channel to communicate, a kind of DoS. The other one is PCA, where the attacker intentionally transmits identical pilot signals (spoofing uplink signals in step 2 of Figure~\ref{fig:5G-eavesdropping-attacks}.d) to contaminate user detection and channel estimation phase of the transmitter (Alice). In the worst case, the transmitter will steer the partial beam towards the attacker in the manipulated downlink direction (step 3 of Figure~\ref{fig:5G-eavesdropping-attacks}.d), which practically degrades a legitimate user's transmission or causes signal leakage. Ultra-massive MIMO and Multi-User MIMO are susceptible to PCA\cite{NWang21}. Besides, a cell-free massive MIMO system can risk itself by exposed location of radio stripes \cite{Ylianttila2020}, i.e., dense antennas are easier to reach by physical attacks.

\begin{figure}
    \centering
    \begin{center}
			\includegraphics[width=1\linewidth]{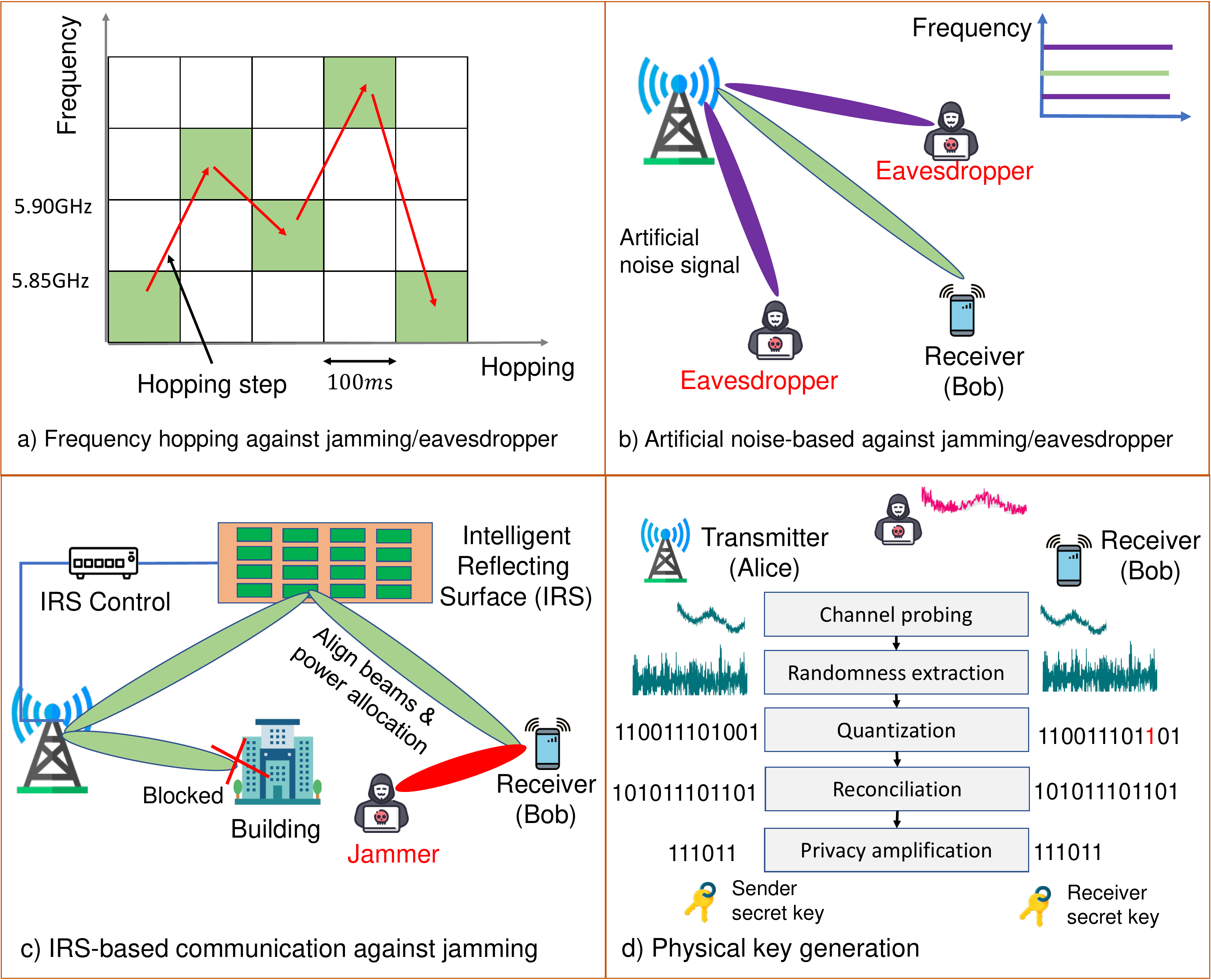}
	\end{center}
   \caption{Illustration of several methods to mitigate eavesdropping/jamming attacks for 6G physical technologies: (a) Frequency hopping; (b) Artificial noise generation; (c) Large Intelligent Surface/Intelligent Reflecting Surface-aided; (d) Physical key generation. }
   \label{fig:6G-PHY-defense-methods}
\end{figure}

Preventing the three aforementioned attack types has attracted much attention recently. In essence, enlarging signal strength between legitimate UEs over an eavesdropper's channels, i.e., maximizing the secrecy rate, is a fundamental approach to preventing eavesdropping and PCA. Key techniques of signal strength-based approaches are to equip secrecy capacity maximization in the precoding process \cite{Liu17}, where the transmitter will spread information signals to the receivers to gain pre-knowledge about the communication channel (i.e., channel state information). The most common idea is to introduce extra randomness into the modulation, which aims to prevent an eavesdropper from predicting the next signal sequence/frequency that the transmitter will use. Many recent studies \cite{Tugnait18,NWang21} rely on this method. For example, Zhang et al. \cite{WZhang18} proposes a method where the sender will create multiple random frequency shifts (as illustrated in Figure~\ref{fig:6G-PHY-defense-methods}.a) in the publicly known pilot sequence or frequency hopping to evade eavesdropping. Another emerging technique is to use covert communication with artificial noise or friendly jamming (as shown in Figure~\ref{fig:6G-PHY-defense-methods}.b), where artificial interference signals will be added in the null-space of a legitimate user channel to confuse the eavesdropper on the real transmission channel\cite{Liu2020,Hong20}. 
Another approach is to use physical key generation (Figure~\ref{fig:6G-PHY-defense-methods}.d), i.e., to exploit the entropy of randomness in the transmit-receive channels such as \ac{CSI} to generate secrecy keys for communications. The authors of \cite{Tang21} propose a physical key exchange between the transmitter and the legitimate users to verify the transmission against untrusted partners. However, integrating the encryption/decryption process into the precoding can impact the performance of the crowded transmission, let alone the threats of internal attacks (i.e., the eavesdropper is one of the legitimate users). An emerging approach is to use AI/ML techniques (e.g., reinforcement learning \cite{Arjoune20}) to enhance \ac{CSI} knowledge and apply proper defense strategies such as channel hopping, although most methods still suffer a setback of high energy consumption. Note that maximizing the secrecy rate can also significantly mitigate jamming attacks. Without obtaining information about particular communication signals between the transmitter and the legitimate receiver (through eavesdropping), it is a challenge for an attacker to jam a communication channel effectively, given the extremely high cost of overwhelming all frequencies in modern times broadband wireless channels. Because of the frequent changes of transmission frequency (frequency hopping), attacking at a fixed frequency also has little impact on the overall performance of the receiver/transmitter. More details on jamming attacks and corresponding anti-jamming methods can be found in the surveys \cite{lichtman20185g,Jameel19}.

\textit{Remaining challenges}

The core issue for MIMO channels is the assumption of the effective precoding process. However, precoding is heavily impacted by fading influence and partial/imperfect \ac{CSI} in practice. Most existing defense approaches, which have heavily relied on the assumption of perfect knowledge of full or partial \ac{CSI} information of the eavesdroppers \cite{Si20}, likely fail at their task in the harsh environment. Utilizing secrecy channels with poor knowledge on the \ac{CSI} of the eavesdropper is then the center of many ongoing efforts. An early study \cite{ZLi21} on the issue suggests a potential solution is to transform the uncertain CSI constraints into deterministic ones through decoupling the legitimate transmission outage probability and the secrecy outage probability. The other open challenge is to detect cooperative eavesdroppers, where several eavesdroppers cooperate to wiretap a wireless channel or perform active attacks (PCA) against the base stations \cite{YLiu17,furqan2020physical}. Finally, besides helping to evade the attacks, the attacker can use covert channels to create a backdoor for leaking data or intrude the system. The common defense methods are detecting dangerous open ports, anomaly signals, or blocking traffic of vulnerable protocols at the upper layers (e.g., ICMP). More detail of covert channel countermeasures can be found at \cite{Zander07}.

\subsection{Security in 6G Large Intelligent Surface}

\ac{LIS}, the other name of Intelligent Reflecting Surface (IRS), is a revolutionary technology for 5G beyond and 6G that uses a planar array of low-cost reflecting elements to dynamically tune the transmission signal phase shift for enhancing communication performance\cite{QWu21}. LIS supports a programmable space through a LIS controller, which is referred to as intelligent control.  Many believe that LIS/IRS will be a critical technology for 6G dense THz networks since deploying so many 6G radio units to overcome the limited coverage of THz communications (presented below) is extremely expensive. By contrast, expanding the use of low-cost LIS/IRS devices (made of metallic or dielectric patches with low-power and low-complexity electronic circuits) to replace several 6G radio units can accomplish the same performance with much lower expenditure. LIS/IRS model benefits security significantly. As illustrated in Figure~\ref{fig:6G-PHY-defense-methods}.c, LIS/IRS can be used to reflect the signals between a base station (Alice) and receivers (e.g., Bob), effectively making propagation channel more favorable. This reflecting model is extremely helpful if the direct link quality (between Alice and Bob) is degraded due to far distance or obstacles. The authors of \cite{Wijewardena21} indicate that optimizing the transmit powers and the phase shift at each element of the LIS/IRS can maximize the sum-secrecy rate, e.g., by destructing the reflected signal power to the eavesdropper or providing different communication links to the receivers. Technically, the idea of using IRS as a data transmitting source to enhance security is similar to that of using multi-path propagation channels to provide different secure communication links to legitimate users \cite{Hafez16}. 

\subsection{Security in NOMA for 6G massive connectivity}

\begin{figure}
    \centering
    \begin{center}
			\includegraphics[width=1\linewidth]{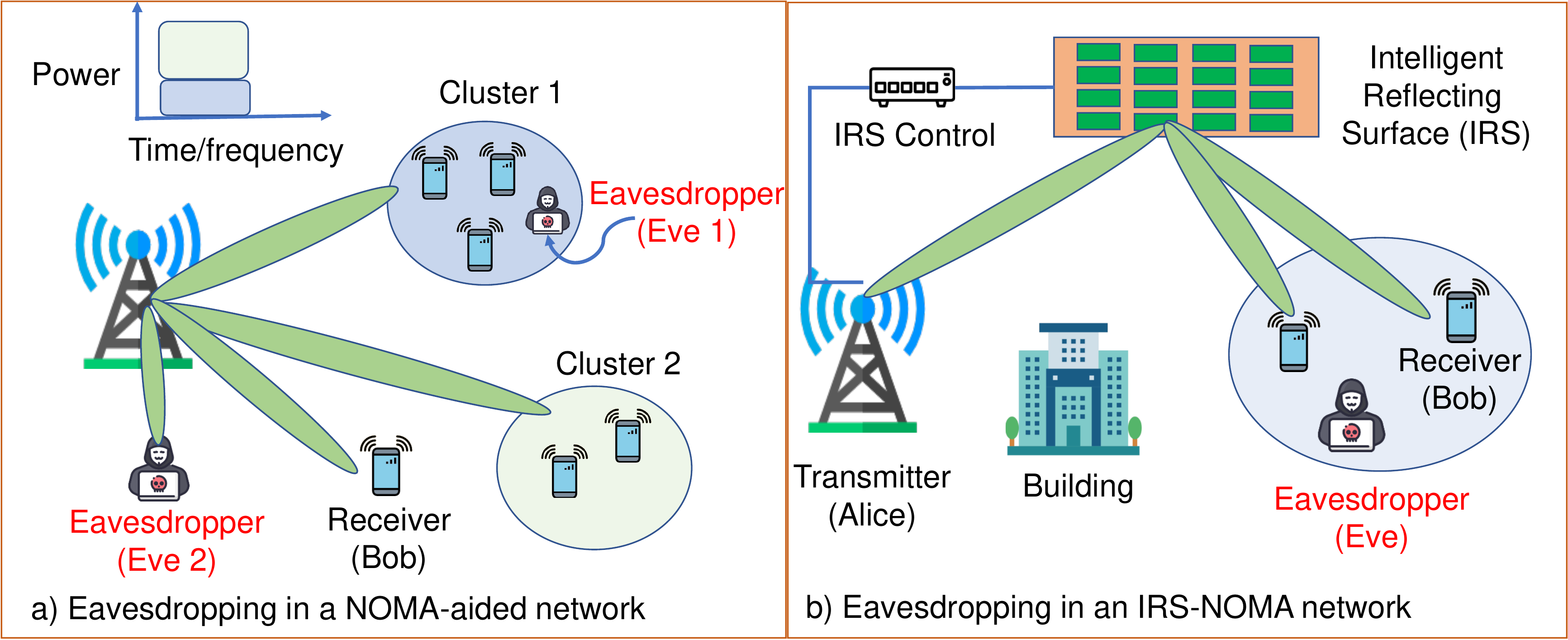}
	\end{center}
   \caption{Illustration of two eavesdropping attack cases in NOMA-aided networks: a) in NOMA-cluster networks and (b) IRS-NOMA networks.}
   \label{fig:5G-NOMA-attacks}
\end{figure}

Besides MIMO, IRS, and beamforming, \ac{NOMA} is another important technology of 5G that can also be used to support 6G massive connectivity. By allocating channel resources for the receivers fairly via assigning more power for ``weak signal'' users and subtracting power of strong signal users \cite{Islam17}, the NOMA-aided transmitters allow more users to gain the channel, which practically increases the number of simultaneously served users. NOMA supports both multicast transmission (to cluster users) and unicast transmission (to specific user). However, NOMA still suffers multiple security threats. For example, as illustrated in Figure~\ref{fig:5G-NOMA-attacks}, an internal eavesdropper can wiretap or launch jamming attacks on the legitimate users in the beam scope of NOMA clusters (Figure~\ref{fig:5G-NOMA-attacks}.a) or reflecting the scope of \ac{IRS} (Figure~\ref{fig:5G-NOMA-attacks}.b). The mentioned techniques on maximizing the transmission secrecy rate in mmWave can be applied for NOMA-aided networks to mitigate the attacks. Moreover, the assessments from channel condition (weak/strong) of legitimate users from NOMA's successive interference cancellation (SIC) process can benefit the CSI knowledge-based defense strategies \cite{Herfeh2020}. However, in an extreme case, several eavesdroppers may cooperate in interfering with the normal operations of a NOMA, e.g., by sending a large number of identical pilot signals to the NOMA-aided transmitter in the ``superimposed messages'' broadcasting stage. As a result, the transmitter may waste much power to allocate for these malicious users. In this case, the cooperation of NOMA systems or NOMA systems with the other partners (e.g., \ac{IRS}) can enhance the security of the systems, i.e., using multi-path channels to provide different communication links to legitimate users. Indeed, enhancing physical layer security for the NOMA-aided networks (i.e., where NOMA is integrated with other technologies such as visible light communication \cite{Peng21} and THz) has been a hot topic recently. A short survey on the relevant matters can be seen at \cite{furqan2020physical}.

\subsection{6G Holographic radio technology with Large Intelligent Surface}

Holographic radio, or holographic beamforming and MIMO, is a new paradigm and disruptive radio technology for 6G indoor/outdoor communications that uses software-defined antenna or photonics-defined antenna arrays than conventional phased arrays or MIMO systems \cite{CHuang20Holo}, i.e., uses no phase shifter or active amplification in the beam-steering process. With the assist of LIS panels, holographic radio can generate the directional beams perfectly through holographic recording and reconstruction. At this point, holographic radio can enable intelligent and reconfigurable wireless environments \cite{Zhang2019} while maintaining low cost, spectrum efficiency, and energy efficiency of network devices greatly. From security, similar to LIS technology, holographic radio will benefit the secrecy rate optimization through canceling out reflections of the base station signals to eavesdroppers \cite{CHuang20Holo}. On the bad side, the electromagnetic waves scatter uncontrollably in the holographic spatial space can cause more concerns of being interfered with or wiretapped \cite{Liaskos18}.

\textit{Remaining challenges}

Holographic radio is heavily under development with many remaining challenges, e.g., few available implementations for hardware design. Similarly, the specific researches on the security aspect of holographic radio technology are still in its infancy.

\subsection{Security in 6G Terahertz communications}

Terahertz is expected to be the central communication technology in 6G. Technically, THz can enable ultra-high data rate (up to terabits per second \cite{Huang11}) for many 6G applications such as tactile Internet and XR/AR services, which 5G technologies such as mmWave will not be able to support \cite{Rappaport19,Saad2020}. However, THz technology is strongly impacted by the surrounding atmospheric conditions, i.e., the spectrum is absorbed by water molecules and spreading loss. THz also has low penetration power against specific obstacles (e.g., thick wall). Because of the high absorption resonance, THz's communication coverage is relatively small, within dozens of meters.

With the low coverage area and high absorption resonance, 6G likely includes dense networks of THz-enabled devices for effective communications. To enhance transmission performance, THz antennas will need to perfectly align signal beams to reduce the angular divergence of transmitted signals. The limited transmission coverage and the high directionality characteristic theoretically make THz much more secure and resilient against attacks, e.g., jamming and eavesdropping \cite{Singh20}. To be successful, an eavesdropper's antenna must be located in the beam scope of transmitting signals, which is even smaller than that of mmWave MIMO antenna, to wiretap a THz link. As a result, it will be much more difficult for an eavesdropper to place a receiver and intercept signals without blocking the receiver's transmission and thereby potentially reveal his attack intention. Performing successful jamming in THz networks to disrupt transmission is not also easy. According to \cite{Singh20}, given the large bandwidths, THz can enable frequency hopping over a large number of sub-channels. The frequency hopping in a wide range can reduce the probability of an adversary detecting and interfering with a particular signal. Moreover, to jam the link successfully, the attacker must generate high power bandwidth to overwhelm the receiver while keeping a short distance, e.g., several meters away from the receiver. The high cost and probably the awareness of users make eavesdropping and jamming less attractive.

\begin{figure}
    \centering
    \begin{center}
			\includegraphics[width=1\linewidth]{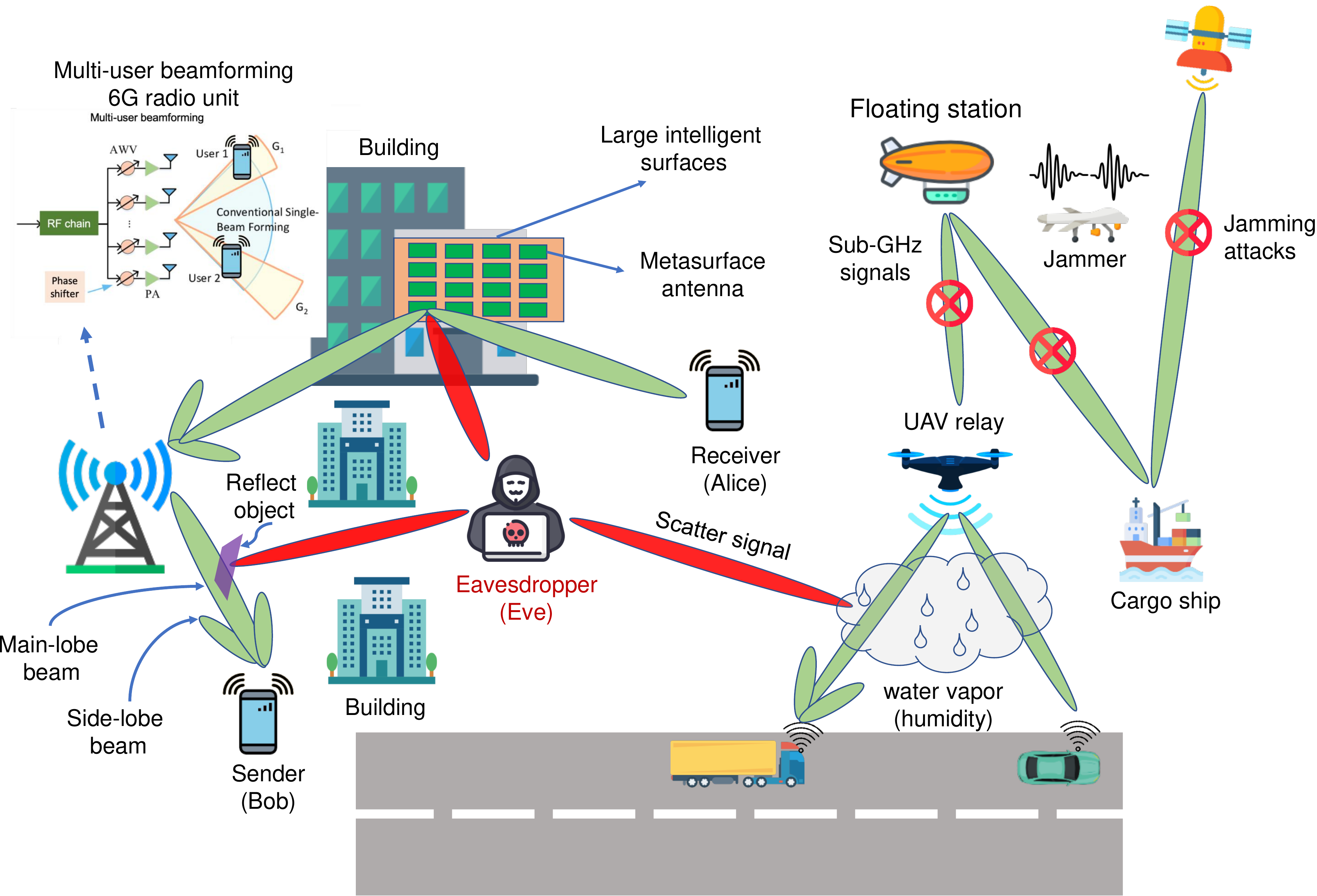}
	\end{center}
   \caption{Illustration of eavesdropping and jamming attacks in 6G heterogeneous networks. The eavesdropper (Eve) can sit behind a building to collect scatter signals from reflectors or the self-creating gaseous cloud.}
   \label{fig:6G-eavesdropping-attacks}
\end{figure}

Despite the high resistance capability against jamming and eavesdropping, THz links can still be attacked in special conditions. The authors of \cite{Ma18} found that wiretapping THz links is still possible in line-of-sight (LOS) transmissions. As illustrated in Figure~\ref{fig:6G-eavesdropping-attacks}, the eavesdropper (Eve) can place a reflector in the LOS beam scope of the sender (Bob) to scatter the radiation towards the attacker's (Eve's) receiver antennas located behind a building. On the other hand, the attacker can exploit the conditions of high humidity by rain/snow or self-creating gaseous cloud to collect the scatter signals of the THz links in an area (e.g., the link between a UAV and the vehicles in Figure~\ref{fig:6G-eavesdropping-attacks}). To prevent this attack type, the authors of \cite{Singh20} suggest that several conventional techniques, such as randomly varying power limits, frequency hopping, and strategic access point placement, can make the communication covert and much more difficult for the attacks to be successful. Ma et al. \cite{Ma18} proposed to share data transmission over multiple paths to enhance covert communication. Porambage et al. \cite{Porambage21} also hint that the multi-path approach can work with the physical key exchange to support sensitive data transmission. Qiao et al. \cite{Qiao21} indicate that \ac{LIS}/\ac{IRS} can also assist in steering signal power to desired user in multiple paths and reduce information leakage in Terahertz Systems. If a multi-path propagation link is not possible, lightweight encryption, beam encryption, and spatial modeling can be used \cite{Singh20}. 

\textit{Remaining challenges}

Despite many defense techniques, jamming and eavesdropping are still open issues. In an extreme case, \cite{Singh20}, cooperation of multiple adversaries to collect scatter signals is still much more challenging to prevent with the mentioned methods, given the complexity of attack re-verification during hand-off of devices. Privacy issues are also particularly concerns in 6G. In essence, the THz spectrum can be used for centimeter-level localization applications. The idea of THz Access Points (APs) following a user's motion to centimeter-level precision for improving link connectivity could expose user locations that can be abused for inappropriate or harmful purposes. In dense networks like 6G, many THz devices can make the problem much more concerned. If the APs are compromised, they likely become potential surveillance devices for adversaries and unauthorized users. Even with potential data anonymization (detailed in Section~\ref{sec:privacy-in-6G}), sensitive or behavioral information about users can still be assembled from mining information of multiple sources \cite{Singh20}. In this vision, maintaining the privacy principles (as GDPR requires) becomes critical but particularly challenging for devices with limited memory, computation power, and energy. Besides, the location exposure of APs can attract physical attacks to shut down the networks. In conclusion, there is a trade-off between high privacy and optimization for THz communications. A potential solution is to accept conditional anonymity and secure access points with hardware security modules to prevent compromise attacks.

\subsection{Security in 6G VLC communications}

Visible light communication (VLC) is a high-speed communication technique to transmit data by using visible light between 400 and 800 THz. According to \cite{NanChi20}, in an in-lab experiment, a free-space VLC system can integrate with a 50-cent off-the-shelf Light Emitting Diode (LED) to provide peak data rate up to 15.73Gbps over a distance of 1m. Similarly, the study reports the peak data rate for an underwater VLC system can reach 16.6Gbps over 5m and 6.6Gbps over 55-m in water. With a promising future of higher transmission rates and cheap deployment, VLC will be an economical alternative approach to complement 6G THz systems in indoor buildings (hospitals, personal rooms), underwater applications, or electromagnetic sensitive areas (nuclear plants). Another variant of VLC is Light Fidelity (LiFi), which is considered as the incorporation of WiFi and VLC to support bidirectional wireless networking with access points. Besides supporting super-high bandwidth, the big advantage of VLC/Li-Fi over other wireless technologies is almost no limitations on capacity \cite{XWu21} because of the large visible light spectrum. However, VLC/LiFi cannot penetrate a wall.

Like THz, VLC/LiFi is more secure than prior wireless communication techniques, e.g., WiFi. VLC/LiFi's coverage is small and cannot penetrate opaque objects like walls. Similar to THz technology, an eavesdropper has to move into the LOS area of the sender-transmitter VLC/LiFi link to intercept signals. There are several methods to mitigate the risk of such attacks in the literature. The goal is to maximize the secrecy rate under the constraints of the open channel. In an early study, Mostafa et al. \cite{Mostafa15} proposed to utilize beamforming signals of the transmitters (e.g., main lobe beams focus on the legitimate users) for gaining a certain secrecy rate. For multiple transmit models, the idea of using spatial modulation aided VLC systems with optical jamming has got much attention recently \cite{Yesilkaya20,Panayirci20}. In this technique, friendly jamming signals will be inserted into the null space of the legitimate user's channel matrix. As a result, the attacker may need to span interference signals to follow and suffer high costs while secrecy channels are protected. Recent studies \cite{Panayirci20} indicate that the combination of spatial modulation with the zero-forcing precoding strategy to utilize beamforming also gives promising results in MIMO-VLC systems. In essence, friendly jamming and beamforming aid are still mainstream techniques for protecting VLC/LiFi communications.

\textit{Remaining challenges}

Most recent studies assume a VLC environment with a single eavesdropper only. However, a large-scale VLC network with the presence of multiple eavesdroppers scattered randomly (conference rooms) can be more common in the future. Proposing innovative methods for protecting VLC systems against the collusion of these collusion attackers can be an important topic in the coming years.

\subsection{Security in 6G Molecular communications}

According to \cite{Moritani10}, molecular communication is defined as a communication technique that uses chemical signals or molecules (instead of electronic and optical signals in traditional communications) as an information carrier for nano/cell-scale entities to communicate with each other. At the end of this decade, advanced nanotechnologies may be able to enable the industry-level manufacture of nanodevices to be used for many fields, e.g., drug delivery in blood vessels \cite{Zhang2019} in healthcare and water/fuel distribution monitoring in industry. The target of molecular communications is to provide connectivity for such nanodevices. 6G specifies the term of the Internet of Nano-Things and extremely low-power communications (ELPC) to indicate the vision of supporting communication among the nanodevices \cite{Zhang2019}. Besides civil applications, molecular communications are useful to provide alternative communication methods or extra covert channels in harsh environments such as water or adversarial networks (high jamming and interception), where wave-based communications often fail or suffer heavy absorption. In short, because of high promising applications, molecular communications can be in the 6G physical layer \cite{Guo21}.

Security and privacy are particular concerns when many nanodevices may be embedded into a human body, e.g., to deliver drugs. A major threat is the potential leak of healthcare information. On the other hand, bio-machines can be remotely attacked if the communication systems are also connected to the Internet. An attacker can exploit classic vulnerabilities of IoT devices to compromise the molecular control system remotely. The authors in \cite{Valeria14} envisaged a case where, by manipulating the configuration in bio-machines, an attacker can force to accelerate/delay the absorption process. In another attack, an attacker floods the environment with particles or kills the molecules to disrupt medical application functions in order to harm or kill the host. The side effects will be severe if a large number of bio-machines are in malfunction. In extreme cases, the immune system may react strongly to the malicious stimulate from the malfunctioned bio-machines and harm the body. The study \cite{DRESSLER2012} also showed a situation in which nano-robots can be manipulated to damage the patient blood vessels instead of repairing them. However, there have been no such real attacks to be recorded to date. Given the vision of implantable devices and nano-robots to be common in 6G, protecting molecular communications against the mentioned threats is still critical.

With different network structures and interactions, protecting security and privacy for molecular communications are different from conventional approaches, i.e., existing solutions to wireless communications cannot be applied to molecular systems. Several preliminary studies \cite{DRESSLER2012} envisage that biochemical cryptography, which uses biological molecules like DNA/RNA information or protein structure to encode information, can protect information integrity in molecular communications. Suppose molecular communications are connected to the Internet. In that case, we believe that a strict access control system or strong firewall at the gateway portal may need to distinguish unauthorized users and filter out malicious traffic.

\textit{Remaining challenges}

The research on preventing security attacks and potential data breaches in molecular communications is still at the early stage. Given the ethics, carrying out a real attack to a host or human body is somewhat less attractive.

\subsection{Other prospective technologies for 6G physical layer security}

The following three key physical-layer-based technologies will benefit many 6G applications in terms of mitigating special attacks, such as spoofing messages and tampering physical data bits, which can occur on the network and application layers.

\subsubsection{Physical layer authentication}

\begin{figure}
    \centering
    \begin{center}
			\includegraphics[width=1\linewidth]{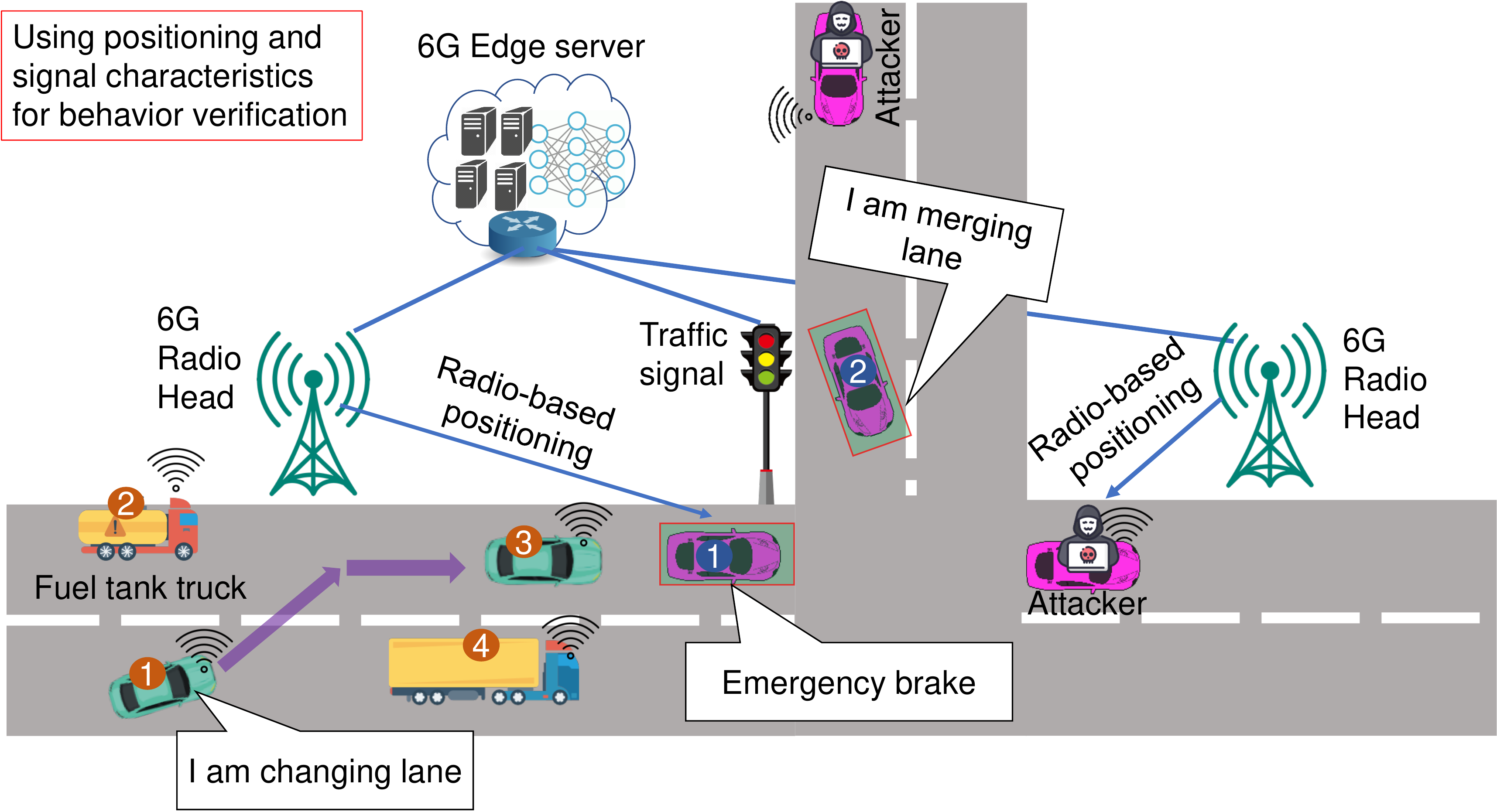}
	\end{center}
   \caption{An illustration of spoofing attacks to disseminate falsification beacon messages to the vehicles behind or approaching ones to an intersection in vehicular networks, which can be very common in 6G.}
   \label{fig:6G-PHY-authentication}
\end{figure}

Physical layer authentication is an emerging technology to crack down spoofing or impersonation attacks. Figure~\ref{fig:6G-PHY-authentication} illustrates several cases in which attackers intentionally send falsification messages in vehicular networks to fool nearby vehicles. This attack type can cause high risks to users when connected and autonomous vehicles become popular in 6G \cite{Wang2020}. The essence of physical layer authentication is to recognize the identities of subscribers by exploiting location-specific or device-specific properties of wireless channels. For example, it is traditionally difficult to detect Sybil attacks at the application layer, where an attacker uses a large number of pseudonymous identities to send spoofing messages (in vehicular networks) or disproportionately influence reputation-based verification systems (misbehavior detection). Using physical attributes (e.g., angle-of-arrival, Received Signal Strength Indicator) in this case has significant advantages to detect the attack successfully since all the spoofing messages, regardless of using many pseudonyms, come from a single signal source and direction \cite{Nguyen2020Tran}. The authors of \cite{Jian20} proposed to learn unchanging, hardware-based characteristics of the transmitter, such as its instantaneous amplitudes, phases, and frequencies, with the reference templates of all known devices to find out the attacker. The authors of \cite{Senigagliesi21} indicate that a deep learning-aided physical layer authentication model can soon outperform conventional statistical methods in terms of detection accuracy, particularly in heavily corrupted channels by noise. 

\textit{Remaining challenges}

CSI estimation errors are the major challenges of physical layer authentication. Since physical layer authentication relies on the uniqueness of physical characteristics over any transmit-receive channels, it is sensitive to incorrect CSI estimation. Theoretically, a correct reference of CSIs for comparison is a must. However, that assumption is not always correct in all contexts. The authors of \cite{Xiao08} found that low SNR (e.g., noise, user movements, distance, etc.) and overlapping CSIs from crowded nearby sources can significantly degrade the performance of the hypothesis testing process. Training data in particular can be compromised with signal patterns from untrusted sources, such as some users colluding with an adversary \cite{NXie20}. Thus, protecting data integrity for hypothesis testing is critical to the success of physical layer authentication. A potential solution is to fuse data from multiple sources or track multiple physical layer attributes, i.e., multi-attribute multi-observation techniques \cite{Fang18}, to mitigate the bias of each and enhance the overall detection performance.

\subsubsection{Physical key generation}

Physical key generation is to protect confidentiality for communications between UEs and stations from eavesdropping. The fundamental idea is to exploit the entropy of randomness in transmit-receive channels such as CSI and Received-Signal-Strength (RSS) to generate secrecy keys for communications. Figure~\ref{fig:6G-PHY-defense-methods}d) illustrates generic physical key generation on a TDD-based wireless channel. In the randomness extraction process, the sender and the receiver measure both CSI and RSS. Theoretically, the measured information is identical when two terminals are connected by the same wireless channel but can be different if the responder (e.g., attacker) is located one-half wavelength away from the sender. The quantization is to generate the extracted randomness into bits encoded to ensure perfect secrecy in the encoding process. To ensure the keys generated on both sides are the same, reconciliation is performed synchronously between the sender and the receiver. Such synchronous privacy amplification then creates the final encryption keys from the generated bit sequences and practically eliminates the threats from an eavesdropper's partial information, e.g., obtained from any previous processes. Existing physical key generation techniques can be CSI-based \cite{XLu21}, RSS-based \cite{HLiu14}, phase-based \cite{Ebrahimi21}, or code-based \cite{Aldaghri20}. 

\begin{table*}[ht]
\caption{Prospective solutions to enhance 6G physical layer security}
\label{tab:security-solution-physical-layers}
\begin{adjustbox}{width=1\textwidth}
\small
\begin{tabular}{lllllll}
\hline
\textbf{6G PHY technologies} &
  \textbf{Reference} &
  \textbf{Security \& privacy issues} &
  \textbf{Key solutions} &
  \textbf{Key points} &
  \textbf{Open problems} \\ \hline
 \begin{tabular}[c]{@{}l@{}}mmWave \\ MIMO beamforming\end{tabular} &
  \begin{tabular}[c]{@{}l@{}} \cite{Liu17}, \cite{Tang21}, \\ \cite{Perazzone21} \\ \end{tabular} &
  \begin{tabular}[c]{@{}l@{}}Eavesdropping\\ Jamming\\ Pilot contamination \\ Location exposure\end{tabular}
   & \begin{tabular}[c]{@{}l@{}}$\bigcdot$ Frequency hopping \\ $\bigcdot$ Injecting artificial noise \\ or friendly jamming\\ $\bigcdot$ Utilize beam alignment\\ $\bigcdot$ Physical key generation\\$\bigcdot$  Physical coding\end{tabular} & \begin{tabular}[c]{@{}l@{}}Secrecy rate maximization \end{tabular} & \begin{tabular}[c]{@{}l@{}}$\bigcdot$ Optimal beam alignment\\ $\bigcdot$ AI-based low-complexity \\ anti-jamming\\ $\bigcdot$ High-performance coding\\ $\bigcdot$ Energy efficient solutions\end{tabular}
   \\ \rowcolor[HTML]{EFEFEF} 
Large Intelligent Surface &
  \begin{tabular}[c]{@{}l@{}}\cite{Hafez16}, \cite{Wijewardena21}\end{tabular} &
  \begin{tabular}[c]{@{}l@{}} Eavesdropping \\ Location exposure \end{tabular} &
  \begin{tabular}[c]{@{}l@{}}$\bigcdot$ Frequency hopping \\ $\bigcdot$ Injecting artificial noise \\ or friendly jamming,\\ $\bigcdot$ Physical key generation\\ $\bigcdot$ Physical coding\end{tabular} & 
 \begin{tabular}[c]{@{}l@{}}Secrecy rate maximization\end{tabular} & \begin{tabular}[c]{@{}l@{}}$\bigcdot$ Optimal LIS deployment\\ $\bigcdot$ AI-enabled LIS\\ $\bigcdot$ Specific LIS applications\\ $\bigcdot$ Energy efficient solutions\end{tabular}
   \\
NOMA &
  \begin{tabular}[c]{@{}l@{}} \cite{Herfeh2020}, \cite{Peng21}, \\ \cite{furqan2020physical} \end{tabular} &
  \begin{tabular}[c]{@{}l@{}}Eavesdropping\\ Power allocation \\ contamination \\ Location exposure\\ Signal space expose\end{tabular} & \begin{tabular}[c]{@{}l@{}}$\bigcdot$ Frequency hopping\\ $\bigcdot$ Physical key generation\\ $\bigcdot$ Physical coding\end{tabular} & \begin{tabular}[c]{@{}l@{}}Secrecy rate maximization \end{tabular}  & \begin{tabular}[c]{@{}l@{}} $\bigcdot$ Security for NOMA-VLC,\\ NOMA-THz, NOMA-LIS\\ networks\end{tabular}
   \\ \rowcolor[HTML]{EFEFEF} 
Holographic radio &
  \begin{tabular}[c]{@{}l@{}} \cite{CHuang20Holo}, \cite{Liaskos18}  \end{tabular} &
   \begin{tabular}[c]{@{}l@{}}Eavesdropping \\ Location exposure \end{tabular}  &
  \begin{tabular}[c]{@{}l@{}}$\bigcdot$ Utilize beam alignment\\ $\bigcdot$ Randomly power limits\\ $\bigcdot$ Access point placement\\ $\bigcdot$ Physical key generation\\ $\bigcdot$ Physical coding\end{tabular} & \begin{tabular}[c]{@{}l@{}}Secrecy rate maximization \end{tabular} & \begin{tabular}[c]{@{}l@{}}$\bigcdot$ Optimal radio management\\ $\bigcdot$ Joint RF and non-RF hardware\\ $\bigcdot$ Holographic radio-LIS \\ integration\\ $\bigcdot$ Energy efficient solutions\end{tabular}
   \\
Terahertz communications &
  \begin{tabular}[c]{@{}l@{}} \cite{Singh20}, \cite{Ma18}, \\ \cite{Qiao21}\end{tabular} &
   \begin{tabular}[c]{@{}l@{}}Eavesdropping\\ Jamming \\ Location exposure\end{tabular}  &
  \begin{tabular}[c]{@{}l@{}}$\bigcdot$ Frequency hopping \\ $\bigcdot$ Randomly power  limits\\ $\bigcdot$ Access point placement\\ $\bigcdot$ Utilize beam alignment\\ $\bigcdot$ Injecting artificial noise \\ or friendly jamming\\ $\bigcdot$ Physical key generation\\ $\bigcdot$ Physical coding\end{tabular} & \begin{tabular}[c]{@{}l@{}}Secrecy rate maximization \end{tabular} &  \begin{tabular}[c]{@{}l@{}} $\bigcdot$ Optimal THz base stations\\ $\bigcdot$ Optimal THz-LIS integration\\ $\bigcdot$ Optimal beam alignment\\ $\bigcdot$ $\bigcdot$ AI-based low-complexity\\ anti-jamming solutions\\ $\bigcdot$ High-performance coding\\ $\bigcdot$ Optimal mmWave-THz links\\ $\bigcdot$ Energy efficient solutions\end{tabular}
   \\ \rowcolor[HTML]{EFEFEF} 
VLC communications &
  \begin{tabular}[c]{@{}l@{}}\cite{NanChi20}, \cite{XWu21}, \\ \cite{Panayirci20} \end{tabular} &
  \begin{tabular}[c]{@{}l@{}}Eavesdropping\\ Obscured attacks\end{tabular}  &
  \begin{tabular}[c]{@{}l@{}}$\bigcdot$ Frequency hopping \\ $\bigcdot$ Injecting artificial noise \\ or friendly jamming\\ $\bigcdot$ Physical key generation\\ $\bigcdot$ Physical coding\end{tabular} &  \begin{tabular}[c]{@{}l@{}}Secrecy rate maximization \end{tabular} & \begin{tabular}[c]{@{}l@{}}$\bigcdot$ NOMA-VLC performance\\ $\bigcdot$ VLC/LiFi deployment\\ $\bigcdot$ Optimal VLC access points\end{tabular} 
   \\
Molecular communications &
  \begin{tabular}[c]{@{}l@{}} \cite{Zhang2019}, \cite{DRESSLER2012}, \\ \cite{Valeria14}  \end{tabular} &
    \begin{tabular}[c]{@{}l@{}}Device configuration\\ manipulation, kills the\\  molecules, attacking\\ bio-machines from \\ Internet environment \\ Data leakage \end{tabular} &
  \begin{tabular}[c]{@{}l@{}}$\bigcdot$ Biochemical cryptography\\ $\bigcdot$ Firewall, IDS to detect\\ attacks from the Internet\end{tabular} & \begin{tabular}[c]{@{}l@{}}Confidentiality \&\\ Integrity\end{tabular} & \begin{tabular}[c]{@{}l@{}} $\bigcdot$ Energy efficient solutions\\ $\bigcdot$ Secure Internet access\end{tabular}
   \\ \rowcolor[HTML]{EFEFEF} 
\begin{tabular}[c]{@{}l@{}}Physical-aided security\end{tabular} &
  \begin{tabular}[c]{@{}l@{}} \cite{Jian20}, \cite{Nguyen2020Tran}, \\ \cite{Senigagliesi21}\end{tabular} &
  \begin{tabular}[c]{@{}l@{}}Sybil attack \\  Physical data  tampering \\ Trajectory tracking\end{tabular}  &
  \begin{tabular}[c]{@{}l@{}}$\bigcdot$ Physical layer authentication\end{tabular} &  \begin{tabular}[c]{@{}l@{}}Exploit physical \\signal attributes\\ to detect special \\ attacks\end{tabular} & \begin{tabular}[c]{@{}l@{}} $\bigcdot$ AI-based low-complexity\\ solution\\ $\bigcdot$ Multi-attribute \\ multi-observation technique\end{tabular}\\ \hline
  
\end{tabular}
\end{adjustbox}
\end{table*}

\textit{Remaining challenges}

Securing wireless communications based on physical key generation has its weakness. First, the key generation process must deal with heavy computation in signal processing and encoding. The key error-correction process in reconciliation requires several extra bits to reconcile bit mismatch that often consumes a significant amount of time and space. Second, due to the limited capacity of wireless channels and reconciliation overhead, physical key generation techniques offer very low key-generation rates. The authors of \cite{HLiu14,PXu16} propose to use group keys to overcome the low-rate problem. The group-key generation model can increase the efficiency of key usages, e.g., all the nodes in a group use an identical key. However, the challenge is to maintain trust in a group. Defining optimal criteria to group the nodes correctly is another open issue. 

\subsubsection{Physical coding for PHY data integrity protection}

Physical coding is the fundamental process in communication technology to maximize data transmission rate among legitimate users. However, enhancing physical coding can significantly mitigate attacks on integrity, such as tampering with messages in transit. For now, several structured coding schemes such as \ac{LDPC} coding and polar coding have demonstrated excellent performance in enhancing both integrity and network capacity in a 5G physical layer \cite{Liu17}. For 6G, advanced channel coding will be crucial for achieving super speed (terabits per second), extremely low latency, and reliability. New coding schemes for 6G, such as space-time coding and quasi-cyclic multi-user LDPC are still under development. The preliminary results of the study in \cite{Zhang2018} indicate that, by controlling the propagation direction and harmonic power distribution simultaneously, the proposed space-time modulated digital coding scheme can significantly improve reliability of data transmission in multi-antenna technologies. The technology can be applied to enhance performance for several 6G technologies such as adaptive beamforming and holographic imaging.

\textit{Remaining challenges} 

Optimal coding requires perfect knowledge of \ac{CSI} from both the receiver and the sender, including their transmission probabilities and channel gains. However, in practice, estimation errors, feedback quantization errors/delays, or channel mobility are challenges to CSI estimation. Fading influence and partial/imperfect CSI are the prime challenges for achieving secrecy capacity with coding techniques. The authors of \cite{Mao18} list several promising studies on using deep learning models to overcome the challenges as well as improving the error-correction and decoding/encoding process of \ac{LDPC} and polar code. However, lack of practical implementation is still a major issue.

\subsection{Summary of lessons learned from physical layer security}

 Table~\ref{tab:security-solution-physical-layers} summarizes key technologies, security attacks, corresponding solutions and open problems in enhancing physical layer security. Eavesdropping is the most common threat over nearly all the technologies while the favorite defense method is to maximize the secrecy rate for the communication channel. Also, when joint communication and radar become real in 6G, because of the high directionality characteristic in 6G communications, location exposure will be likely the major privacy concern for subscribers. Three lessons learned from the survey for 6G physical layer security are as follows.
 \begin{enumerate}

     \item \textit{Monitoring the surrounding atmospheric conditions and the objects present in the transmitter-receiver path is critical for ensuring reliable and secure 6G physical communications}. Since the harsh environment conditions (rain/fog) or thick obstacles like building can cause a THz/VLC link to scatter signals to surrounding areas, making the link vulnerable to eavesdropping. In this way, an attacker may release gaseous particles or molecules sensitive to the target THz links to cause inference to signals. The attacker can then adjust this physical interference to cause scattering of signals, i.e., signals can get deviated from their trajectories to non-uniform paths after passing through a medium, aiding eavesdropping of the signal. And then, a multi-attribute multi-observation technique to sense the surrounding environment and adjust beamforming and routing strategy can significantly reinforce the security capability for THz/VLC links and mitigate the impact of anti-jamming and eavesdropping.
     
     \item \textit{For AI-aided physical layer security methods, dataset is an Achilles' heel}. Lack of a rigorous public dataset for testing is a major obstacle to a breakthrough in performance detection. Thus, generating/collecting a qualified physical-layer dataset (e.g., benign, attack logs) will be critical and beneficial for 6G-PHY-related research. Also, if the jammers are equipped with intelligent capability, the mentioned conventional approaches likely fail to gain effective defense.
    
     
      \item \textit{Physical layer security will be a breakthrough for 6G security}. When millions of devices are connected in 6G, securing communications by cryptography alone is somewhat insufficient, given the growing threats of many physical attacks. And then 6G physical layer security will be the key technology to satisfy the requirement, a feature many earlier generations wanted but has achieved little success so far. However, based on our survey, neither 6G enabling physical technologies (THz, VLC, LIS/NOMA) nor any state-of-the-art solution of the seven protection approaches mentioned above, has the capability of resisting all physical attacks, such as jamming and eavesdropping. Further security enhancements for the technologies are \textit{imperative} in the coming years to reach the goal.


 \end{enumerate}

%% file: Section_VI_Connection.tex
The connection layer can be seen as a combination of network and transport layers. Security in this layer addresses a broad range of communication security issues between a UE and its requested services, particularly the network segments of access networks and core networks. The network segments include access networks (radio units, gNB, ground stations), and core networks (endpoint gateways, authentication servers, edge servers). For years, communications in the connection layer have been subject to many notorious attacks. For example, the lack of integrity protection of signalling data traffic between UE and the gNB or AMF can let an attacker compromise and alter data or advance a spoofing attack. Worse, exploiting the paging procedures, an attacker can launch massive signalling DoS attacks from millions of injected mobile UEs \cite{Hussain21} to overload core networks (authentication servers) and degrade/block access of legitimate subscribers. The following subsections discuss key security concerns and prospective technologies that enhance 6G connection layer security. Note that network operators are the main stakeholders to take in charge of security protection in this layer.

\subsection{6G authentication and key management (6G-AKA) for mutual authentication between the subscriber and the network: Network access control}
\label{subsec:6G-AKA-authentication}

Like the upgrade of 4G EPS-AKA to 5G AKA, 6G AKA will certainly require significant upgrades of 5G AKA in order to satisfy highly personalized services and requirements of novel applications such as holographic telepresence. Since many components of 6G networks have not yet been formalized, it is unclear how the final 6G AKA shape will look. However, based on the open security problems of 5G, 6G AKA expects to solve the following issues.
 
\begin{figure}
    \centering
    \begin{center}
			\includegraphics[width=1\linewidth]{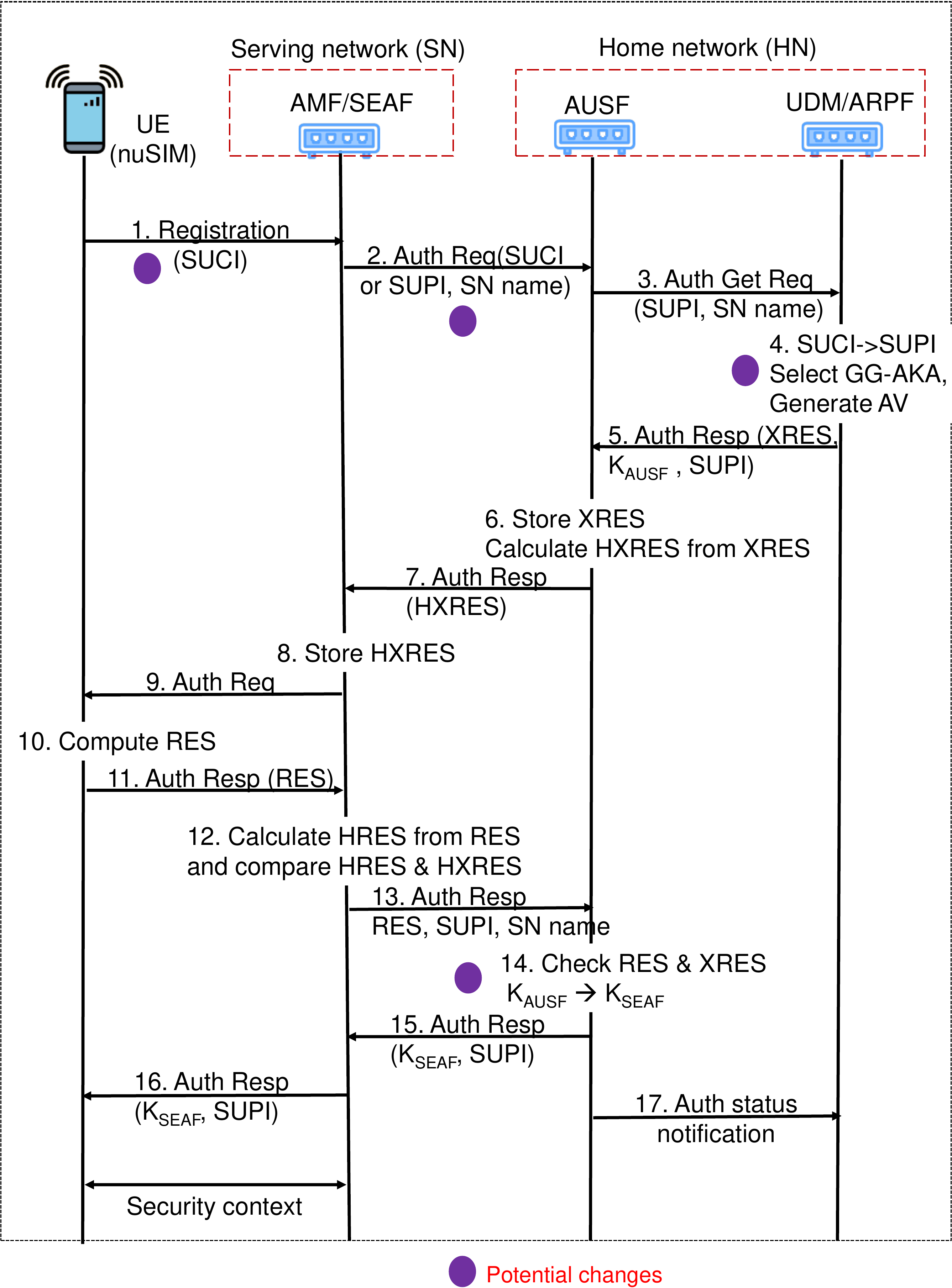}
	\end{center}
   \caption{An illustration of 5G-AKA for reference on the potential changes of 6G AKA. For now, the authentication occurs in the serving network (SN) against the subscribers is weak. }
   \label{fig:6g-AKA-authentication}
\end{figure}

 \begin{itemize}
     \item Stronger authentication between the serving network (SN) and the subscribers as well as between the SN and the home network (HN) should be seriously considered. For example, as illustrated in Figure~\ref{fig:6g-AKA-authentication}, AUSF and ARPF of the HN take the main responsibility for authentication on the subscribers. The problem is the subscriber's SUPI and $K_{SEAF}$ only appear together in a one-way response (steps 15, 16). Due to this limitation, if there are two concurrent authentication sessions on the SN and the HN, $K_{SEAF}$ in step 15 can be linked into another SUPI of the concurrent session. Consequently, the SN may associate the session key $K_{SEAF}$ to a wrong SUPI. Exploiting the vulnerability, an attacker can transfer his network bill to someone else for what he used on the SN \cite{Basin18}. Note that SUPI can be tracked by using rogue stations \cite{Jover19}. Although there is no attack recorded on this vulnerability, 6G AKA can prevent the threat by enabling the stronger binding between the HN and the SN.
     
     \item Closing the gap for the security capabilities of the network operators and dual authentication model is critical. Like 5G, 6G can be a huge heterogeneous network where many operators with different security capabilities co-exist. Some small operators may provide 5G or older generations in the rural areas or sparse populations. The dual authentication to support both old-generation and new-generation subscribers along with weak security infrastructure of the small operators can double trouble for security protection. The attacker can carry out bidding down attacks \cite{ETSI.TS133.501} to fool the small HN that the subscriber does not have a security capability and select the weaker authentication model (step 4 in Figure~\ref{fig:6g-AKA-authentication}). Efficient authorization and authentication to protect the subscribers, regardless of the operators' capability, is critical. For supporting a secure united authentication model in 6G space-air-ground-sea networks, non-3GPP authentication protocols for open interfaces can be implemented with quantum-resistant algorithms (present below).
     
     \item A new subscriber identifier privacy model (refer more to Section~\ref{subsec:vision-6G-security-architecture}) can require a new design for 6G AKA. Privacy is still a problem in 5G for now. 3GPP recommends using SUCI to enhance anonymity during the network access (e.g., as step 1 in Figure~\ref{fig:6g-AKA-authentication}). However, a fake home network injector can track users, similar to IMSI catching \cite{Jover19}. To protect privacy, a better solution is to let the home network's AUSF publish its certificate to registering UEs, which then encrypts the SUPI to prevent it from being identified by the SN. The challenge of this approach is to determine who are trusted partners. Another is a requirement to support lawful interceptions in the case of satisfying data recovery from authorized agencies for crime investigation. If a new identity model (e.g., non-ID) becomes reality, privacy preservation in 6G AKA likely needs a reform of attaching identifier information in authentication requests.

 \end{itemize}

\subsection{Quantum-safe algorithms and quantum communication networks for 6G secure communication}
\label{subsec:communication-security}

Preventing unauthorized interceptors from accessing communications is an essential requirement of secure data transmission in cellular networks. For example, 3GPP suggests using cipher algorithms such 128-NEA1(SNOW 3G)/128-NEA2(AES-128 CTR)/128-NEA3(128-bit ZUC) to protect the confidentiality of user data \cite{Verizon5GPrivacy} and 128-NIA1(SNOW 3G)/128-NIA2(AES-128 CMAC)/128-NIA3(128-bit ZUC) for integrity checking in the 5G networks. However, quantum computing appears to threaten the security status quo pretty soon \cite{Carames20}. With the capability of searching and factoring much faster than a classical computer, a quantum computer can theoretically break any cryptosystem built on top of the mathematical complexities of integer factoring and discrete logarithms by running Shor's algorithms \cite{ETSIQuantum}. Accordingly, most public-key cryptography such as RSA and ECC and related security protocols (e.g., SSH, IPSec, TLS) are vulnerable to quantum attacks. By contrast, symmetric key algorithms with a proper key length (e.g., AES-256, SNOW 3G-256) or good hash functions (e.g., SHA-2, SHA-3) are safe from quantum attacks. Although there are no known attacks that successfully break existing public cryptographic schemes \cite{Bernstein17}, many cryptographers are still pursuing futuristic cryptographic schemes to prepare for when such attacks become reality. Adopting quantum-safe cryptographic schemes is also vital for protecting long-lived sensitive information (e.g., finance transactions). This is to prevent an adversary from obtaining the documents and decrying them later when quantum computing is available.

\begin{figure}[ht]
    \centering
    \begin{center}
			\includegraphics[width=1\linewidth]{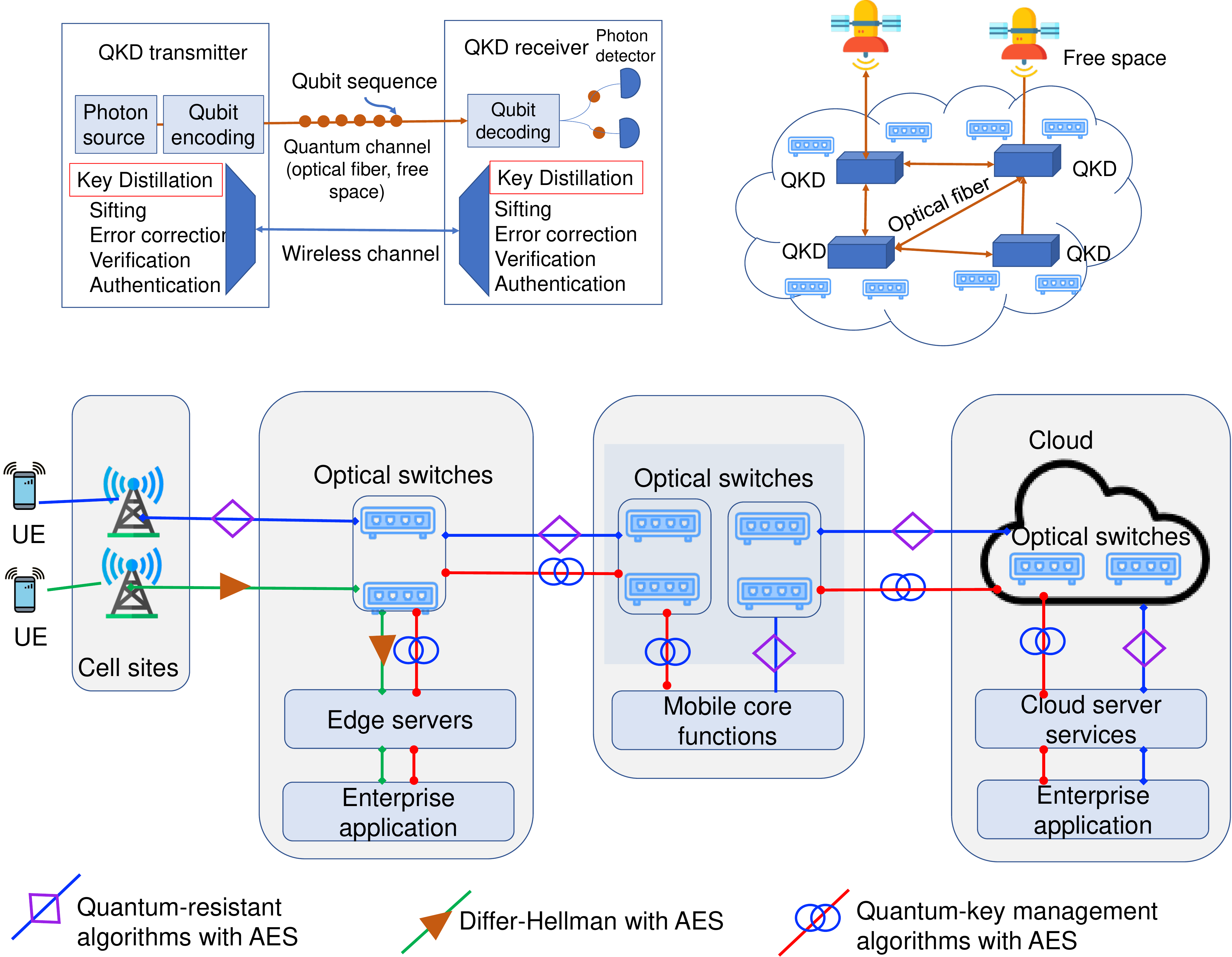}
	\end{center}
   \caption{The illustration of a QKD system and the deployment of using quantum-safe communications for various enterprises. Quantum-resistant algorithms with AES will likely be used for many purposes, while quantum-key models may be applied to high sensitive applications first.}
   \label{fig:quantum-key-management}
\end{figure}

There are two approaches to build quantum-safe algorithms: (1) in the near term, enhancing existing ciphersuites and related protocols to support a certain quantum resistance, and (2) in the long term, using post-quantum algorithms. In the first approach, cryptographic algorithms can extend their key length to enhance the resistance to quantum attacks. However, increasing the key sizes of nearly all public-key ciphers is infeasible to adapt to a constant increase of quantum computing power every year \cite{ETSIQuantum}. For the long-term target, a potential enhancement is to use a public-key quantum-resistant algorithm like lattice-based cryptography (e.g., NTRU) to replace RSA/ECC. Another prospective technology to enhance the quantum resistance in the public-key cryptography model is to use quantum key distribution (QKD) (as illustrated in Figure~\ref{fig:quantum-key-management}). QKD enables security based on fundamental laws in quantum physics (transmitting a string of photons in real time) and quantum information theory (cannot be eavesdropped without detecting). However, applying QKD for long-distance transmission is still a technical challenge due to the difficulty of developing repeater systems for QKD networks \cite{ETSIQuantum}. In the first quarter of 2021, researchers demonstrated the state-of-the-art prototypes of QKD networks to support transmission over 4,600km if use satellites \cite{ChenY2021} or 511km on the ground if using optical fiber networks \cite{ChenJ2021}. However, due to the high cost, it is unclear how to deploy such a QKD worldwide. Another emerging method is to use the quantum-safe hybrid key exchange mechanisms, which is based on the theory that the cryptosystem will remain secure if one of its key exchange methods remains secure \cite{ETSI.TS103.744}. Following this direction, some researchers propose to combine a classic key exchange method like Elliptic-curve Diffie–Hellman (ECDH) and a quantum-safe key-encapsulation mechanism (KEM), e.g., ECDH with NIST P-256, Kyber512, and SHA-256.  

The most feasible plan securing 6G communication is gradual transformation and coexistence of the current ciphersuites. A quantum-safe cryptographic model like QKD will be deployed on the market demands and the progress of standardization. Currently, NIST has been hosting a contest since 2016 \footnote{https://csrc.nist.gov/projects/post-quantum-cryptography} to find the optimal quantum-safe cryptographic standard for the U.S., which will likely be applied worldwide. Figure~\ref{fig:quantum-key-management} illustrates the coexistence of the ciphersuites among network nodes. The QKD model based on post-quantum algorithms and symmetric cryptography (e.g., QKD-AES) will be applied to enterprise nodes, highly sensitive applications, and key network elements in 6G networks. By contrast, the advanced standard Diffie–Hellman key exchange with AES, or quantum-resistant algorithms (QRA) based on NTRU/AES could be more suitable for legacy networks \cite{Wright21} and regular applications.

\textit{Remaining challenges}

\begin{figure}
    \centering
    \begin{center}
			\includegraphics[width=1\linewidth]{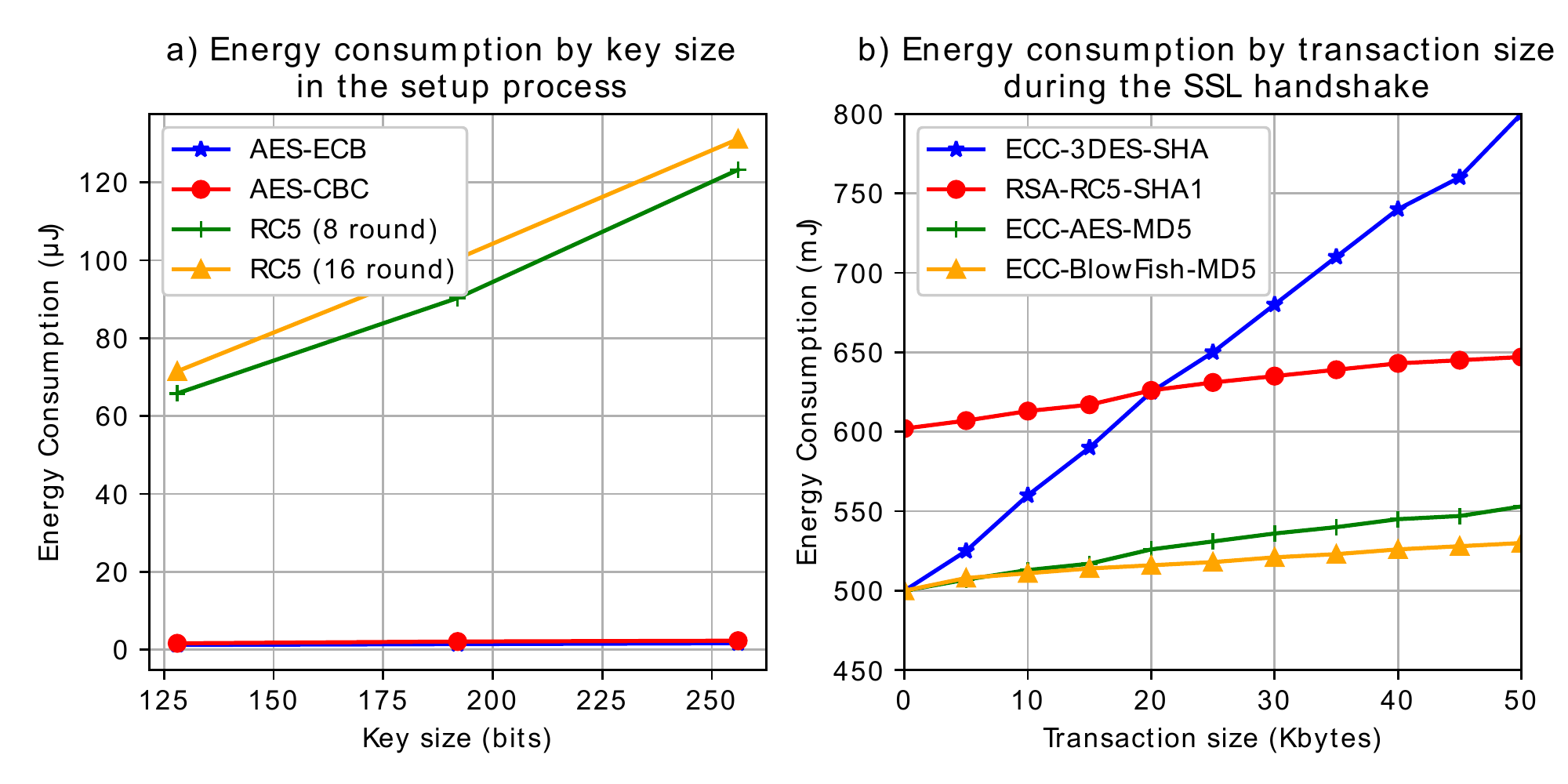}
	\end{center}
   \caption{Energy consumption of the cryptographic algorithms, according to \cite{Potlapally06}. Energy consumption of AES and related cipher suites is least in all cryptographic algorithms.  }
   \label{fig:energy-consumption-encryption}
\end{figure}

Major challenges for 6G communication security are end-to-end encryption and reducing the expense of security (e.g., energy consumption, deployment cost). Since user traffic is booming in recent mobile networks, implementing end-to-end encryption may involve too much overhead for data transmission in 6G. 3GPP and standardization bodies currently prefer the optional use of security measures according to their ability to meet required services. If 6G requires mandatory end-to-end encryption, it is unclear how to satisfy this requirement. Also, upgrading all current security protocols to support quantum-safe standards will be a multi-year effort, given the time consuming of standardization and commercial testings. Finally, the expansion keyspace, if any, also has a significant effect on energy consumption and storage size. According to \cite{Potlapally06}, nearly all encryption algorithms' energy consumption will rapidly soar if increasing their key size or data amount for processing (transaction size), as illustrated in Figure~\ref{fig:energy-consumption-encryption}. Given the goal of energy efficiency in 6G up to 10-100 times, compared with 5G \cite{DOCOMO2020}, it is unclear how to satisfy the requirement if all 6G devices are forced to equip quantum-resistant algorithms. Due to the constraints, we believe that many IoT networks may not quickly jump to the end-to-end encryption or fully quantum-resistant goal but use them in context (e.g., in the network segments integrated with dedicated energy). The trade-offs of performance and cost/time-to-market at expense of security can cause some operators to even ignore several mandatory security features of expensive implementation.

\subsection{Enhanced Security Edge Protection Proxy (SEPP) for securing interconnect between 6G networks: Roaming Security}

In 5G security architecture model, Security Edge Protection Proxy (SEPP) is the main force to protect interconnections between the home network and serving/visiting networks (as illustrated in Figure~\ref{fig:6G-security-architecture-visual}). SEPPs support end-to-end authentication, integrity and confidentiality protection via signatures and encryption of all HTTP/2 roaming messages \cite{3GPP33501}. If there are no IP exchange (IPX) entities, SEPP will use TLS protocols to protect communications; otherwise, an application layer security protocol over the N32 layer called PRINS shall be used. According to \cite{3GPP33501}, SEPPs use JSON Web Encryption described in IETF RFC 7516 for protecting exchange messages between the home network and serving network (via N32 interface) against eavesdropping and replay attacks. During the transit, if the IP exchange (IPX) service providers need to carry out modifications (for mediation services), the standard JSON Web Signatures (defined in IETF RFC 7515) will be used to sign for the modifications\cite{ENISA5GSecurity}. In 6G, these TLS-based protocols can be upgraded with potential enhancements on cryptographic algorithms (e.g., support quantum-safe standards) or to support high-performance TCP/IP transmissions on gigabit networks.

\textit{Remaining challenges}

Due to the wide usage of TLS in many secure communication protocols, finding and addressing potential vulnerabilities of future TLS (e.g., protocol downgrade attacks \cite{TLS}, in which the entities are lured to communicate with previous versions of TLS that are notoriously insecure) will be major challenges.

\subsection{Blockchain and distributed ledger technologies for a vision of 6G trust networks}
\label{subsubsec:trust-network}

\begin{figure}
    \centering
    \begin{center}
			\includegraphics[width=1\linewidth]{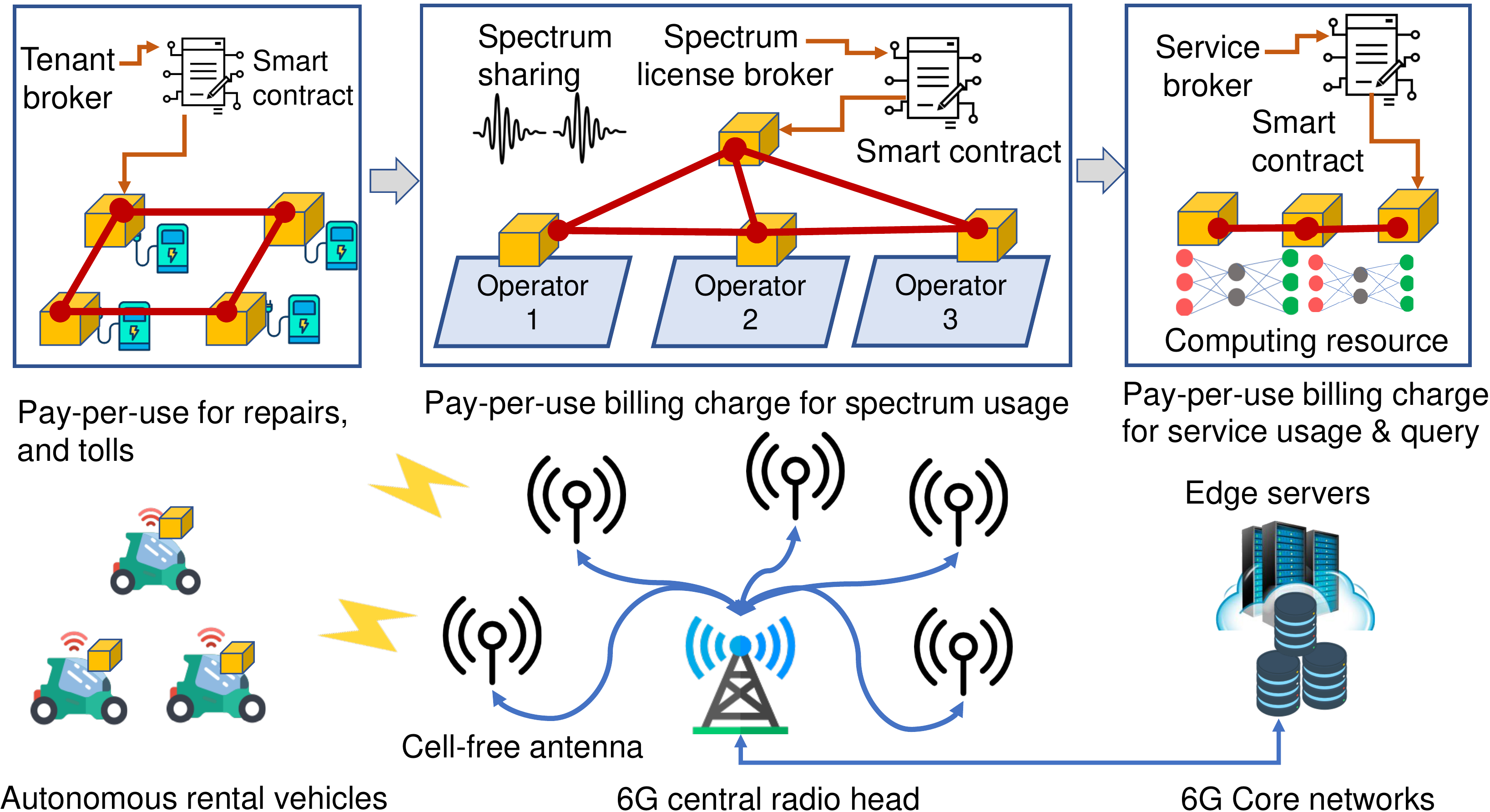}
	\end{center}
   \caption{Illustration of using blockchain technologies for autonomous rental vehicles, spectrum sharing, or dynamic resource allocation in 6G.}
   \label{fig:6G-blockchain-networks}
\end{figure}

Trust networks and services are key expectations of 6G \cite{6GSumit2019}. By definition, trust assumes a risk in an interaction \cite{Kantola2020TrustNF}, and implies the assumption that the communication party of an entity will act consistently and faithfully \cite{TrustITU3052}. Many assume that embedding trust into the 6G networks will include the following key features: (1) maintaining the worth of information sharing while preventing fake/misbehaving sources, (2) guaranteeing that the likelihood of any undesirable events is extremely low, and (3) avoiding the single-point-of-failure. However, satisfying all of these is a challenge. From a design perspective, trust is commonly accomplished by various cryptographic schemes, such as digital signatures and certificates. Other than quantum-safe encryption mentioned above, blockchain and distributed ledger technologies (DLT), which are renowned for being used in cryptocurrency and financial transactions, are possible solutions to evolve to build a trust network in 6G \cite{Nguyen2020}. Theoretically, the peer-reviewed ability of blockchain and DLT will guarantee key security and privacy features such as immutability, transparency, verifiability, anonymity and pseudonymity, data integrity, traceability, authentication, and monitoring. For 6G vision, US Federal Communications Commission (FCC) eyes blockchain to provide a more efficient tool to track and monitor growing wireless spectrums, given the complexity and high cost of the current spectrum auction and administration model \cite{FCC_blockchain}. Further, blockchain/distributed ledgers are expected to be used for many other applications such as pay-per-use energy sharing and computing infrastructure sharing \cite{Maksymyuka2020}, as illustrated in Figure~\ref{fig:6G-blockchain-networks}. The top expectations of blockchain and DLT in 6G are (1) extending to apply blockchain and DLT for enhancing specific applications such as UAV and autonomous driving, (2) enhancing the security of smart contracts and reliability of consensus protocols, and (3) combining with artificial intelligence for enhancing the analytics on computing nodes ( e.g., to detect 51\% attacks) and then exploiting the blockchain smart contracts to automate the synchronization process. 

\textit{Remaining challenges}

Blockchain and DLT implementation are still at an early stage. Many of their fundamental components are still under active development. The burden on computation and communications are the main concerns of these technologies. Further, lack of clarity on how the technology is governed, uncertainty around regulation, and the energy-intensive nature of the technology, are among many reasons that it will take more years to deploy in practice. Despite many available mitigating solutions \cite{Nguyen2020}, security risks such as 51\% attacks and transaction privacy leakage are still open issues.

\subsection{SD-WAN security: 6G network management control}

SDN is expected to work at full capacity in 6G. Two future versions of SDN are \ac{SD-WAN} and \ac{SD-LAN} \cite{Troia20,Dulinski20}. Like SDN, SD-WAN attempts to enhance network control performance and intelligence by separating the packet forwarding process (data plane) from the routing process (control plane). The largest threats against SDN/SD-WAN are DoS/DDoS attacks and insider adversaries. To protect SDN/SD-WAN, many detection and mitigation methods have been developed, such as using abnormal \ac{IDS} \cite{Chica2020} and \ac{MTD} \cite{Sengupta20}. If an anomaly is detected, the detection system can instruct the SDN controller how to reprogram the data plane (programmable switches) in order to mitigate the attack magnitude. There is a growing trend to use ML/DL to enhance detection engines \cite{Abdou18,Chica2020}. MTD-based systems protect networks by periodically hiding or changing key properties of networks (e.g., real IPs) to evade DDoS attacks and reconnaissance scanning. Supporting security implementation for SDN architecture, e.g., FlowVisor, FlowChecker, and FlowGuard \cite{Chica2020}, is another common approach to fixing its design flaw. Secure access service-edge (SASE) architecture, a term introduced by Gartner \cite{WOOD20206}, can be a way of providing cloud-native security service for SD-WAN in mobile networks. 

\textit{Remaining challenges}

Since SD-WAN is still at the early development stage, it is unclear whether many pure SDN protection methods also work with SD-WAN. The convergence of SD-WAN security and cloud security to work as a unified framework also needs more evaluation. SD-WAN security will likely involve IPsec, VPN tunnels, enhanced firewalls, and micro-segmentation of application traffic \cite{Craven20}.

\subsection{Deep slicing and Open RAN for 6G network security isolation}

\subsubsection{RAN/Core network slicing}

\begin{figure}
    \centering
    \begin{center}
			\includegraphics[width=1\linewidth]{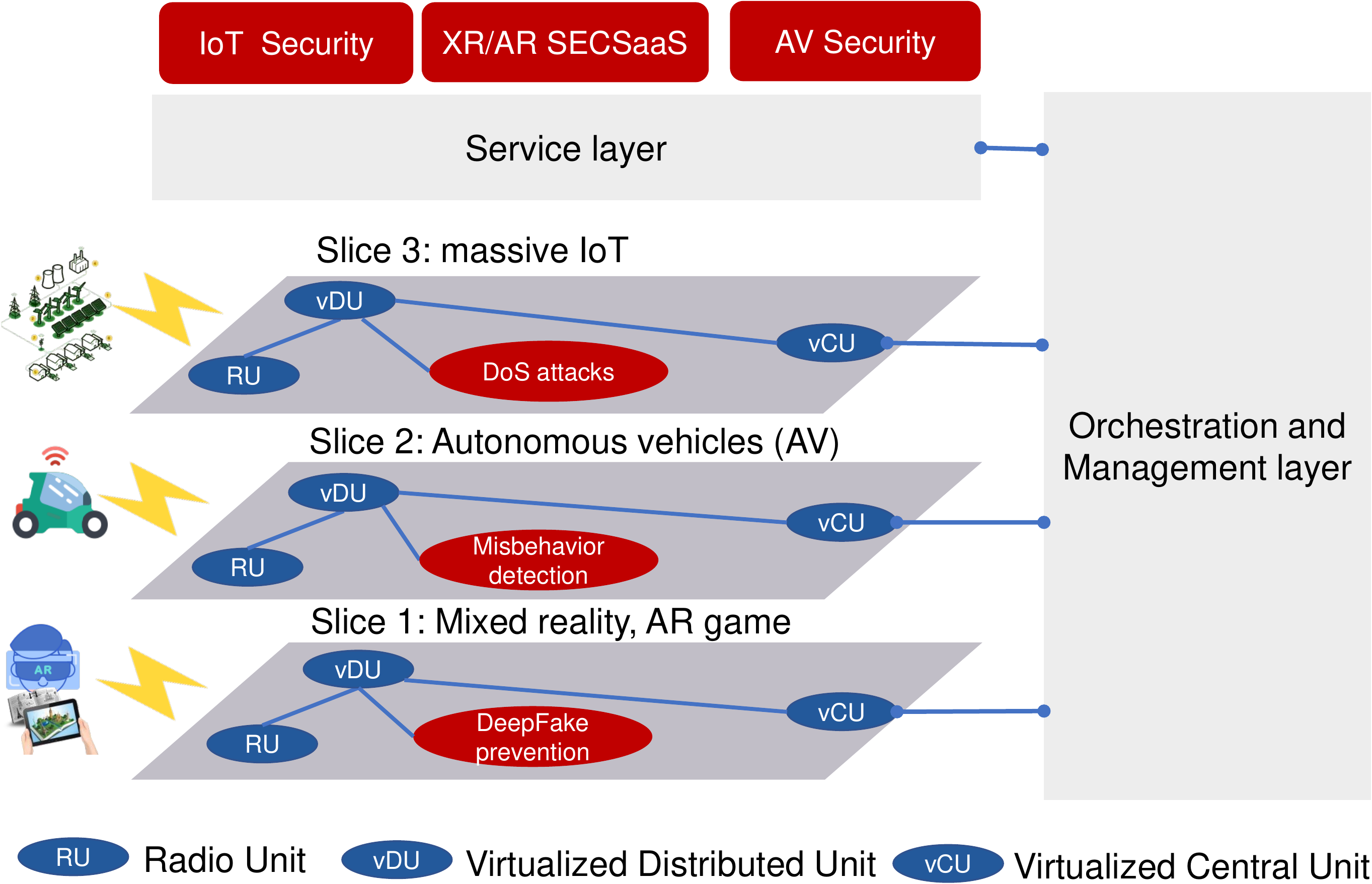}
	\end{center}
   \caption{An illustration of security isolation through full slicing in conjunction with RAN and core network slicing. Implementing slicing at the radio level and guaranteeing stringent Service-Level Agreement (SLA) are challenges for now because of the cost. }
   \label{fig:6G-PHY-radio-slicing}
\end{figure}

Security enhancement by isolation through network slicing is critical in 5G. In 6G, network slicing promises to be fully utilized at all levels, physical, network and service layers \cite{Khan20,Bi19,Zong2019}. Figure~\ref{fig:6G-PHY-radio-slicing} illustrates how RAN and core network slices logically help to isolate security problems for various 6G applications. Each RAN/Core slice is logically isolated from the others, which are set by specific KPIs \cite{Chahbar2020,HZhang2020,TR33.811}. Since cross-talk between slices is prohibited in network slicing; inter-slice communications must be made through their respective interfaces. Following this model, a potential security interference/breach that occurs in a slice, if any, will not affect the others. Ideally, end-to-end isolation will enhance 6G security significantly.

\textit{Remaining challenges}

The primary challenge to develop a security-isolation solution via network slicing is to satisfy all KPI requirements per slice while powerfully and independently enforcing the security policies for massive slices. Lack of specifications on how to perform end-to-end slicing and develop a framework for automatic deployment is thus an open issue. Defining isolation attributes for each slice, setting the KPI requirements, and enforcing them play a critical role in network slicing. A network slice manager, such as Network Slice Management Function (NSMF) \cite{TR33.811}, will be responsible for handling the functions for abstract virtual networks within its administrative domains. This means the NSMF must remain accessible all the time. However, a large-volume DoS attack \cite{Lal17} can make it extremely hard to contain. Despite several existing protection methods such as Slice Management Service Authorization Procedure and mutual authentication \cite{TR33.811}, real implementation is a challenge, given the stringent requirements of processing and response time.

\subsubsection{Virtualized RAN, Cloud-RAN, and Open RAN}
\label{subsub:openRAN}

Virtualized radio access networks (vRAN) and Open RAN are prospective technologies for improving security in 6G \cite{6GSumit2019}. vRAN enables the same functions as conventional RAN but is virtualized -- for example, virtualized baseband units -- on commodity servers instead of physical parts on vendor-specific hardware. Two more essential advantages of the vRAN and open RAN models for security are improved modularity, and reduced inter-dependencies \cite{CiscoRedHatvRAN}. The modularity allows more granular security attestation. By contrast, reducing dependencies on unique software makes it less risky for operators to update live networks. These can enable more controls for the operators over their security infrastructure, particularly with the rapid expansion of the threats on the network in the future. Reducing dependencies can also enable operators to select best-of-class vendors -- a more robust supply chain-- that meet their requirements in security. Note that depending on a single vendor may have massive repercussions. This nightmare occurred in 2018 \cite{EricssonNetworkOutage}, when expired software certificates on equipment caused a large-scale network outage of O2 and Softbank, two Tier 1 operators in UK and Japan.

\begin{table*}[ht]
\caption{Prospective solutions to enhance 6G connection layer security}
\label{tab:6g-5g-network-security-changes}
\begin{adjustbox}{width=1\textwidth}
\begin{tabular}{lllllll}
\hline
\rowcolor[HTML]{EFEFEF} 
\cellcolor[HTML]{EFEFEF} &
  \cellcolor[HTML]{EFEFEF} &
  \cellcolor[HTML]{EFEFEF} &
  \multicolumn{2}{c}{\cellcolor[HTML]{EFEFEF}\textbf{Prospective security solutions}} &
  \cellcolor[HTML]{EFEFEF} \\ \cline{4-5}
\rowcolor[HTML]{EFEFEF} 
\multirow{-2}{*}{\cellcolor[HTML]{EFEFEF}\textbf{Security domain}} &
  \multirow{-2}{*}{\cellcolor[HTML]{EFEFEF}\textbf{Reference}} &
  \multirow{-2}{*}{\cellcolor[HTML]{EFEFEF}\textbf{Security \& privacy issues}} &
  \textbf{5G} &
  \textbf{6G} &
  \multirow{-2}{*}{\cellcolor[HTML]{EFEFEF}\textbf{Open challenges}} \\ \hline

  \begin{tabular}[c]{@{}l@{}} Network access \\ authentication\end{tabular} & \begin{tabular}[c]{@{}l@{}}  \cite{ETSI.TS133.501}, \cite{Jover19}, \cite{Basin18} \end{tabular} &
  \begin{tabular}[c]{@{}l@{}}Impersonation attacks \\ SUPI/identifier exposure \\ \end{tabular} &
  \begin{tabular}[c]{@{}l@{}}3GPP: 5G-AKA\\ Non-3GPP: EAP-TLS \\ 5G USIM, SUCI/SUPI\end{tabular} &
  \begin{tabular}[c]{@{}l@{}}3GPP: 6G-AKA\\ Non-3GPP: Quantume-safe EAP-TLS \\ 6G nuSIM/non-ID\end{tabular} &
  \begin{tabular}[c]{@{}l@{}}$\blacktriangleright$ Many components of 6G  remain undefined \\ so no clear relationship among stakeholders. \\ $\blacktriangleright$ System-on-Chip SIM (nuSIM) integration and\\ non-SIM model are still under development \end{tabular}  \\ \rowcolor[HTML]{EFEFEF} 
 
  \begin{tabular}[c]{@{}l@{}}Signalling data \\ encryption\end{tabular} & \begin{tabular}[c]{@{}l@{}} \cite{Verizon5GPrivacy}, \cite{Carames20}, \cite{Wright21} \end{tabular} &
  \begin{tabular}[c]{@{}l@{}}Man-in-the-middle \\ Eavesdropping \\ Tampering traffic \\ Data leakage \end{tabular} & \begin{tabular}[c]{@{}l@{}}128-NEA1/128-NEA2/128-NEA3 \\ 128-NIA1/128-NIA2/128-NIA3 \end{tabular} &\begin{tabular}[c]{@{}l@{}}256-NEA1/256-NEA2/256-NEA3 \\ 256-NIA1/256-NIA2/256-NIA3  \\(Quantum-safe support) \end{tabular} &
  \begin{tabular}[c]{@{}l@{}}$\blacktriangleright$ Heavy computation, energy consumption\end{tabular} \\  

  \begin{tabular}[c]{@{}l@{}}Transport security protocol\end{tabular} & 
  \begin{tabular}[c]{@{}l@{}} \cite{ETSIQuantum}, \cite{ChenJ2021}, \cite{ChenY2021} \end{tabular} &
  \begin{tabular}[c]{@{}l@{}} Man-in-the-middle \\ Data leakage \end{tabular} & TLS 1.2/1.3 & \begin{tabular}[c]{@{}l@{}}Quantum-safe TLS \\ (AES-256) \\ Quantum key distribution (QKD) \end{tabular} &
  \begin{tabular}[c]{@{}l@{}}$\blacktriangleright$ Heavy computing if using for user data plane \\ $\blacktriangleright$ Quantum-based technology remains no \\ explicit economical gain for now.\end{tabular} \\ \rowcolor[HTML]{EFEFEF}
  
  \begin{tabular}[c]{@{}l@{}}Interconnection security\end{tabular} & \begin{tabular}[c]{@{}l@{}} \cite{3GPP33501}, \cite{ENISA2020} \end{tabular} &
  \begin{tabular}[c]{@{}l@{}}Man-in-the-middle \\ Data leakage \\ \end{tabular} & SEPP with HTTP/2 and TLS 1.3 & \begin{tabular}[c]{@{}l@{}}SEPP with HTTP/3 and \\ Quantum-safe TLS \end{tabular} &
  \begin{tabular}[c]{@{}l@{}}$\blacktriangleright$ Heavy computing if using for user data plane \\ $\blacktriangleright$ Quantum-based technology remains no \\ explicit economical gain for now.\end{tabular} \\
  
   \begin{tabular}[c]{@{}l@{}}Trust networks \end{tabular} & 
   \begin{tabular}[c]{@{}l@{}} \cite{Ylianttila2020}, \cite{Nguyen2020}, \cite{Kantola2020TrustNF} \end{tabular} &
  \begin{tabular}[c]{@{}l@{}}Compromised/insider attacks \\ Data leakage \\ \end{tabular} & \begin{tabular}[c]{@{}l@{}}Blockchain/Distributed Ledgers \\ are supported in several applications\end{tabular} & \begin{tabular}[c]{@{}l@{}} Blockchain/Distributed Ledgers \\ are widely used in many applications \end{tabular} &
  \begin{tabular}[c]{@{}l@{}}$\blacktriangleright$ High energy consumption \\ $\blacktriangleright$ High complexity \\ $\blacktriangleright$ Vulnerable to 51\% attacks\end{tabular} \\ 

   Network management & 
   \begin{tabular}[c]{@{}l@{}} \cite{Troia20}, \cite{Dulinski20}, \cite{Chica2020} \end{tabular} &
  \begin{tabular}[c]{@{}l@{}} DoS attacks \\ Network topology leakage \\ \end{tabular} &
  SDN security &
  SD-WAN security &
  $\blacktriangleright$ The risk of centralized SDN control  \\ \rowcolor[HTML]{EFEFEF} 

  \begin{tabular}[c]{@{}l@{}}Network isolation\end{tabular} & 
  \begin{tabular}[c]{@{}l@{}} \cite{Lal17}, \cite{Chahbar2020}, \cite{TR33.811} \end{tabular} &
  \begin{tabular}[c]{@{}l@{}} DoS attacks \\ \end{tabular} & Network slicing & \begin{tabular}[c]{@{}l@{}}Deep slicing \end{tabular} &
  \begin{tabular}[c]{@{}l@{}}$\blacktriangleright$ Heavy computing to manage massive slices \\ $\blacktriangleright$ High expenditure and energy consumption.\end{tabular} \\ 
  
  \begin{tabular}[c]{@{}l@{}}Endpoint/network nodes \end{tabular} & 
  \begin{tabular}[c]{@{}l@{}} \cite{ALDWEESH2020105124},\cite{Chaabouni19}  \end{tabular} &
  \begin{tabular}[c]{@{}l@{}} DDos attacks \\ Adversarial attacks \\ Traffic meta profile \end{tabular} &
  Firewall/IDS/MTD & \begin{tabular}[c]{@{}l@{}}AI-empowered Firewall/IDS/MTD \end{tabular} &
  \begin{tabular}[c]{@{}l@{}}$\blacktriangleright$ Breakthroughs in AI \\ $\blacktriangleright$ High computing \\ Adversarial defense\end{tabular} \\ \hline
  
  \hline
\end{tabular}
\end{adjustbox}
\end{table*}

\textit{Remaining challenges}

 Most of the drawbacks of vRAN are on the physical signal spectrum management. First, vRAN creates a massive amount of data and complicated computation for beam spaces that require faster baseband processing hardware. As capacity and bandwidth increase, computational requirements will increase. This requirement incidentally will demand high CAPEX investment to virtualize RAN fully. Second, since fade and attenuation of signal processing are more demanding in a high-frequency spectrum, managing the efficiency of data transmission and overheating phenomenon on compact processors is another technical challenge \cite{Sexton17}. Finally, the distributed workload environment that vRAN employs through the complete separation of hardware and software can lead to unknown latency between workloads. These challenges forced the industry to coalesce around to virtualize 5G networks in the low-band mmWave spectrum. It is unclear how vRAN will do when 6G is pushed to even higher frequencies, i.e., THz. 
 
 Open RAN model and software-driven RAN approaches have their shortcomings. First, publishing open-source code is intended for developers to get constructive feedback from the community. However, suppose the source code is not appropriately designed and inspected intensively by security experts. In that case, it may become vulnerable, and hackers can also easily look for potential vulnerabilities to exploit without reverse engineering. Second, vulnerabilities are frequently propagated if a vulnerable source code is reused (e.g., as a library) for developing other codes.

\subsection{Next-generation firewalls/intrusion detection for 6G network endpoint and multi-access edge security}

Endpoints are entities in the borders of the core network (e.g., perimeter routers, IP core network gateways). Core networks are traditionally required to have security gateways to protect against external attacks. A security gateway can inspect bi-directional traffic against the operation policy to prevent unauthorized traffic from core network elements. The types of security gateways include network firewalls, web application firewall (WAF), IDS, service-oriented architecture API protection, antivirus program, VPN, and so on. In 5G system architecture, a security gateway located at the \ac{AMF} side, normally an Internet Security and Acceleration (ISA) Server, is responsible for inspecting all traffic between RAN and AMF \cite{3GPP33501,ETSI.TS133.501}. In 6G, such gateways will need to upgrade their capacity significantly. Many predict that the enhanced capabilities will consist of (1) in-line deep packet inspection (DPI), (2) TLS encrypted traffic inspection, (3) integrated intrusion prevention, (4) inspection for antivirus, and (5) third-party identity management integration (i.e., LDAP). Current AI-driven engines (e.g., using deep learning) will need a significant upgrade in detection capability such as more ability for online traffic training and less impact by the imbalanced datasets as well as robust for protecting heterogeneous networks. A promising solution is to improve the generative learning ability of deep learning models, as the suggestion of artificial general intelligence in \cite{ALDWEESH2020105124,SILVER2021103535}. 

\begin{figure}
    \centering
    \begin{center}
			\includegraphics[width=1\linewidth]{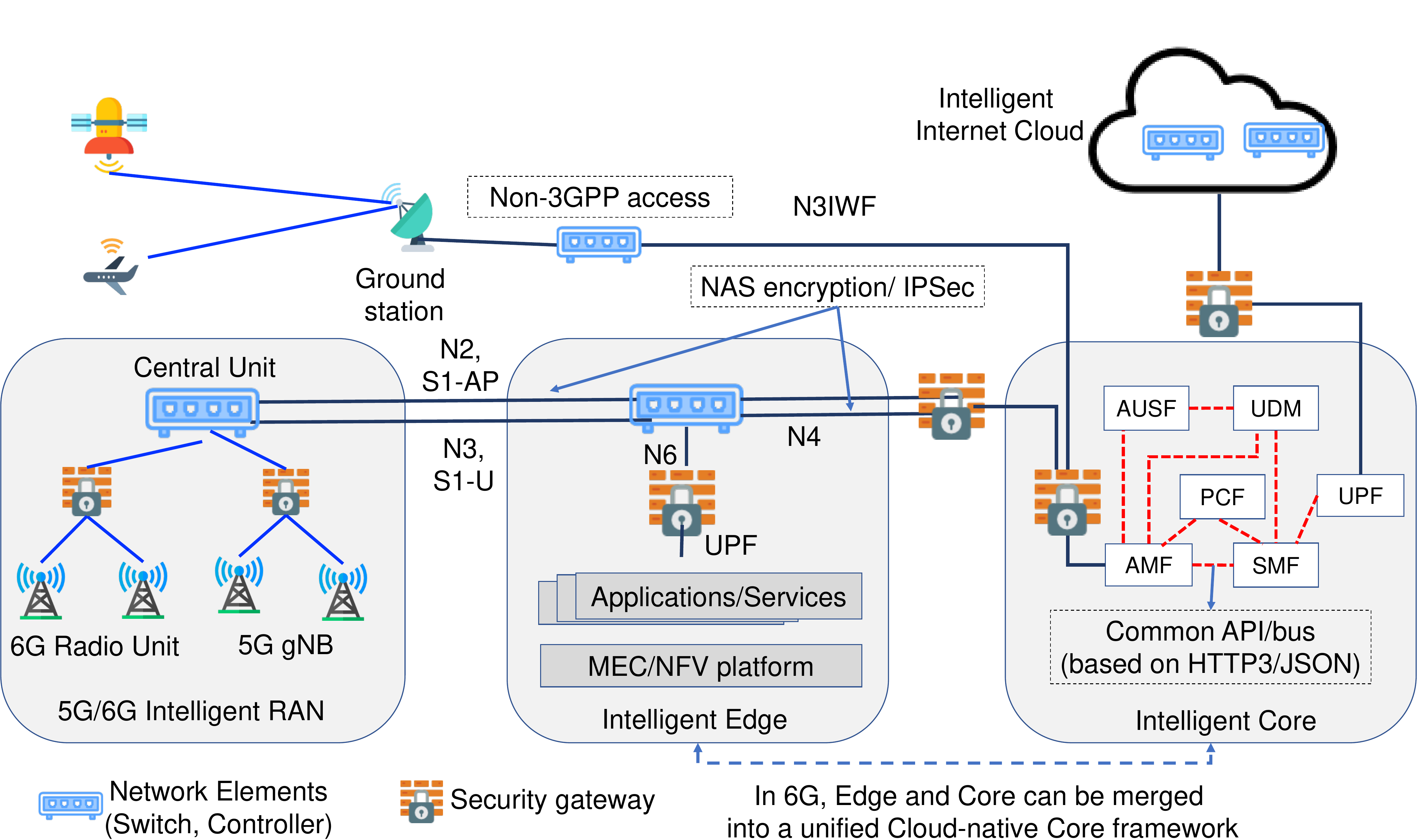}
	\end{center}
   \caption{The illustration of security endpoints located in the RAN/Core mobile networks. The big upgrade of 6G endpoint security is the wide use of AI-driven engines and in-line deep packet inspection (DPI) in firewalls and network intrusion detection systems.}
  \label{fig:security-gateway}
\end{figure}

\textit{Remaining challenges}

The main challenges are that many features of next-generation security gateways are still only concepts. Security automation is likely a mandatory feature to enable efficient protection for 6G ultra-dense networks. However, many of such AI-driven technologies need further enhancements (addressed further in Section~\ref{sec:security-ai}). 

\subsection{Summary of lessons learned from connection layer security}

This section reviews several prospective solutions to protect 6G connection (network) layer such as quantum-safe communications and SD-WAN security.  Table~\ref{tab:6g-5g-network-security-changes} summarizes our vision on the potential changes of 6G security and privacy in the connection layer, compared with those of 5G. In conclusion, three key lessons learned from 6G connection layer security are as follows.
 \begin{enumerate}

     \item The proposed quantum-safe cryptographic schemes have not yet been standardized or reached a community agreement on the shape. Key generation time, signing time, verification time, encryption time, decryption time, key size, quantum resistance, and well-compatibility with existing security protocols are eight of many important factors to determine the winners. For now, QKD is one of the most prospective candidates for quantum-safe cryptographic schemes. However, the transition from the current non-quantum-safe algorithms (RSA, ECC) to the quantum-safe schemes will be a multi-year process. Accordingly, the transition speed will follow the demands from the market, commercial viability testing, and the readiness of the standards. Meanwhile, exploring the enhancements for the existing cryptographic schemes to mitigate the risk of quantum attacks, such as extending the key length, will still have a shot to maintain many 6G applications, e.g., which cannot afford the high expenditure of post-quantum cryptographic schemes.

     \item \textit{Distributed ledgers and blockchain can be the game changers in 6G but their hungry energy consumption can be a trouble for wide usage}. With the help of these two prospective technologies, 6G can be the first generation to be implemented as a grid of trusted networks. However, such a vision will likely be overlooked if distributed ledger technologies' overhead computation and security vulnerabilities are not fixed.

      \item \textit{Firewalls, IDS, MTD systems will not lose their roles in 6G, but new capabilities will be required}. These platforms have proven their reputations in protecting the networks against many attacks and network intrusions for years. They are still key players in 6G security. However, these legacy technologies need further upgrades in both automation and predictive capabilities so as to maintain their detection efficiency in a complicated environment with many connection technologies. A potential approach is to equip their core detection engines with AI. However, it is unclear whether AI can achieve a significant improvement, given many existing issues of AI-based models (see details in Section~\ref{sec:security-ai}).

 \end{enumerate}

%% file: Section_VII_Service.tex
The service layer consists of edge/fog/cloud technologies that aim to provide middleware for serving third-party value-added services. Although upgrading the service layer is supposed to be independent of the timeline of mobile generations, the birth of new applications and hungry KPI requirements in this layer are the main motivations driving the lower layers' evolution. For example, 5G was rushed to deployment because it was challenging to satisfy the low latency of 1ms and gigabit throughput for industrial applications with 4G technologies. Other than active studies to enhance the physical layer, top network providers and the industry recently started to accelerate network transformation towards 6G by equipping the power of AI and the flexibility of cloud-based technologies for the lower layers, such as network/edge intelligence.

In essence, protecting the service layer infrastructure requires a combination of many tasks: authentication, data encryption, application security protocols, firewalls, hardware security, service identity access management, operation/kernel systems reinforcement, data-center network protection, and so on. Protection should be continuous from the host, operation systems, virtual machines, containers, applications to API services. In 6G, the protection systems may have significant changes in functional capabilities, such as intelligence and automation. The following subsections summarize prospective technologies to mitigate attacks in the service layer and envision the remaining challenges for 6G security research.

\subsection{6G application authentication: Distributed PKI and blockchain-based PKI}
\label{subsec:6G-PKI-quantum-safe-TLS}

\begin{figure}[ht]
    \centering
    \begin{center}
			\includegraphics[width=1\linewidth]{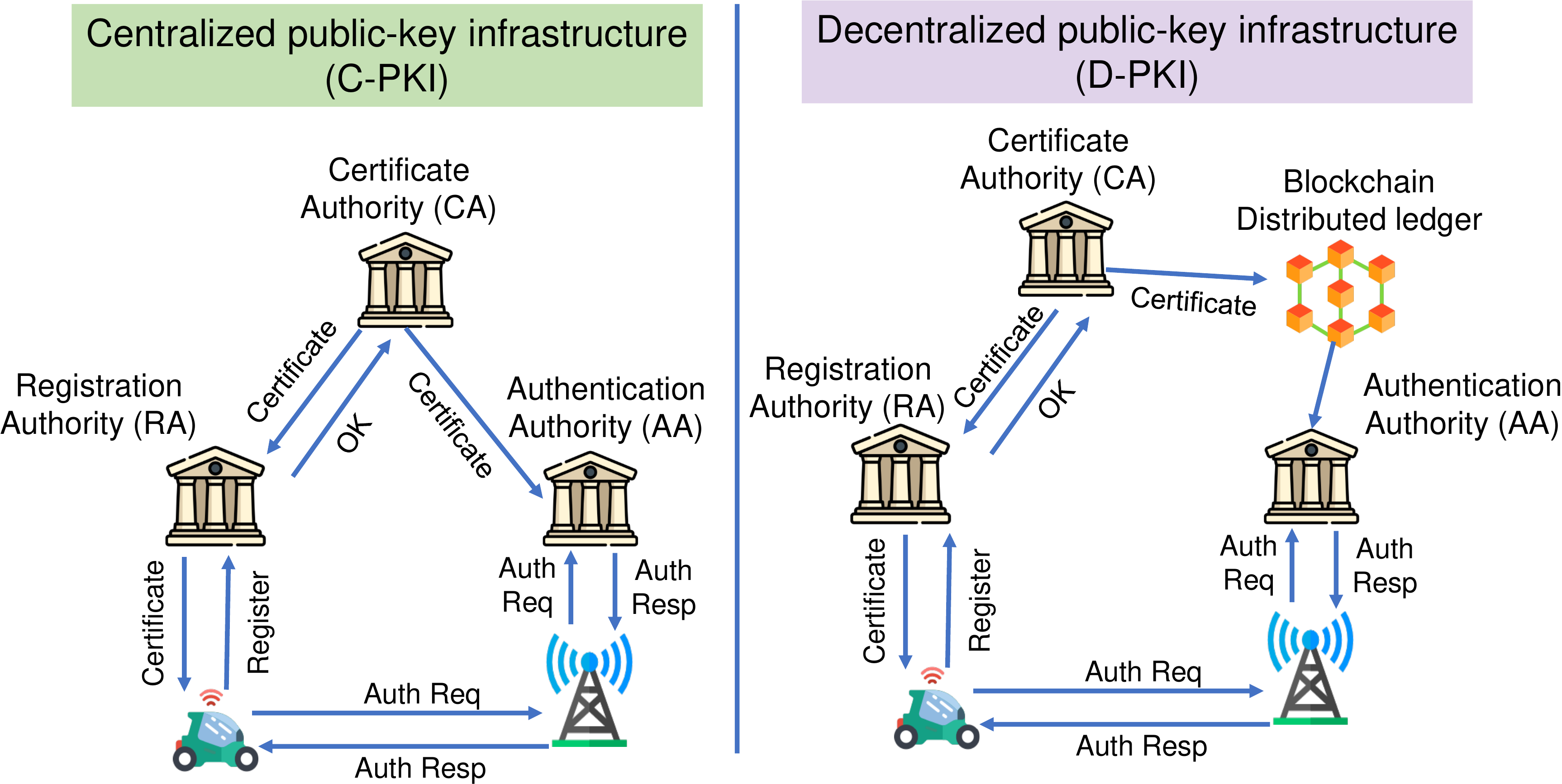}
	\end{center}
    \caption{An illustration of centralized and distributed blockchain-based PKIs for future vehicular networks. Blockchain-based PKIs can eliminate the risks of trusting the CAs only in the centralized PKI model while enhancing both scalability and reliability for many upcoming 6G applications.}
    \label{fig:6G-PKI-vehicular-networks}
\end{figure}

The public-key infrastructure (PKI) is a fundamental function to support user and application authentication. Compared with 5G PKI, 6G PKI likely upgrades its core cryptographic algorithms to quantum-safe mode. Another promising upgrade is to decentralize the PKI. Figure~\ref{fig:6G-PKI-vehicular-networks} illustrates a typical example of decentralized PKI by using blockchain/distributed ledgers \cite{yanan2020}.  Note that the single point of failure at the Certificate Authority (CA) and CA's supreme role (without any formal oversight) in the centralized PKI architecture has been the concern for years. The centralized CAs are also well-known targets for hackers. By breaching the CAs, hackers can issue many fraudulent certificates to break many applications of public-key cryptography. A blockchain-based PKI model, which leverages the strength of verifiable peer-to-peer networks, can effectively eliminate the risks of trusting the CAs only while enhancing both scalability and reliability for many upcoming 6G applications. Recently, Lin et al. \cite{lin2020bcppa} presented a proof-of-concept of PKI based on Ethereum (a public blockchain) to facilitate secure communication in vehicular networks. Another advantage of a blockchain-based PKI is to satisfy high privacy of users (anonymity) and transparency (everyone can know ``who did'',``what'', ``when'' on their record update), which the conventional PKI models do not support. For privacy enhancement, the centralized PKI like X.509 must accomplish through via a complicated model (e.g., using pseudonym certificate generation and Certificate Revocation List (CRL) \cite{rfc5280}).

\textit{Remaining challenges}

Expanding the key length in 6G PKI's core cryptographic algorithm will certainly require more computation and energy to run. While end-users may see little impact, service providers may prioritize the upgrade to high-value applications at the vision of potential high expense. Besides, massive processing in blockchain-based PKI verification can also increase significant energy and resource consumption. This constraint, along with remaining security issues of blockchain technology (e.g., 51\% attacks), likely limits its application scope. There is still a long way to realize large-scale blockchain-based PKI, let alone finding a suitable application context.

\subsection{Using service access authentication (6G AKA) for application authentication}
\label{subsec:6G-AKA-application-authentication}

In prior generations, authentication between a UE and an application server is typically based on credentials such as usernames/passwords and tokens/certificates. In this model, credentials such as session keys will be maintained at both the UE and the application server to protect integrity and confidentiality. A common point of the credential-based authentication approach is to rely on provisioning of pre-shared keys or certificate management. However, managing a large number of pre-shared keys or certificates can be a grave challenge for some application providers against the risks of data breaches. New technology is proposed from 5G is Authentication and Key Management for Applications (AKMA) \cite{TS33.535}. While AKMA is not common in prior generations due to the lack of cooperation between the service providers and network operators, the growth of heterogeneous networks and service-based architecture can open the door for AKMA, e.g., as the form of unified application protocol models to satisfy new business cases (IoT devices). At this point, AKMA exploits the pre-existing cellular authentication and key management procedure to support service authentication instead of managing its own credential management. This approach brings convenience and reduces the complexity of building a new authentication, particularly for small application providers. We believe 6G will further enhance AKMA architecture for reducing delay in edge applications such as XR/AR. Indeed, AKMA can be an alternative solution for single sign-on (SSO) schemes and is particularly suitable for applications located in the network operators' computing infrastructure.

\textit{Remaining challenges}

Since AKMA is still at the early stage of implementation, the lack of application models and business cases is the most challenging. Unlike conventional authentication, to perform AKMA, close cooperation between the network operators and application providers is critical. However, such a relationship is limited for now, given the competition of OAuth or SSO schemes. The security issues of maintaining AKMA in the interconnection environment (roaming) is also an important issue, but 3GPP has not yet been fully addressed \cite{TS33.535}.

\subsection{6G biometric authentication for 6G-enabled IoT and implantable devices}
\label{subsec:biometric-authentication}

\begin{figure}[ht]
    \centering
    \begin{center}
			\includegraphics[width=1\linewidth]{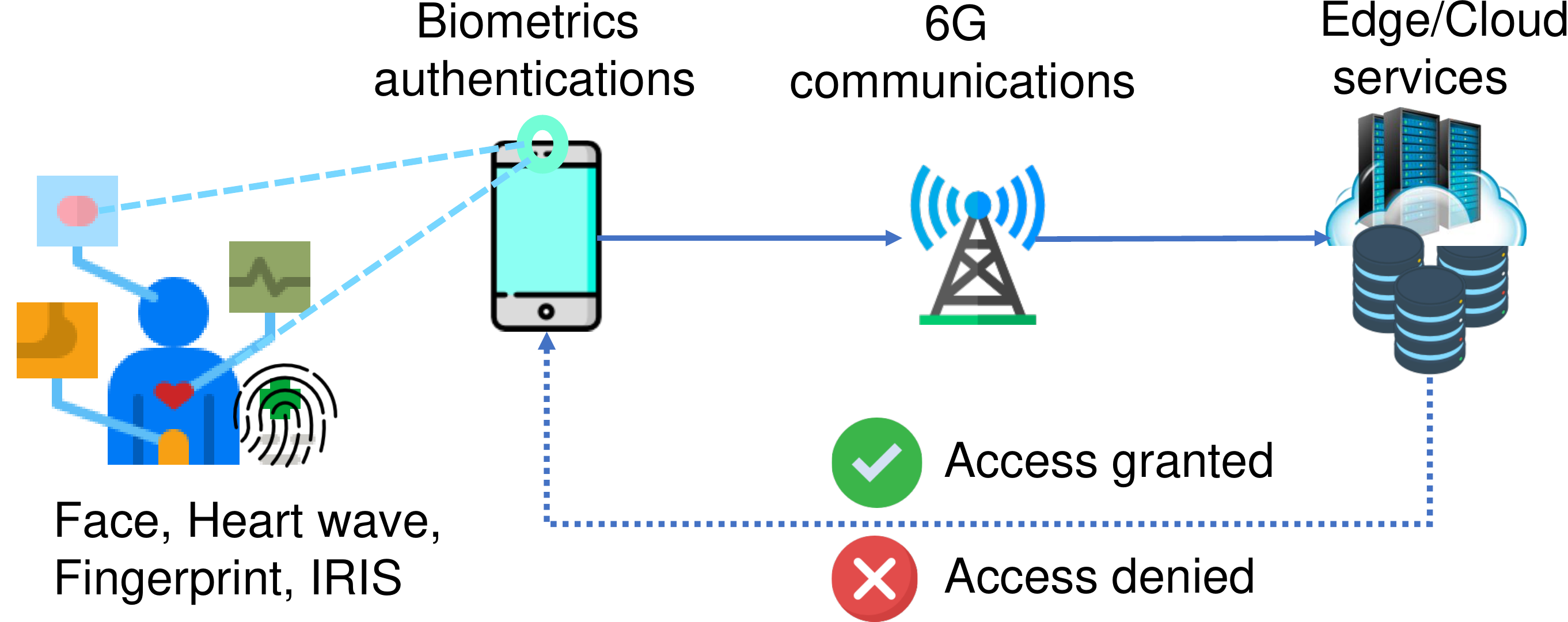}
	\end{center}
    \caption{An illustration of biometric authentication for accessing 6G edge/cloud services. In the future, since the hassles of passwords are expected to be eliminated, biometric authentication will play a key role for 6G service access authentication.}
    \label{fig:6G-biometric-authentication}
\end{figure}

Biometric and behavioral authentication are expanding as prospective technologies for 6G. Biometric authentication systems will directly benefit many 6G applications such as wearable devices and implantable equipment. Biometric authentication can be done without keying complicated codes or memorizing username/passwords, and therefore will benefit many applications and users, including people with disabilities. Besides being applied to the service layer, this approach also has much potential for non-SIM-based access control in 6G core networks. Figure~\ref{fig:6G-biometric-authentication} illustrates a case of using biometric authentication for verifying the unique biological characteristics of a user to grant/deny service access. These characteristics have been used in some commercial applications (e.g., traveller/migrant/passenger identification) and in public security (e.g., criminal/suspect identification). However, only recently, this model is supposed to have been implemented for the service access \cite{HAGHIGHAT20157905}.

On the other hand, when 6G THz imaging technologies with penetration depth capability go into operation, they will significantly enhance biometrics security. By identifying superficial skin traits or faces, THz imaging-based scanning can differentiate real from artificial fingers. Biological characteristics in combinations (multimodal biometric \cite{Talreja21}) can provide higher accuracy, and more flexibility than a single form \cite{Yuan21}. Biometrics based on brain (electroencephalogram) and heart (electrocardiogram) signals have recently also emerged \cite{Arteaga-Falconi16,Arnau21}. It is no longer imagination to identify people from a distance, e.g., 200m, by analyzing their heartbeat. Like a fingerprint, an individual's cardiac signature is unique and cannot be altered or disguised. The futuristic technologies such as brain/heart-signal-based authentication are expected to be much more fraud-resistant than conventional methods like using fingerprints or usernames/passwords.

\textit{Remaining challenges}

Personal information leakage, the safety of technology, and ethical issues are the main concerns of biometric authentication. Despite the many mitigation methods proposed \cite{Ratha01,Yuan21}, protection against various types of biometric spoofing attacks, e.g., using forged or synthetic face/iris samples, will unlikely be sufficient. It will be a nightmare if highly sensitive information of individuals such as fingerprints becomes massively exposed and then abused for surveillance. Unlike traditional identifiers such as username/password, once a biometric identifier is stolen, there is no way to change and replace it with a newer version. Addressing the challenge, some recent studies propose to use secure enclave equipment, cancellable biometric models, or pseudo-biometric identities  \cite{FLin20}. However, such implementation may result in poor performance for objective reasons, such as aging or worn-out fingerprints. The danger to owners of secured items is another great concern. A thief may assault an owner's property to get access to an asset if it is secured with a biometric device. In this case, the irreversible damage to the owner may cost more than just the secured property \cite{BiometricWiki}.

\subsection{OAuth 3.0: new authorization protocol for 6G applications and network function services}

OAuth 2.0 is a widely used protocol for end-users to authorize an application to access the data in another application without exposing the passwords. OAuth 2.0 appears in nearly all API-enabled applications and many mobile apps. Currently, TLS 1.3 and OAuth 2.0 are also used for authorization of network function service access in 5G service-based architecture \cite{3GPP33501}. However, complex ecosystems like distributed ledgers and deep slicing in 6G offer a unique means of identification and verification that OAuth 2.0, the version developed in 2012, cannot support. The initiatives to build OAuth 3.0 are in progress with major expectations about new features such as key proofing mechanisms, multi-user delegation, and multi-device processing \cite{RFCOAuth}. We believe that upgrading OAuth 2.0 to OAuth 3.0 is mandatory to support authorization in 6G end-to-end service-based architecture, where spectrum sharing allocation/networking/security can be quickly deployed as a service through on-demand authorization.

\textit{Remaining challenges}

OAuth 3.0 is still at the stage of concept proposal and feature consideration \cite{RFCOAuth}. The detail of the development progress can be found at \url{https://oauth.net/3}

\subsection{Enhanced HTTP/3 over QUIC for secure data exchange in 6G low-latency applications}
\label{subsec:6G-HTTP2-QUIC}

\begin{figure}[ht]
    \centering
    \begin{center}
			\includegraphics[width=1\linewidth]{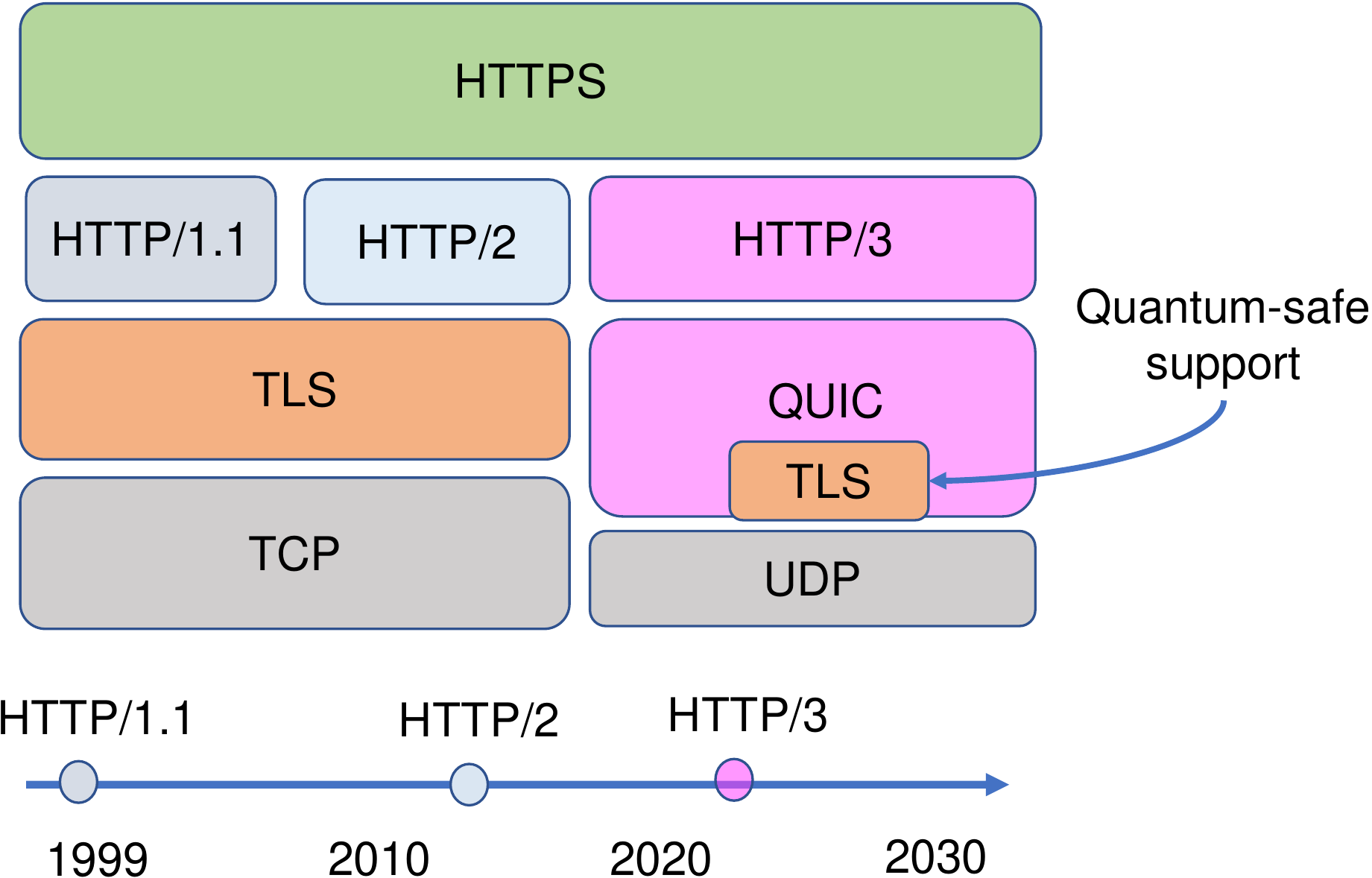}
	\end{center}
    \caption{An illustration of HTTP/3 over QUIC with the support of TLS handshake. HTTP/3 and future enhanced versions will be critical for many 5G/6G low-latency applications, e.g., virtual/extended reality, and interactive microservices in core networks. }
    \label{fig:6G-HTTP3}
\end{figure}

Hypertext Transfer Protocol Version 3 (HTTP/3) is the upcoming major version of HTTP for exchanging information on Web applications and mobile platforms, alongside HTTP/2 (RFC 7540 in 2015). In 5G, HTTP/2 and TLS 1.3 have also been used to support secure communication on the inter-exchange/roaming links among the serving network and the home network \cite{3GPP33501}. Unlike HTTP/1.1 and HTTP/2, which use TCP as their transport, as illustrated in Figure~\ref{fig:6G-HTTP3}, HTTP/3 is built on top of Quick UDP Internet Connections (QUIC), a transport layer protocol to handle congestion control over UDP \cite{ietf-quic-http-34}. The switch to QUIC can eliminate a major problem of HTTP/2 called ``head-of-line blocking'', where a lost or reordered packet can stall all object transactions, even those that are irrelevant to the lost packet. Since QUIC offers per-object error and congestion control over UDP, lost packets only impact the transactions with lost packets. Also, with the support of TLS 1.3 handshake and many fields, including packet flags encrypted in QUIC, HTTP/3 over QUIC can practically prevent pervasive monitoring attacks and protect sensitive data against gathering behavior of protocol artifacts and metadata. In the future, the handshake protocol TLS in QUIC will likely be upgraded along with the development of quantum computing, e.g., support quantum-safe cryptographic algorithms.

With all the enhancements and the advantage of running on the multicast protocol UDP, HTTP/3 can enable faster and more reliable transmission than HTTP/2 does \cite{ietf-quic-http-34}. These features are significant for many 5G/6G low-latency applications, e.g., virtual/extended reality (which often demands fast transmission to render intricate details of a virtual scene), real-time applications (online game streaming services), and broadcasting. Several browsers initially support HTTP/3, although the standard is still officially an Internet-Draft \cite{ietf-quic-http-34} at the time of this work. Given the life cycle of the prior generations, HTTP/3 will continue to be enhanced and eventually dominate in major applications at the time of 6G rolling out. Some new potential implementations are to replace HTTP/2 in carrying roaming messages or support interactive microservices in 6G core networks.

\textit{Remaining challenges}

The change from TCP to UDP in the transport layer protocol of HTTP/3 can be problematic in security. For example, that the change may negatively impact the filters of many deployed security infrastructure such as load balancers or firewalls to parse and inspect application traffic because UDP traffic may be blocked by default in highly secure networks. HTTP/3 over QUIC is also vulnerable to several attacks, e.g., replay attacks if the server and clients have a mismatch on 0-RTT configuration \cite{ietf-quic-http-34,Lychev15}. Since HTTP/3 promises to be widely used in the coming years, more thorough studies on the potential attacks and defense techniques are needed.

\subsection{Quantum homomorphic encryption for secure computation}
\label{subsec:homorphic-encryption}

Secure computation is a fundamental feature for protecting confidentiality and integrity in computing nodes like edge servers. Unlike those in the connection layer, the encryption methods in such service computing nodes may vary and be specified by the providers\cite{Ren2012}. For storage information protection in the computing nodes, common approaches include (1) identity-based encryption, (2) attributed-based encryption, and (3) homomorphic encryption \cite{PYang20}. Identity-based encryption denotes a public-key encryption method in which a user's public key is derived from their well-known identity (e.g., email). A trusted authority is required to generate secret keys for every user over a secure channel, but it may be vulnerable to DoS or compromising attacks. Different from identity-based encryption, identity in the attribute-based encryption is replaced by a set of attributes (e.g., living location, age) and only users, whose private keys (related to certain access policies or attributes) match the attributes or access policy of the ciphertext, can access the data. The most prospective technology of the three is homomorphic encryption, which allows to carry out algebraic operations on ciphertext directly without decryption. The model of operations without revealing personal data in homomorphic encryption is a huge step for privacy preservation in the centralized computing nodes. This is extremely meaningful for secure computation and privacy preservation in the 6G era, when outsourcing data storage and computations to the edge servers for low-latency applications become common and encryption in a shared environment is essential. A notable recent upgrade for homomorphic encryption, as presented in the work of \cite{Zeuner21}, is to support quantum-safe standards. We believe the transition to quantum homomorphic encryption will be easy since an operator can carry out the upgrade on their centralized platforms, reducing security management overhead for individual companies or end-users.

\textit{Remaining challenges}

Enforcing highly secure encryption standards for every service provider is a challenge. Because of cost constraints, not every provider will implement the strongest protection, let alone satisfy all the security standard recommendations in their platforms. Any fault in the configurations, outdated firmware, employee missteps, or lack of strong encryption can cause risks of data breaches or leave systems open to attacks. Second, with millions of users being served, satisfying both performance and the highest encryption standards, e.g., searching over encrypted data efficiently, is extremely difficult. As the authors in \cite{ZFu18, XGe21} suggested, despite many steps forward, many challenging problems remain, e.g., searchable symmetric encryption, secure multi-keyword semantic search, and even secure range query. More studies are needed on applying suitable quantum-safe encryption to prevent adversaries from using the quantum computing power to run data mining and identify information about the individuals.

\subsection{Liquid software security: a step to 6G platform-agnostic security}

The concept of liquid software \cite{Peltonen21}, which allows data and applications to flow from one node to the others, is not new. Currently, many universal apps have partially supported this feature to run on different device types (e.g., tablets, smartphones, wearable devices). In 6G, this platform-agnostic strategy will be enhanced to support computing nodes (e.g., edge servers). However, while applying the strategy for mobile devices in the same provider ecosystem can be carried out easily, due to the coexistence of many ecosystems and multiple network technologies, doing that on the computing nodes of the different network operators is more challenging. A solution to build such platform-agnostic systems is enhancing containerization architecture (e.g., Kubernetes) and cloudization of edge/fog nodes to accommodate multiple applications and enable interactions through API calls. In container-based systems, for security protection, an \textit{extensive security service} such as Docker Trusted Registry will scan container images in advance to detect potential injections and then enforce access policies accordingly. However, according to \cite{YWang20, Gao21}, the imperfection of resource isolation mechanism and shared kernel in multi-tenancy container-based systems can be the potential source of meltdown and spectre attacks that may lead to information leakage of co-resident containers. The authors of \cite{YWang20} introduced ContainerGuard, a non-intrusive variational autoencoders-based method to collect performance events data of processes to detect the attacks.  
 
 \textit{Remaining challenges}
 
The topics of platform-agnostic security and related attacks, particularly for multiple devices and in combination with hardware solutions such as Trusted Platform Module (TPM) and Hardware Security Module (HSM), are still at the early stage of development and deserve more research efforts.
 
  \begin{table*}[ht]
\caption{Prospective solutions to enhance 6G service layer security}
\label{tab:security-solution-service-layers}
\begin{adjustbox}{width=1\textwidth}
\begin{tabular}{llllll}
\hline
\rowcolor[HTML]{EFEFEF} 
\cellcolor[HTML]{EFEFEF} &
  \cellcolor[HTML]{EFEFEF} &
  \cellcolor[HTML]{EFEFEF} &
  \multicolumn{2}{c}{\cellcolor[HTML]{EFEFEF}\textbf{Prospective security solutions}} &
  \cellcolor[HTML]{EFEFEF} \\ \cline{4-5}
\rowcolor[HTML]{EFEFEF} 
\multirow{-2}{*}{\cellcolor[HTML]{EFEFEF}\textbf{Security domain}} &
  \multirow{-2}{*}{\cellcolor[HTML]{EFEFEF}\textbf{Reference}} &
  \multirow{-2}{*}{\cellcolor[HTML]{EFEFEF}\textbf{Security \& privacy issues}} &
  \textbf{5G} &
  \textbf{6G} &
  \multirow{-2}{*}{\cellcolor[HTML]{EFEFEF}\textbf{Open challenges}} \\ \hline
  & \begin{tabular}[c]{@{}l@{}}\cite{yanan2020}, \cite{lin2020bcppa}, \\ \cite{rfc5280} \end{tabular} & Credential exposure &  Public key infrastructure (PKI)  &
  \begin{tabular}[c]{@{}l@{}}PKI with quantum-safe algorithms \\ PKI with blockchain\end{tabular} &
  \begin{tabular}[c]{@{}l@{}}Under trials, no standard till now\end{tabular} \\
  & \cite{TS33.535} & \begin{tabular}[c]{@{}l@{}}Unauthorized access, \\personal info leakage \end{tabular} &
  5G AKA for applications & 
  6G AKA for applications &
  \begin{tabular}[c]{@{}l@{}}Efficient cooperation between \\ network operators and service \\  providers.\end{tabular} \\
\multirow{-3}{*}{Service authentication} & \begin{tabular}[c]{@{}l@{}}\cite{Talreja21}, \cite{Yuan21}, \\ \cite{Arnau21} \end{tabular} &  \begin{tabular}[c]{@{}l@{}}Impersonation \\ Biometric data leakage \end{tabular}  & 
  Face ID, Touch ID &
  \begin{tabular}[c]{@{}l@{}}Face ID, Touch ID, IRIS\\ Heart rate, brain signal ID\\ (Biometric authentication)\end{tabular} &
  \begin{tabular}[c]{@{}l@{}}Biometric data protection\end{tabular} \\ \rowcolor[HTML]{EFEFEF} 
  Application protocol & \cite{ietf-quic-http-34}, \cite{3GPP33501} &  \begin{tabular}[c]{@{}l@{}}Man-in-the-middle \\ Fingerprinting a \\ specific client \end{tabular}  &
  \begin{tabular}[c]{@{}l@{}}HTTP/2 over TLS 1.2/1.3 \\ HTTP/3 over QUIC \end{tabular} &
  \begin{tabular}[c]{@{}l@{}}Enhanced HTTP/3 over QUIC \end{tabular} &
  \begin{tabular}[c]{@{}l@{}}Update many deployed security \\ infrastructure such as load balancers \end{tabular} \\

Service authorization & \cite{3GPP33501}, \cite{RFCOAuth} & \begin{tabular}[c]{@{}l@{}}Flawed redirect \\ Access code leakage \end{tabular} &
  \begin{tabular}[c]{@{}l@{}} OAuth 2.0 \end{tabular} &
  \begin{tabular}[c]{@{}l@{}} OAuth 3.0 \end{tabular} &
  \begin{tabular}[c]{@{}l@{}} The proof-of-concept is still \\ under development\end{tabular} \\
Software security & \cite{Peltonen21}, \cite{YWang20} & \begin{tabular}[c]{@{}l@{}}API vulnerabilities, \\Data breach\end{tabular} &
  \begin{tabular}[c]{@{}l@{}} Container-based security \end{tabular} &
  \begin{tabular}[c]{@{}l@{}} Platform-agnostic security \end{tabular} &
  \begin{tabular}[c]{@{}l@{}} Security features can synchronize \\ to support different devices\end{tabular} \\ \rowcolor[HTML]{EFEFEF} 
  
  Secure computation & \cite{PYang20}, \cite{Zeuner21} & \begin{tabular}[c]{@{}l@{}}Data breach\end{tabular} &
  \begin{tabular}[c]{@{}l@{}} Homomorphic encryption \end{tabular} &
  \begin{tabular}[c]{@{}l@{}} Quantum homomorphic encryption \end{tabular} &
  \begin{tabular}[c]{@{}l@{}} High computation, data mining \\ performance degradation\end{tabular} \\
  
  Security service & \cite{IQBAL201698}, \cite{5GIA2020} & \begin{tabular}[c]{@{}l@{}}Malware/Virus/spam \\ Deepfake\end{tabular} &
  \begin{tabular}[c]{@{}l@{}} Cloud security-as-a-service (SECaaS)  \end{tabular} &
  \begin{tabular}[c]{@{}l@{}} Enhanced AI-empowered \\ SECaaS Everywhere\end{tabular} &
  \begin{tabular}[c]{@{}l@{}} Support interoperability\end{tabular} \\\hline
\end{tabular}
\end{adjustbox}
\end{table*}

\subsection{AI-empowered security-as-a-service transition for 6G ``Service Everywhere'' architecture}

In 5G, many forecast that time-sensitive applications such as virtual reality will likely drive network operators and service providers to equip more computing capability near end-users, e.g., edge servers, to reduce the latency. In 6G, the appearance of holographic telepresence, massive IoT applications, and autonomous driving, the trend of allocating computing power will not only occur at the edge but at every hop of 6G communications \cite{Saad2020,5GIA2020}. The service-based architecture will evolve into the era of end-to-end application or Service Everywhere. The increasing growth of applications will require a better organization from spectrum allocation, memory access to interoperability among the services (e.g., through API interfaces). However, detecting malware/intrusion behaviors and preventing data leakage in such a ``Service Everywhere'' environment is more challenging. For example, security services need to be able to plugin and detach quickly into a massive amount of microservices in distributed computing nodes. The attacks from many dimensions and different network technologies also require more intelligent defense capability, which traditional IDSs and passive deep learning-based detection models \cite{IQBAL201698} such as CNN does not support. In short, we believe that there will be two major changes to 6G security-as-a-service. One is the expansion and easy deployment of the security-as-a-service (SECaaS) model to the computing nodes. In this model, service tenants or individuals on any network node can outsource anti-virus, anti-malware/spyware, intrusion detection, or penetration testing to the Internet security providers. The other is the upgrade of AI models to support proactive learning and reaction (super-intelligent AI) against multiple threats (see detail in Section~\ref{sec:security-ai}), which are hot topics in academia.

\textit{Remaining challenges}
 
 The most technical challenge for service security is to detect insider attacks. While service providers have clear advantages of providing users with advanced security protection and secure storage, data abuse is possible and then leaked by authorized service provider staff. To tackle these threats, deploying robust distributed backup facilities can be a good strategy. Besides, implementing strong auditing mechanisms on data access with digital time stamping and signatures on data could help to reduce the risks of the abuse of authorized insiders. 

\subsection{Summary of lessons learned from service layer security}

Table~\ref{tab:security-solution-service-layers} summarizes prospective solutions to enhance 6G service layer security, compared with those of 5G. From the review, application authentication (e.g., 6G PKI, OAuth 3.0), intelligent security-as-a-service, and platform-agnostic security are the primary targets of major upgrades for 6G and the central of ongoing research efforts. In summary, three key lessons learned from the review on security issues and defense methods for 6G service layer are as follows.
 \begin{enumerate}

     \item \textit{Platform-agnostic security with edge/fog intelligence is a key issue to enhance 6G service layer security}. When 6G involves many heterogeneous network technologies and ecosystems, the platform-agnostic systems will be an important strategy to simplify the complexity of implementing security solutions and delivering security updates among distributed computing nodes as fast as possible. At this point, softwarization and cloudization-based security solutions are likely the enablers to realize the target.

     \item \textit{Supporting open source for security is likely a good approach but not the answer for many enterprise applications}. On the plus side, open-source security can enable network providers to be free from specific vendors. However, the success of open-source platforms in 6G may still need substantial funds from enterprises to attract quality feedback and innovation features. It is otherwise unclear how to maintain security for live business at scale. The best practice would be to develop open standards and encourage competition between multiple vendors. 
     
     \item \textit{6G service layer security boosts 6G privacy preservation at best, but more regulations are needed to enable}. Although personal information leakage can occur in the physical and connection layers, the service layer is at greater risk of massive data breaches as a result of its accessibility from the Internet and centralized application data storage. Secure computation and service access control (authentication) in this layer are then vital for enhancing privacy preservation. However, strong or weak implementation to support those features relies on the companies holding the data. Reasonable policing and regulations can encourage these keepers to adopt highly secure standards.

 \end{enumerate}

%% file: Section_VIII_Sec_AI.tex
A key difference between 5G and 6G is intelligence. Artificial Intelligence (AI) creates new opportunities in 6G networks for innovation and business models powered by various machine learning techniques. By definition, machine learning allows a system to learn representations and procedures to perform human tasks in an automatic manner. In other words, machine learning can learn, predict and make improvements all by itself and is a major sub-field of AI. The ability of AI and security in 6G are key success factors in future AI-empowered wireless networks \cite{Challita2020WhenML}. Due to serving a large number of paid subscribers (e.g., mobile users, enterprise, industry), 6G network operators may have more motivation to enhance their security interest by adopting the state-of-the-art achievements of general AI. Accordingly, the evolution of general AI will then benefit overall 6G-AI-empowered security systems. In the following subsections, we discuss three ways of how AI can change the nature of 6G security in each layer from three aspects: (1) AI as guardians (2) AI as a target, and (3) AI as weapons.

\subsection{AI as a guardian: AI for enhancing 6G security}

Defending against security attacks has been the task of many traditional solutions such as firewalls and intrusion detection systems, but AI makes such systems more capable and intelligent. While legacy security mechanisms (e.g., signature-based intrusion detection) have been extensively used, they have limitations in handling complex attacks in a 6G environment. Several studies \cite{Zhang2020,Kaloudi_2020} have surveyed a large number of security applications powered by AI, such as intrusion detection. They found that AI techniques are suitable for 6G security enhancement. Table~\ref{tab:security-AI-as-guardians} summarizes the AI techniques that aim to improve security for enabling technologies in each layer, along with prospective approaches to enhance the related AI models in the coming years. In the physical layer, open unsecured wireless communication is vulnerable to many attacks such as eavesdropping and jamming attacks. AI can significantly assist security defense by enhancing the performance of the detection engines. For example, the authors of \cite{Ebrahimi21} proposed to enhance randomness in physical layer secret key generation by deep reinforcement learning (DRL). The authors of \cite{Xie21PLA,Perazzone21} explored to use a CNN/RNN-based channel state estimation to enhance physical layer authentication. Using DRL-based models for anti-jamming is a common approach \cite{Han20,Erpek2019DeepLF}.

 \begin{table*}[ht]
\caption{Key AI solutions to enhance security technologies}
\label{tab:security-AI-as-guardians}
\begin{adjustbox}{width=1\textwidth}
\small
\begin{tabular}{|l|l|l|l|l|l|l|}
\hline  \rowcolor[HTML]{EFEFEF}
\textbf{Layer} &  \cellcolor[HTML]{EFEFEF}\textbf{Reference} &  \cellcolor[HTML]{EFEFEF}\textbf{Security \& privacy issues} &
  \cellcolor[HTML]{EFEFEF}\begin{tabular}[c]{@{}l@{}}\textbf{AI-based defense methods}\end{tabular} &
 
  \cellcolor[HTML]{EFEFEF}\textbf{5G} &
  \textbf{6G (vision)} &
  \cellcolor[HTML]{EFEFEF}\textbf{Open challenges} \\ \hline
Physical layer & \begin{tabular}[c]{@{}l@{}}\cite{Mao18}, \cite{Ye2019}, \\ \cite{WKIM20}, \cite{Ebrahimi21},\\ \cite{Xie21PLA}, \cite{Perazzone21},\\ \cite{MYan19,JLiu20}\end{tabular} & \begin{tabular}[c]{@{}l@{}} $\bigcdot$  Eavesdropping, jamming \\ $\bigcdot$ Location tracking \\ $\bigcdot$ Compromised IoT devices \end{tabular} &
  \begin{tabular}[c]{@{}l@{}}$\blacktriangleright$ Channel coding\\$\bigcdot$ Signal detection in PLS\\$\bigcdot$ CSI estimation in PLS\\$\bigcdot$ Beamforming alignment \\$\bigcdot$ Misbehavior detection \\ $\bigcdot$ Anti-jamming \\ $\bigcdot$ Physical layer authentication\end{tabular} &
  
  \begin{tabular}[c]{@{}l@{}}SVM, CNN, LSTM, DNN, \\ RL, DRL, Autoencoder, \\ Deep autoencoder, RNN, \\ RBM\end{tabular} &
   &
  \begin{tabular}[c]{@{}l@{}}$\blacktriangleright$ High computing/training cost\\$\blacktriangleright$ Lack of physical-based datasets\\$\blacktriangleright$ Energy efficiency \\$\blacktriangleright$ Realtime processing \\ $\blacktriangleright$ Reliable signal generation\end{tabular} \\  \cline{1-5} \cline{7-7} 
Connection layer & 
  \begin{tabular}[c]{@{}l@{}}\cite{Talreja21}, \cite{ALDWEESH2020105124}, \\ \cite{JChen21} \cite{Sengupta20},\end{tabular} & \begin{tabular}[c]{@{}l@{}}  $\bigcdot$ Man-in-the-middle \\ $\bigcdot$ DoS, DDoS attacks \\ $\bigcdot$ IP Spoofing \\ $\bigcdot$ SDN controller attacks \\ $\bigcdot$ Traffic trace \end{tabular} &
  \begin{tabular}[c]{@{}l@{}}$\blacktriangleright$ Risk-based authentication\\$\bigcdot$ Network intrusion detection\\ $\bigcdot$ Deep packet inspection (DPI)\\ $\bigcdot$ Protocol vulnerability detection\\$\bigcdot$ Encrypted traffic inspection\\$\bigcdot$ Proactive intrusion prevention\end{tabular}  &
  \begin{tabular}[c]{@{}l@{}}CNN, DNN, RBN, \\ Autoencoder, LSTM,\\ RBN, DBN, RL\end{tabular} &
   &
  \begin{tabular}[c]{@{}l@{}}$\blacktriangleright$ High computing/training cost \\$\blacktriangleright$ Online learning\\$\blacktriangleright$ Real-time processing \\ $\blacktriangleright$ High generative learning\end{tabular} \\  \cline{1-5} \cline{7-7} 
Service layer & \begin{tabular}[c]{@{}l@{}}\cite{HAGHIGHAT20157905}, \cite{Talreja21}\\ \cite{Hwang21}, \cite{Sengupta20}\end{tabular} & \begin{tabular}[c]{@{}l@{}} $\bigcdot$  Malware/virus/spam \\ $\bigcdot$ NFV and VNF attacks \\$\bigcdot$ Malicious microservices  \\ $\bigcdot$ Data breach\end{tabular} &
  \begin{tabular}[c]{@{}l@{}}$\blacktriangleright$ Biometric authentication\\$\bigcdot$ Anti-virus/malware detection\\ $\bigcdot$ Trusted program verification\\$\bigcdot$ Trusted updates verification\\$\bigcdot$ Edge/Cloud control verification\\$\bigcdot$ Container/Runtime protection\end{tabular} &
  
  \begin{tabular}[c]{@{}l@{}}CNN, DNN, LSTM, DNN,\\ DBM, RBM, Autoencoder\\ Deep RL\end{tabular} &
  \multirow{-3}[40]{*}{\begin{tabular}[c]{@{}l@{}}$\blacktriangleright$ More generative learning \\ Meta learning \\ Deep RL, Experienced DRL\\ Deep Convolutional GAN\\ Causal Learning\\ $\blacktriangleright$ More large-scale learning \\ Distributed Learning\\ Federated Learning \\ Transfer Learning \\ $\blacktriangleright$ More explainable learning \\ $\blacktriangleright$ Toward end-to-end learning \\ Deep autoencoder \\ $\blacktriangleright$ AI-building AI \\ Security design by AI \\ Machine creativity\end{tabular}} &
  \begin{tabular}[c]{@{}l@{}}$\blacktriangleright$ High computing/training cost\\$\blacktriangleright$ Massive surveillance\\$\blacktriangleright$ Bias learning\\$\blacktriangleright$ Lightweight model for IoT devices\\$\blacktriangleright$ Vulnerable to AI-targeted attacks \\ $\blacktriangleright$ High generative learning\end{tabular} \\ \hline
\end{tabular}
\end{adjustbox}
\begin{tablenotes}
	\item Physical layer security (PLS), Support Vector Machine (SVM), Convolutional Neural Network (CNN), Long-Short-Term Memory (LSTM), Reinforcement Learning (RL), Autoencoder, Deep autoencoder, Recurrent Neural Network (RNN), Restricted Boltzmann machine (RBM), Deep Neural Network (DNN)
\end{tablenotes}
\end{table*}

AI is also a favored technique for enhancing system performance in many enabling technologies for 6G at the network layer. With the advantages of big data analysis and pattern recognition, AI has been applied to several key technologies (but not limited to) as follows:
\begin{itemize}
    \item Verifying node behavior for detecting insider attacks in maintaining trusted networks (CNN/RBN-based \cite{ALDWEESH2020105124})
    \item Predicting attacks in networks to redirect traffic, make intelligent recommendations for network changes, and isolate suspicious services in SD-WAN/SDN networks (DRL-based \cite{Troia20})
    \item Optimizing radio and computing control policies in vRAN/Open RAN (Deep autoencoder-based \cite{Ayala-Romero19})
    \item Determining prioritization of equipment recovery and isolating failed VNFs (DRL-based \cite{Ishigaki20,JChen21})
    \item Inspecting traffic/network access behavior to predict attack events and filter/remove malicious traffic (DNN \cite{Sengupta20} CNN, LSTM, DBN, RBM \cite{Mao18}, Autoencoder \cite{ALDWEESH2020105124})
\end{itemize} 

The service layer can be the first place to apply AI as a result of the automation requirement for large-scale data inspection. For example, the authors of \cite{Talreja21} proposed to use CNN-based models to check access behavior, device fingerprinting, time, and context usage in risk-based authentication. Since biometric authentication will be futuristic technology for service access networks in 6G, applying AI-based techniques to enhance its performance is an attractive topic \cite{Edwards21}. Hwang et al. \cite{Hwang21} explored a CNN and LSTM-based model to eliminate the necessity of feature selection and extraction tasks while increasing the robustness and performance of biometric verification systems.

Like general AI, most current studies of AI for security are still at the stage of exploring AI to enhance conventional defense approaches or expand their detection capability. Some prospective approaches for AI security in the coming years, particularly for 6G applications, can be as follows. The first target is to enhance the generative learning capability of the AI-based models, where the intrusion detection engines can auto-learn from the environment and operate correctly on previously unseen inputs. In this direction, DRL, meta-learning, and the combination of DRL and GAN (experienced DRL) will be top approaches \cite{Zhang2020,Yang2020ArtificialIntelligenceEnabledI6,caroline2020artificial}, given their strength in learning from very large inputs and automatically optimizing the decisions based on continuous feedback from the environment. Generative learning will be an important step forward to the vision of ``AI-building AI'', causal AI or ``security design by AI'' in 6G. Besides, since the training cost for AI-based models (computing hardware and energy consumption) and large-scale dataset collection increase substantially, large-scale learning needs a new approach. In this way, distributed learning and federated learning \cite{Liu2020FederatedLF}, which can coordinate the learning process on millions of distributed devices (local FL models) to improve the quality of the centralized learning model (global FL model), are likely the top candidates for many 6G applications such as misbehavior detection in autonomous driving. According to \cite{Lim2020}, edge-based applications will be the best places to apply such a learning strategy. To reduce the cost of labelling large-scale datasets, end-to-end unsupervised models like autoencoders \cite{Muhammad20} can be a promising approach, where their implementation can run directly on the online networks with raw traffic collection. Further, to save training cost, transfer learning is also an emerging approach to enhance service authentication applications (e.g., person re-identification \cite{Peng16}) by exploiting the power of trained models that were often carried out on high-performance computing platforms and related large-scale datasets. Finally, the blackbox of how a deep learning model works under specific conditions has been a concern for applying AI in many applications such as biometric authentication, given the existence of bias and potential flaws of imperfect datasets. To avoid the danger of AI making unjustifiable decisions, e.g., blocking a suspected application or targeting wrong criminals, explainable AI models have been another top target for ongoing efforts in cyber trust. Explainable AI like the studies of \cite{ribeiro-etal-2016-trust} and \cite{pmlr-v119-vidal20a} aims to provide an interpretable and faithful manner to predict the results, by proposing the decision trees, rule lists, or Bayesian networks to study the decision of the DL models in the context of live running scenarios.

\textit{Remaining challenges}

Despite many expectations, there are several challenges to apply AI for enhancing 6G security. First, AI can become the target of adversarial attacks (see details below). If AI-based programs control the key components of network systems, successful attacks can cause devastating damage, creating chaos in packet forwarding or bypassing specific malicious traffic. Second, online processing in AI is still an open issue. Heavy computation and substantial time for training in AI (training cost) make it less friendly for IDS/IPS in energy-constrained devices, a large class in 6G.  Moreover, the current AI has no creativity. Therefore, to gain  ``security design by AI'' and ``AI-building AI'' in 6G, more breakthroughs are needed. The ethics of AI is also important, given potential biases in AI systems. Finally, AI is probably abused as a weapon when AI expects to be more intelligent in 6G era. With incredible capabilities, AI-empowered attacks would be a nightmare for the vision of safe networks. The best practice is to apply a high standard for developing AI-related solutions. 

\subsection{AI as a target: Security attacks against 6G AI-empowered engines and defense approaches}

\begin{figure}[ht]
    \centering
   \begin{center}
			\includegraphics[width=1\linewidth]{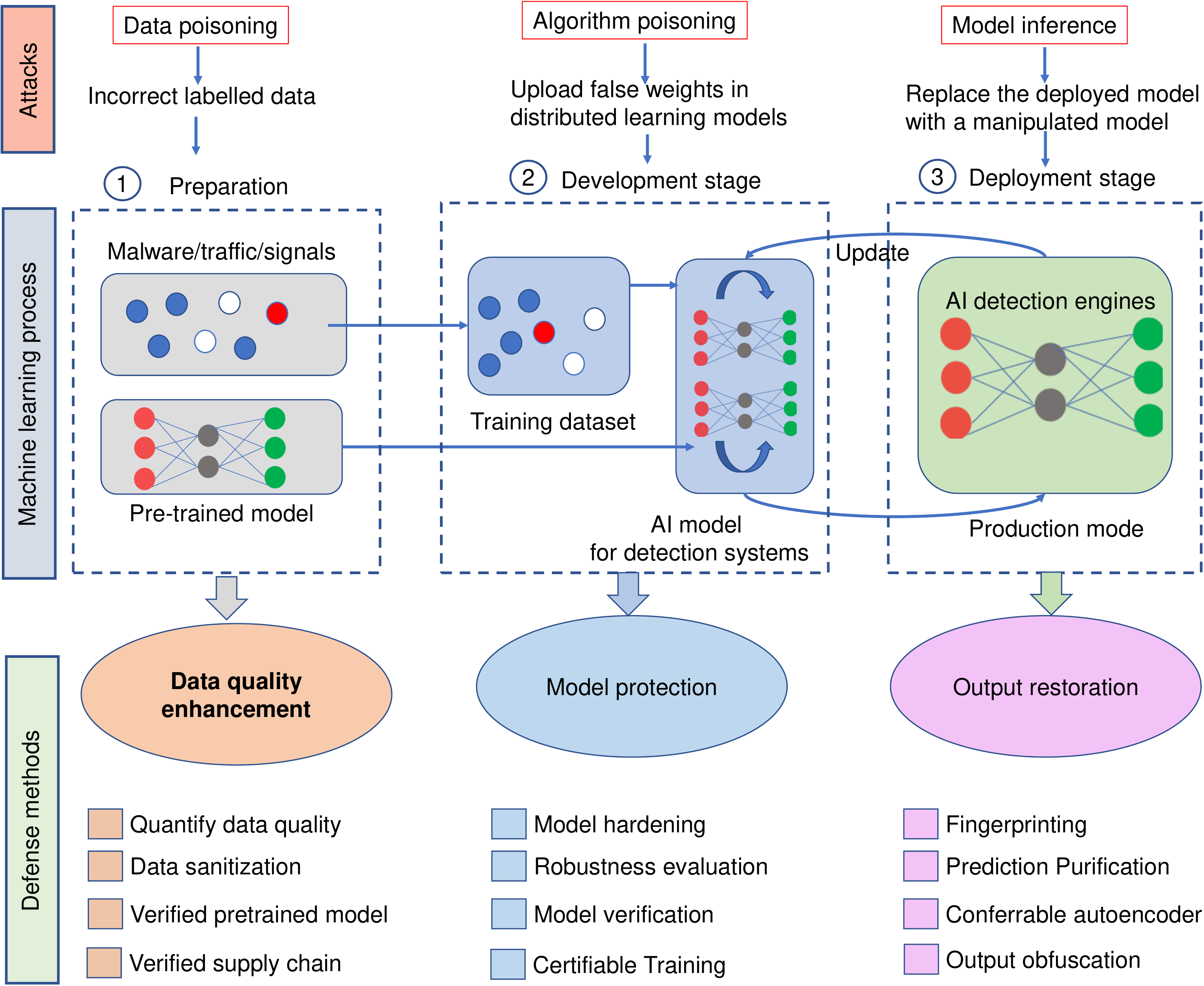}
	\end{center}
   \caption{An illustration of a machine learning-based system, three attack methods against it and potential defense methods. When major 6G applications rely on AI to operate, protecting data supply chain, AI models, and the integrity of output is essential.}
   	\label{fig:6G-AI-models}
\end{figure}

\begin{figure}
    \centering
   \begin{center}
			\includegraphics[width=1\linewidth]{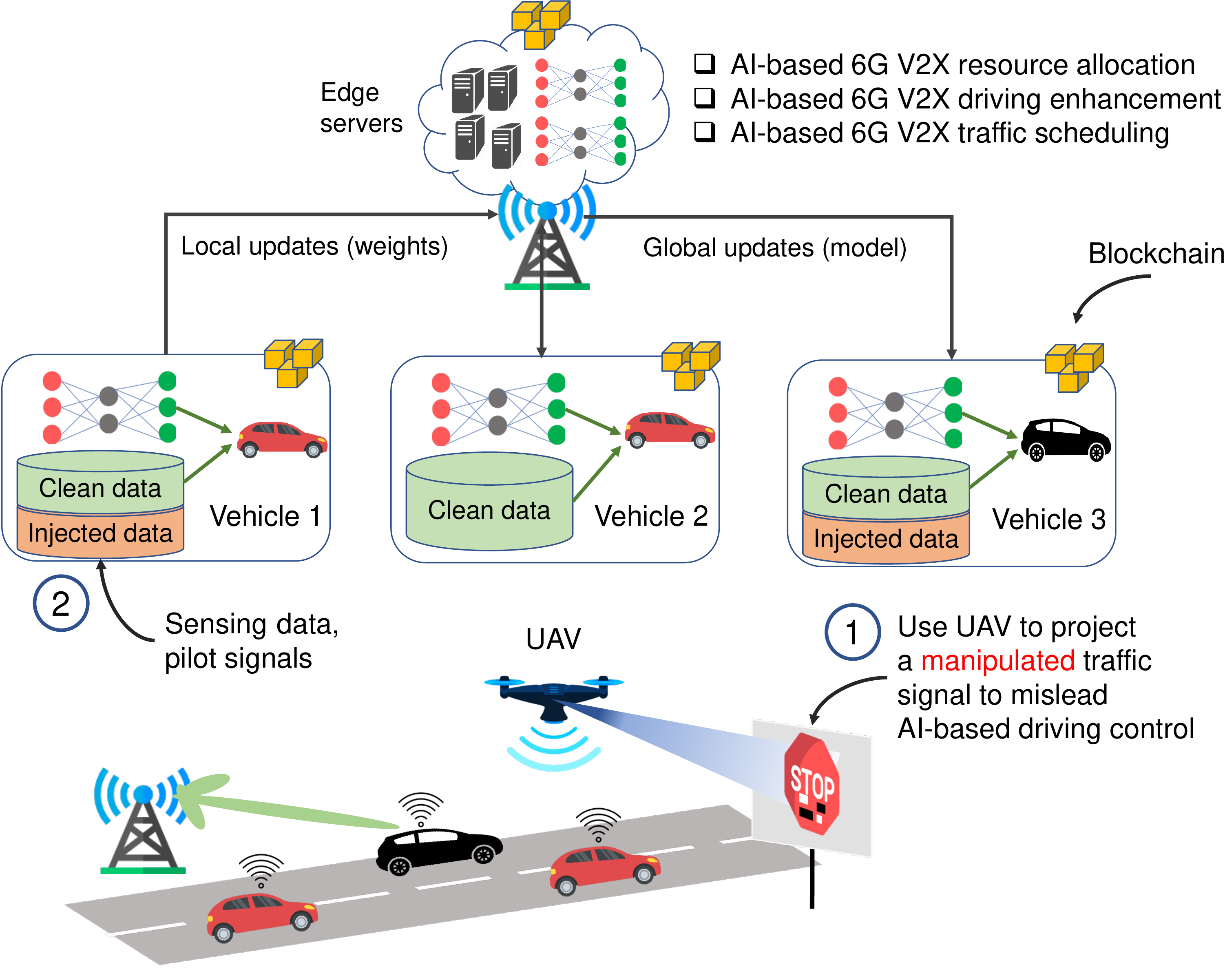}
	\end{center}
   \caption{An illustration of \textcircled{\raisebox{-0.9pt}{1}} the poisoning attacks (physical attacks) by using a UAV to project a manipulated traffic light image on road banner to mislead AI-based driving control in autonomous vehicles and \textcircled{\raisebox{-0.9pt}{2}} algorithm attacks through influencing the global model of AI-based systems at the edge (e.g., for resource allocation, traffic scheduling) with falsification data updates (sensing info, pilot signals). With the expectation of AI popularity in 6G access control (spectrum, resource blocks for transmission) and applications (object detection/tracking in autonomous driving, extended reality), these attacks can pose severe threats.}
   	\label{fig:AI-attacks}
\end{figure}

AI is a double-edged sword. Other than the good side of enhancing 6G capabilities, AI may itself become a target of attacks; AI is particularly vulnerable to adversarial attacks. Hackers may conduct a white-box, gray-box or black-box attack depending on how much knowledge they have of a machine learning system. The fundamental configurations are training data, learning algorithms, and hyper-parameters used to control the learning process. Many studies \cite{Comiter19,Sadeghi2020AST} suggested such processes can be exploited to manipulate AI systems, e.g., exploiting the high linearity of AI models. Figure~\ref{fig:6G-AI-models} summarizes three main attacks that targets an AI-based system in the literature \cite{Comiter19,Sadeghi2020AST,Kaloudi_2020,caroline2020artificial}: (1) \textit{data poisoning} aims to insert wrong labelled data in the datasets or change input objects to mislead machine learning algorithms, (2) \textit{algorithm poisoning} to influence the distributed learning process of an algorithm by uploading manipulated weights in local learning models, and (3) \textit{model poisoning} to replace the deployed model with a malicious one. In three attack types, \textit{data poisoning} is a major challenge since most input objects in the outdoor environment are accessible, and the attacker can easily carry out sophisticated editing. Figure~\ref{fig:AI-attacks} illustrates a case where an attacker uses a UAV to project a manipulated traffic light image on a road banner to mislead AI-based driving control in autonomous vehicles. The attacker can also use data poisoning approaches such as injecting confusing data in transfer learning/federated learning to disrupt AI-based resource allocation systems. AI-based face recognition can also be deceived to accept an attacker's appearance (Bob) as that of a victim (Alice) \cite{Comiter19}. In 6G, when major applications rely on AI to operate, e.g., autonomous driving, the risks of the attacks cannot be underestimated.

To defeat adversary models against AI systems, defense strategies can vary. At the time of 6G, several technologies such as high-performance computing and blockchain may assist in addressing the security issues in AI, e.g., enhancing vulnerability assessment speed or protecting the integrity of local data/AI models against adversarial attacks. Another workable vision of 6G for AI protection is that current AI protection methods will be significantly enhanced. Figure~\ref{fig:AI-attacks} summarizes most prospective defense methods from the studies in \cite{Kaloudi_2020,caroline2020artificial,Sadeghi2020AST}. In essence, there are three approaches: (1) enhance data quality, (2) model protection, and (3) output integrity restoration. For example, the AI designer may modify an ML system with the changes (e.g., reducing noise, altering data features) on adversarial samples or remove contaminated samples from training data (data sanitization) during the training process or ML algorithm. Data collected from verified sources (supply chain) are also preferred, e.g., using blockchain, mutual authentication. For model protection, the AI designer can add a specialized detector in front of an ML system to block any attack in progress or perform multiple evaluations to verify the model.

\begin{figure}
    \centering
   \begin{center}
			\includegraphics[width=1\linewidth]{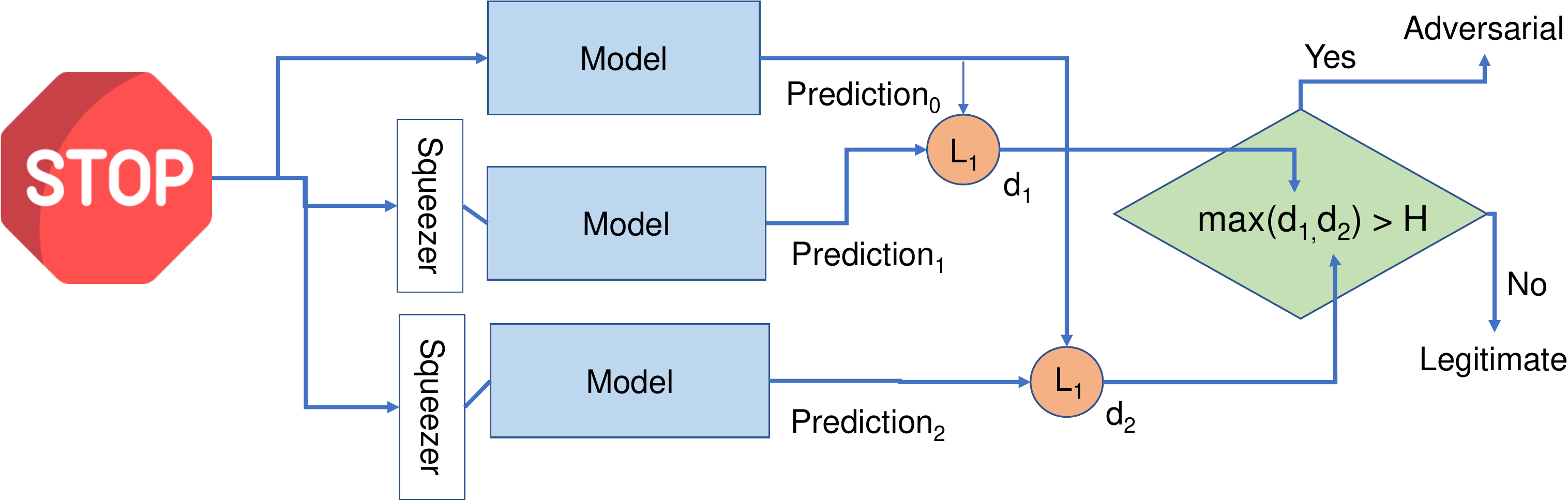}
	\end{center}
   \caption{An illustration of a defense method for AI, where the AI model is trained with multiple types of datasets to find out whether the algorithm is adversarial.}
   	\label{fig:6G-AI-defense}
\end{figure}

Figure~\ref{fig:6G-AI-defense} illustrates a simple defense method against the algorithm poisoning attack from the study \cite{REN2020346}, where the adversarial model can be detected by comparing the model predictions on the original and squeezed data inputs. If the results of two predictions are substantially different from each other, the original input seems to be contaminated (adversarial samples). To protect output integrity (i.e., in the deployment stage), many methods can be used such as output obfuscation and prediction purification \cite{ETSISAI005}. 6G likely enhances specialized techniques for detecting AI-empowered technologies' weaknesses, e.g., AI model assessment, API for scanning vulnerabilities in AI-empowered services. More detail of adversarial attacks and defenses in deep learning can be found in \cite{ETSISAI005,REN2020346,Sadeghi2020AST,Bar2021TheVO}. In conclusion, the field of AI security protection keeps capturing the interests of academic and industrial societies. Further research and discussion need to be carried out in the future.

\subsection{AI as a weapon: Ethical AI/Superintelligent AI/AI Regulation}

It is widely agreed that AI can be abused as a weapon or tool to evade defense systems. Shen et al. \cite{Shen2020DriftWD} presented FusionRipper to defeat a multi-sensor fusion design and carry out two autonomous vehicle-specific attack goals, off-road and wrong-way attacks. The FusionRipper attack design consists of (1) vulnerability profiling to predict when vulnerable periods are and (2) aggressive spoofing to exploit the take-over vulnerability with exponential spoofing based on vulnerability profiling scheduling. The lesson learned from FusionRipper is that we face a new challenge, the adversarial use of AI to create sophisticated types of attacks. AI-powered attacks improve the efficacy of conventional attacks; in other words, attackers can use AI to conduct rapid and effective reconnaissance on the target network to learn and prioritize vulnerabilities that can be exploited. DeepLocker, developed by IBM, uses AI techniques to hide malware within a legitimate application (e.g., video conferencing software) and activates the malware when it reaches specific targets \cite{stoecklin2018deeplocker}. DeepLocker is able to recognize the victim based on facial/voice recognition or geolocation information. Many AI-based attacks can be found in the literature \cite{Comiter19,Kaloudi_2020,Yamin2021WeaponizedAF}. These authors investigated AI-based attack technology and mitigating strategies, which can help to understand how AI is weaponized for attacks and what mitigating mechanisms can be implemented.

Although AI-based attacks have been extensively studied in network applications, research for such attacks in 6G communications is still in its early stages. Hereby ethics of AI is extremely necessary to avoid potential AI abuse as weapons or harm caused by AI systems. Note that the goal of AI is to make \textit{beneficial machines} to benefit humans, instead of creating terminators that can overwhelm and rule humans. Because of such concerns, many companies and governments recently started to promote global initiatives to build a platform about AI \footnote{https://ethicsinaction.ieee.org/} and guidelines for practicing AI technologies at best, including potential legislation and policing. 

\subsection{Summary of lessons learned from AI's impact on 6G security}

Table~\ref{tab:security-AI-as-guardians} summarizes the aspects of technologies in all three layers where AI can assist, from physical coding in the physical layer and radio control policy optimization in the connection layer, to access control in the service layer. Such developments provide a basis for further AI studies that will enhance 6G security technologies, regardless of which layer is targeted. AI can help to enhance the technologies from two perspectives: (1) system detection performance, e.g., the accuracy of channel state estimation in physical coding or physical layer authentication) and (2) automation, e.g., auto-learning on abnormal behavior in live traffic. In conclusion, three key lessons learned from AI’s impact on 6G security are as follows.

 \begin{enumerate}
     \item \textit{The potential for using AI to enhance 6G security is overwhelming but not a magic wand for every security issue}. First, AI can help to enhance system performance (e.g., accuracy) of many tasks of security systems, regardless of the layers. Because of the complexity and diversity of applications, this feature is vital in 6G. Second, AI enables automation for detecting and filtering malicious traffic as a result of self-learning abilities. However, there are some areas where non-AI innovations still get much support from the industry and are potentially the main driving forces for success of 6G deployment, such as quantum-safe encryption algorithms, secure communication protocols, and network slicing for security isolation.
     
     \item \textit{AI's early achievements in assisting security enhancements have demonstrated that getting to the bottom of what causes attacks or how to fix them is challenging}. For now, the capability of AI is limited at detecting whether there is an attack in a given action/traffic pattern or predicting the probability of a specific attack type. AI cannot help to explain why attacks occur or the causal relationships of discrete events to potential intrusions. In the next decade, AI needs to assist security protection to get the bottom of what causes attacks, predicting such causation, and suggesting how to fix and prevent future exploitation at best (security design by AI). To attain such assistance capability, AI needs huge amounts of available data for its training and significant improvements in learning models. Meta-learning and reinforcement learning are potential candidates which can gain fundamental knowledge of generative learning about the causation of attacks and failed defenses by analyzing invariant criteria across data sets. However, data collection and cleaning for such AI models are laborious tasks. As a result, efficient end-to-end learning and causal learning models will be appealing topics in the coming years.
     
     \item \textit{AI is not always a positive actor}. As we highlighted above, AI can be used to evade detection efforts in an IDS or run effective reconnaissance on a victim's networks. Worsen, AI can be abused to create autonomous weapons to attack a designed target chain (e.g., with face recognition). In the end, AI will be a nightmare if a rigid code of ethics does not constrain wild developments. Legislation and policing for AI are required.
 \end{enumerate}

%% file: Section_IX_Privacy.tex
Security and privacy are hand-in-hand technologies. Without security, an attacker can gain access to a victim's networks and steal personal data. By definition, in GDPR Article 4 \cite{GDPR}, personal data can be any information directly or indirectly related to an identified or identifiable person, such as a name, an identification number, subscriber's location, and social identity. Figure~\ref{fig:6G-data-privacy} illustrates three typical personal data types that can be illegally collected or abused for monitoring mobile subscribers. Ensuring \textit{confidentiality}, \textit{integrity}, and \textit{availability} of data in the security design can work as a base for privacy. Privacy preservation is defined as having the ability to protect sensitive information of a specific entity, managed throughout the various stages of data life cycle, such as data in collection, data in processing, and data in use. Different from security, according to \cite{Pleva12}, three principles specified for privacy preservation are: \textit{linkability}, \textit{identifiability} and \textit{traceability} (LIT). \textit{Linkability} means the feasibility of linking consecutive activities of the same identity in sequence. \textit{Identifiability} denotes the possibility of recognizing the true identity of a party in a system through the collected information. \textit{Traceability} describes the possibility of tracking the activities of a specific identity. 
 
 Since collecting and sharing data are essential in today's digital and network economy, any misuse and dissemination of collected data can pose significant threats to users. An adversary can use obtained information to bully or blackmail subscribers. Those risks will increase when 6G networks become more complicated to manage data privacy compliance requirements. To gain public trust, many organizations and companies have recently started to pay serious attention to implementing advanced protection of customer data. The question is, ``what are the new challenges and prospective approaches for privacy preservation in 6G, compared to current techniques?''. The next section is our vision of data privacy matters in 6G, challenges, and prospective solutions to address those issues. 

\subsection{Why data privacy matters in 6G}

\begin{figure}
    \centering
   \begin{center}
			\includegraphics[width=1\linewidth]{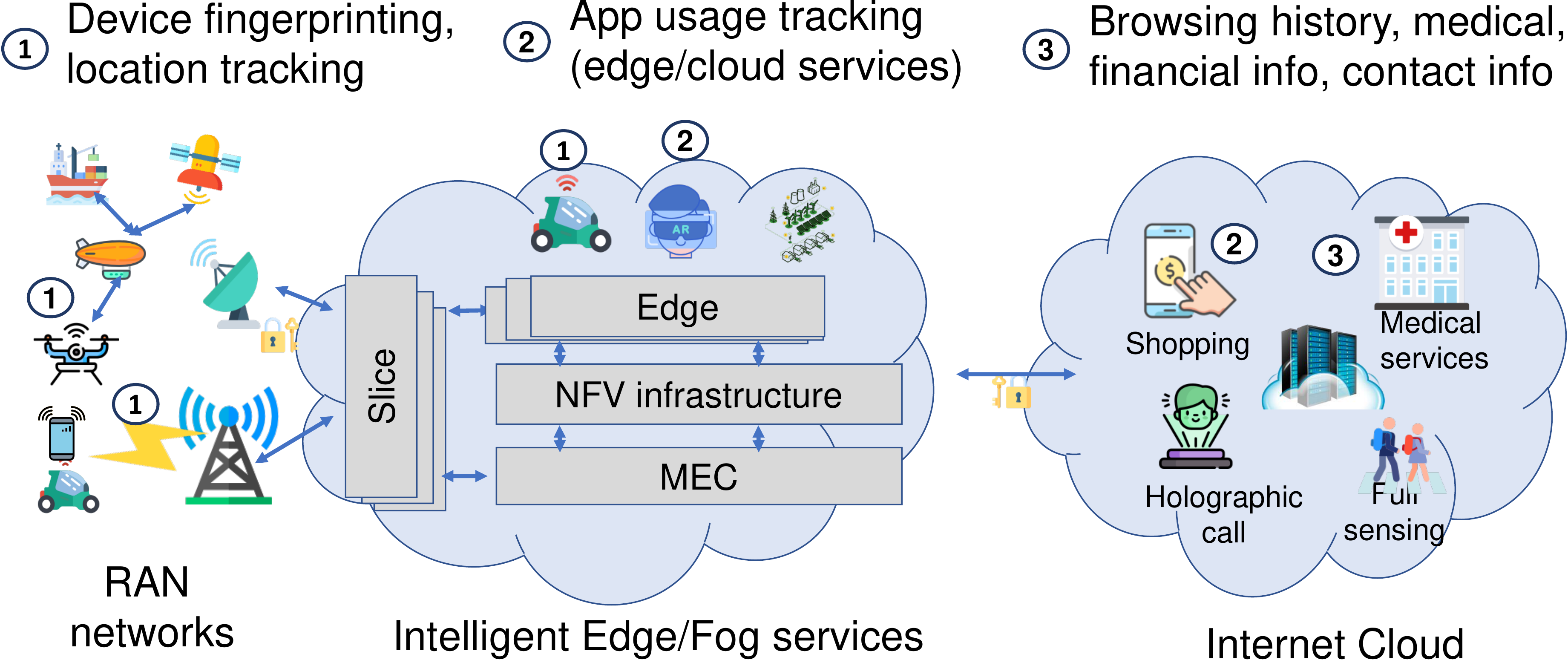}
	\end{center}
   \caption{An illustration of personal data types illegally collected or abused for monitoring mobile subscribers in each layer. As a result of cloudization in data storage and services, large-scale data breaches in the application layer are the most concerns. }
   	\label{fig:6G-data-privacy}
\end{figure}

Data privacy has been a concern for years. Table~\ref{tab:security-architecture-vulnerabilities} in Section~\ref{sec:overview-core-security-architecture} and the summary in the previous sections summarize several potential privacy issues across 6G networks, e.g., location tracking, as illustrated in Figure~\ref{fig:6G-data-privacy}. However, there are many reasons that privacy preservation is more urgent in 6G. First, protecting personal information in an era of supercomputing and smart agents is challenging. With a gigantic network such as 6G that connects things and humans, the demands for AI-enabled smart applications are expected to grow exponentially \cite{Sun2020}. These AI-powered applications can dig out more context-related information of a specific individual and his/her environmental context. By using personal or other confidential data, AI can provide more precise and smarter personalized services that users may enjoy, such as recommendations of points of interest, films, and routes. However, having that kind of experience also impacts a users' privacy. Users may not be aware of being targets of massive data collections for unsolicited advertisements; even worse would be stalking and extortion powered by AI. There is a trade-off between two conflicting goals: (1) high privacy preservation for individuals as well as their right to be forgotten, and (2) mining personal data to maximize the accuracy of recommendations/guidelines for users. The border between providing useful information and being abused for monetization is fragile, particularly in a hungry, data-driven industry. 

 Second, in 6G, more sensitive information of users is expected to be available from key applications such as smart clothes, wearable devices, and implant cyborgs. On the positive side, these applications can help to improve human lives, such as reducing the risk of fatal accidents, enhancing good sleep, or assisting the rehabilitation of people with disabilities. The opposite side is that the physical and medical data for control systems to coordinate these connected applications can be collected illegally and abused. These threats may not be unique but will worsen in 6G. Third, the approach of cloudizing many core components and applications in 6G is also flawed. By migrating workloads to the cloud, a shared infrastructure, customer's personal information faces an increased risk of unauthorized access and exposure, including illegal leakage by unauthorized employees.

 Finally, the more accurate localization through telecommunication in dense networks is a critical concern. The idea of THz Access Points following a user's motion to centimeter-level precision for improving link connectivity will raise severe concerns that it can be used for surveillance. Many countries have begun to tighten the rules for protecting users' personal information. Privacy preservation will be no longer an optional feature as it has been but will be required by law. The U.S. Federal Privacy Act (1974) \cite{USFederatedPrivacyAct} and EU GDPR (2016) \cite{GDPR} was established to provide statutes and fair information practices for preventing potential abuses in using citizens' data. A company can face billions of dollars in fines if a massive amount of its client data is illegally exposed \cite{Sherr18}. The high penalty can be the end of a business in such matters, and personnel can also be prosecuted if the damage is qualified. Again, the liability is not new, but record penalties are likely.

\subsection{State-of-the-art privacy preservation techniques and vision for 6G}

There have been many surveys into privacy concerns and preservation technologies in different technologies: 6G IoT networks \cite{WLin21}, 6G AI-enabled applications \cite{Sun2020}, autonomous driving \cite{Wu_2021}, big data and cloud \cite{Silva21}. The techniques all meet common goals: (1) reduce identity linkage in data collection; (2) increase secure data storage; and (3) control over data sharing and use. Their goals are consistent with privacy preservation principles in law \cite{GDPR}: privacy by design, privacy policies, and privacy engineering. Several mechanisms, such as \ac{PET} \cite{Kaaniche_2020,Cha_2019}, have recently been proposed to provide the best practices for privacy preservation in collecting, processing , and using data that meet different requirements by laws. PETs specifically employ principles such as minimizing personal data, maximizing data security, and empowering individuals. Although PETs are a general approach for data privacy, many principles should still be considered for enhancing privacy in mobile networks, given the convergence of network infrastructure ecosystems and universal data sharing. Based on PETs, Table~\ref{tab:privacy-preservation} categorizes privacy preservation technologies in the literature as three models: trusted, untrusted, and semi-trusted.  

Trusted privacy preservation methods assume that communication parties trust an external entity in data processing. The external entity can be a central authority (CA) which can link and revoke user certificates used in secure communication. The trusted techniques are widely applied in applications of connection and service layers. Access control (authentication), VPN/tunnel encryption (TLS), anonymization, pseudonymization, are typical examples\cite{Kaaniche_2020}. By definition, access control maintains the functions of restricting subscribers from obtaining data or placing data onto storage devices. In the service layer, access control is implemented via service authentication. Tunnel encryption is also a commonly applied protocol for protecting control and user data in core networks. Because of their critical role in restricting data access for unauthorized users, legacy technologies such as authentication and VPNs will continue to be the cornerstone of protecting data privacy in 6G. 

Another important trusted privacy preservation method is pseudonymization. Pseudonymization provides a method of replacing identifiable information fields (e.g., locations or names of individuals) in data records with artificial identifiers, namely pseudonyms \cite{Petit15}. Pseudonymization techniques include scrambling, encryption, masking, tokenization, and so on. The most popular pseudonymization technology in mobile networks is to use IMSI/GUTI/SUCI to hide the real identity of a subscriber. The other futuristic technique for enhancing privacy, which is likely to be key technology in 6G, is to use blockchain networks \cite{Dai19,Nguyen2020,Xu2020bchain}, which are a specific type of distributed ledgers. Blockchain protects user privacy by using a hashing address (wallet) to represent their identity, known as a pseudonymous credential. By using the hashing address to sign and verify all the transactions, users’ identities are not revealed. Many 6G applications such as autonomous driving, health care, fintech, and the energy industry are expected to use blockchain technologies.

\begin{table*}
\caption{Key solutions to enhance privacy preservation}
\label{tab:privacy-preservation}
\begin{adjustbox}{width=1\textwidth}
\small
\begin{tabular}{lllllll}
\hline
\cellcolor[HTML]{EFEFEF}\textbf{Layer} & \cellcolor[HTML]{EFEFEF} \textbf{Reference} & \cellcolor[HTML]{EFEFEF}\textbf{Feature method} & \cellcolor[HTML]{EFEFEF}\textbf{Model} &  \cellcolor[HTML]{EFEFEF}\textbf{5G} & \cellcolor[HTML]{EFEFEF} \textbf{6G (vision)} & \cellcolor[HTML]{EFEFEF}\textbf{Open challenges}  \\ \hline
 Physical layer & \cite{HAGHIGHAT20157905,Yuan21} & Authentication & Untrusted  & \begin{tabular}[c]{@{}l@{}}$\bigcdot$ Radio fingerprinting \end{tabular} & \begin{tabular}[c]{@{}l@{}}$\blacktriangleright$ Physical layer authentication \\ (AI-empowered) \end{tabular} & \begin{tabular}[c]{@{}l@{}}Location exposure \\ Experience degradation\end{tabular}  \\ \hline
  & \cite{Kaaniche_2020} &  Communication anonymization & Trusted & $\bigcdot$ Proxy servers & \begin{tabular}[c]{@{}l@{}} $\blacktriangleright$ Proxy servers \end{tabular} & \begin{tabular}[c]{@{}l@{}} IP exposure at fake proxy \end{tabular}  \\
  & \cite{Dai19,Nguyen2020,Xu2020bchain} & Pseudonymization & Trusted & \begin{tabular}[c]{@{}l@{}}$\bigcdot$ IMSI/GUTI/SUCI \\ $\bigcdot$ Blockchain networks \\ \end{tabular} & \begin{tabular}[c]{@{}l@{}} $\blacktriangleright$ SUCI, Non-ID \\ $\blacktriangleright$ Blockchain/Distributed ledger \end{tabular} & \begin{tabular}[c]{@{}l@{}} Complicated management \\ Energy consumption \end{tabular}  \\
\multirow{-5}{*}[10pt]{ Connection layer} & \cite{Silva21} & Data anonymization & Trusted & \begin{tabular}[c]{@{}l@{}} $\bigcdot$ Randomization \\ $\bigcdot$ Generalization \end{tabular} & \begin{tabular}[c]{@{}l@{}} $\blacktriangleright$ Enhanced anonymization \\ (AI-empowered) \end{tabular} & \begin{tabular}[c]{@{}l@{}} Complexity to implement for \\ large-volume data \end{tabular}  \\ \hline
  
 & \cite{BJian21,Hassan20} & Differential privacy & Semi-trusted  & \begin{tabular}[c]{@{}l@{}} $\bigcdot$ Laplace mechanism \\  \end{tabular}& \begin{tabular}[c]{@{}l@{}}$\blacktriangleright$ Enhanced differential privacy \\ (AI-empowered) \end{tabular} & \begin{tabular}[c]{@{}l@{}}Challenge to detect the changes \\ of particular values \end{tabular}  \\
  & \cite{Rahulamathavan21} & Homomorphic encryption & Semi-trusted  & \begin{tabular}[c]{@{}l@{}}$\bigcdot$ Homomorphic encryption\end{tabular} & \begin{tabular}[c]{@{}l@{}}$\blacktriangleright$ Quantum homomorphic  encryption\end{tabular} & \begin{tabular}[c]{@{}l@{}}Complexity, high computation\end{tabular}  \\
  & \cite{Kaaniche_2020} & Group-based signatures & Semi-trusted  & \begin{tabular}[c]{@{}l@{}}$\bigcdot$ Attribute-based signatures\end{tabular} & \begin{tabular}[c]{@{}l@{}}$\blacktriangleright$ Group-based signatures\end{tabular} & \begin{tabular}[c]{@{}l@{}} Can identify user if few participants \end{tabular}  \\
 & \cite{Kaaniche_2020,Khan20}  & Self-destructing data & Semi-trusted  & \begin{tabular}[c]{@{}l@{}}$\bigcdot$ Self-destructing data \end{tabular}& \begin{tabular}[c]{@{}l@{}} $\blacktriangleright$ Enhanced self-destructing data \end{tabular} & \begin{tabular}[c]{@{}l@{}}Only apply for specific applications \end{tabular}   \\
  & \cite{Kaaniche_2020,Khan20} & Data masking & Semi-trusted  & \begin{tabular}[c]{@{}l@{}}$\bigcdot$ Substitution, Shuffling, \\ Nulling out, Encryption, \\ Character scrambling  \end{tabular} & \begin{tabular}[c]{@{}l@{}} $\blacktriangleright$ Enhanced data masking \\ (AI-empowered) \end{tabular} & \begin{tabular}[c]{@{}l@{}}Privacy-data utility trade-off, \\ traffic overhead \end{tabular}  \\
  & \cite{Xiao2020,Liu2020FederatedLF} & Secure Multi-Party Computing & Untrusted  & \begin{tabular}[c]{@{}l@{}}$\bigcdot$ Federated learning  \end{tabular}& \begin{tabular}[c]{@{}l@{}} $\blacktriangleright$ Federated learning \\ $\blacktriangleright$ Distributed learning\end{tabular} & \begin{tabular}[c]{@{}l@{}}All participants need to be present, \\ High vulnerable to collusion attacks\end{tabular}  \\
\multirow{-8}{*}[20pt]{ Service layer} &  \cite{Kaaniche_2020} & Data perturbation & Untrusted & \begin{tabular}[c]{@{}l@{}}$\bigcdot$ Probability distribution \\$\bigcdot$ Value distortion \end{tabular} & \begin{tabular}[c]{@{}l@{}} $\blacktriangleright$ Enhanced data perturbation \\ (AI-empowered)\end{tabular} & Traffic overhead  \\ \hline
\end{tabular}   
\end{adjustbox}
\end{table*}

By contrast, the idea behind untrusted privacy preservation methods is that subscribers trust themselves only. Subscribers take on the role of protecting their own privacy. Secure Multi-Party Computing (SMC), also known as multi-party computation (MPC) or privacy-preserving computation, is a typical example of the untrusted model \cite{Kaaniche_2020}. According to \cite{Kaaniche_2020}, SMC aims to protect a distributed computation model from the inputs of communication parties while keeping those inputs private. Participants' communications are also encrypted and protected by cryptographic protocols where each participant cannot say that they learn nothing. A state-of-the-art technique, which is likely a prospective technology for many 6G applications, is federated learning. Many scholars have recently started exploring tweaking federated learning-based techniques for SMC, for example, the work in \cite{QZhang21,YLi20}. Federated learning can be solution to enhance privacy and power of distributed learning models in many 6G cooperative applications, e.g., misbehavior behavior for connected vehicles, 6G mixed reality. However, the risks of adversarial attacks, system induced bias in training data from different capabilities of devices, the ability of monitoring /debugging problems in a wide range, or slow convergence learning speed have been open challenges to federated learning. Data perturbation, known as obfuscation techniques, is another typical technique of an untrusted privacy preservation model. Data perturbation secures data exchange by adding ``noise'' (e.g., false or irrelevant data, scrambling user names) to the data source and then renders this into a form that unauthorized users cannot read or understand. This technique has been used in protecting electronic medical records from prying eyes \cite{Silva21}, which can become more common in 6G health applications. The drawback of data perturbation is that it reduces the ability of data mining/DL-based tools to access information since the noise values may not be meaningful but negative for training accurately.

Unlike the above two models, a semi-trusted model uses a distributed trust model where a communication party's trust is maintained through particular protocols \cite{Kaaniche_2020}. In this model, a data owner does not completely trust other users, including the service provider, but can maintain a level of trust through majority voting. The key assumption is that peer users are honest or consistent in sharing their information. Homomorphic encryption, differential privacy, group-based/attribute-based signatures, self-destructing data, and data masking are typical examples. As defined in Section~\ref{sec:security-service-layer}, homomorphic encryption allows performing operations, such as search and query, on encrypted data directly without decryption. Subscribers can thus send their data to a third party (e.g., operators, cloud providers) for storage or processing. Homomorphic encryption is expected to be a key technology for protecting data confidentiality and privacy in 6G \cite{Khan20}.

\begin{figure}
    \centering
   \begin{center}
			\includegraphics[width=1\linewidth]{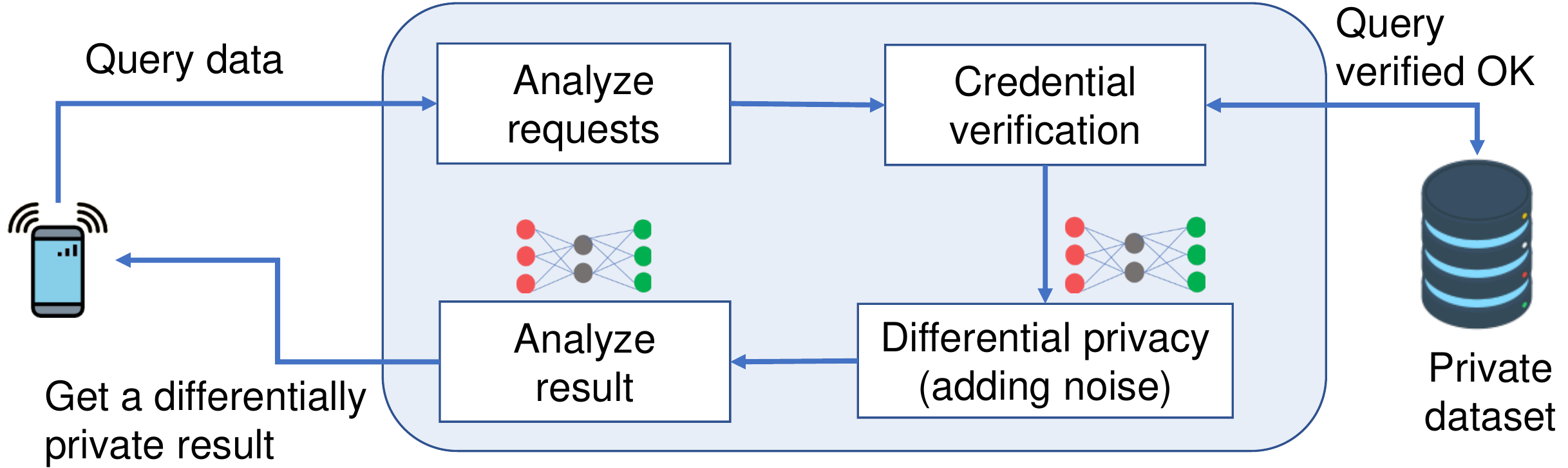}
	\end{center}
   \caption{Illustration of how differential privacy works. Differential privacy will be widely used in 5G/6G for processing private-data-related queries, where all retrieved results will be processed to eliminate the privacy-related data.}
   	\label{fig:6G-differential-privacy}
\end{figure}

Differential privacy is another prospective technology for privacy preservation in coming years \cite{BJian21,Hassan20}. As illustrated in Figure~\ref{fig:6G-differential-privacy}, the idea behind differential privacy is to create aggregate information within a dataset (e.g., by group), so that no query to any single individual or private data can be carried out directly without a filter (adding noise, randomness to the result retrieved), thereby providing privacy \cite{Sun2020}. However, the weakness of differential privacy is that it is a challenge to distinguish whether a particular value has been changed. Self-destructing data means an entity’s encrypted information exists only for a period of time and is valid for decryption if the private key has not expired over that duration, thus protecting privacy. However, this technique’s drawback is that data storage is temporary; thus, it may only be applied for some particular applications, such as to secrete instant messages. Finally, another name of data obfuscation, data masking protects personal data by hiding identifiable information with modified content. Some state-the-art studies \cite{Rahulamathavan21} found that the combination of homomorphic encryption and data masking can be a perfect way to provide a high degree of security against quantum attacks, which are taken to appear in 6G, while still maintain privacy at best. 

\subsection{Privacy challenges}

Implementing perfect privacy preservation mechanisms, at least to satisfy laws/regulations, is more challenging than just talk. Some factors that may have to be considered are as follows. First, implementation could burden the finances of an organization in order to maintain data privacy. Such cost will come from substantial investments in end-to-end protection equipment for data encryption and anonymization, let alone software customization to fully comply with legal requirements. The challenge is, not many small enterprises have the resources for such investments. In an era of data explosion like 6G, processing massive data may require expensive commodity devices that can overwhelm business expenditure. 

Protecting privacy in massive IoT devices with limited resources is another challenge. Wearable devices and low-cost IoT sensors tracking user location and health information are rarely equipped with strong authentication and security mechanisms. Lack of high-end protection can ease the shield for massive data collection from attacks. In 6G, when smart things and humans are supposed to be connected, the challenges of securing networks will increase exponentially.

Mobile users may regrettably accept risks to use services and are unaware of potential threats until real damage occurs. Another common pitfall is that collecting personal information does not matter since many users think data poses no risk or has little value. Worse, many data-driven businesses may not warn or may mislead users about potential consequences following the severity or the range of data collection in all ways. For example, a company may want sensitive data about their employees, arguing it may impact job performance. Social platforms seek information to help them run personalized ads and improve user experience by obtaining relevant information. Many laws \cite{USFederatedPrivacyAct} have tried to close such loopholes by imposing a high responsibility on data collectors or a limit on the kind of data they can collect. However, the challenge is that the oversight of privacy preservation practices is often limited. Users have little control over what companies can do regarding the kind of data gathered or how much data is collected. Penalties can only be imposed after massive data breaches have occurred. It is also unclear whether any campaign exploited collected data to target a specific object. 

\subsection{Summary of lessons learned from privacy enhancing technologies }

Table~\ref{tab:privacy-preservation} summarizes key state-of-the-art solutions and their pros and cons for enhancing privacy preservation in all three layers (physical/connection/service). Privacy preservation in 6G will likely significantly inherit from these key methods. Some prospective privacy enhancing technologies that will impact 6G consist of blockchain, federated learning/distributed learning, quantum homomorphic encryption, and differential privacy. However, to achieve feasible implementation for 6G, many of their flaws and weaknesses need to be addressed, such as the capability to resist insider attacks. Three key lessons learned from the privacy-enhancing technologies are as follows.

 \begin{enumerate}
     
     
     \item \textit{6G privacy matters, but preserving it properly needs much input}. Privacy preservation issues are not new and have been researched for years. However, the technologies have achieved little to resolve the realities of massive data breaches, which take place almost every day. Bad practices by data collectors in storing/processing user information and lack of advanced data protection are two of many reasons that contribute to worsening the chance of addressing these issues thoroughly. If policy and laws do not keep up, solving all of these problems may take years, even longer.
      
      \item \textit{A good privacy preservation solution should take care more about conditional anonymity to meet QoS of applications and less cost to implement than perfect anonymity at any price}. History shows that none of the state-of-the-art security and privacy preservation methods such as end-to-end encryption and blockchain can quickly be applied at large if their implementations result in the high complexity and high latency. A reasonable approach would be to link privacy preservation design to the anonymity requirements of applications, deployment/maintenance cost, and, importantly, less to changes in the security architecture. Also, besides enhancing high-performance computing and networking technologies to accelerate processing/transmission capability, security solutions that can balance the anonymity and QoS of applications such as distributed learning may have a greater chance of being implemented in many 6G applications.  
      
     \item \textit{Community awareness about the right to privacy and well-organized regulations are important factors for accelerating privacy-enhancing technologies}. While few data-driven companies will voluntarily comply with GDPR and ISO/IEC's privacy principles if not forced to do, individuals may be able to protect themselves by avoiding sharing their personal data to use a service at any cost. The stringent requirements from end-users and high penalties by law are the strongest motivations for businesses to enhance data privacy.

 \end{enumerate}

%% file: Section_X_Discussion.tex
\begin{figure}[ht]
    \centering
    \begin{center}
			\includegraphics[width=1\linewidth]{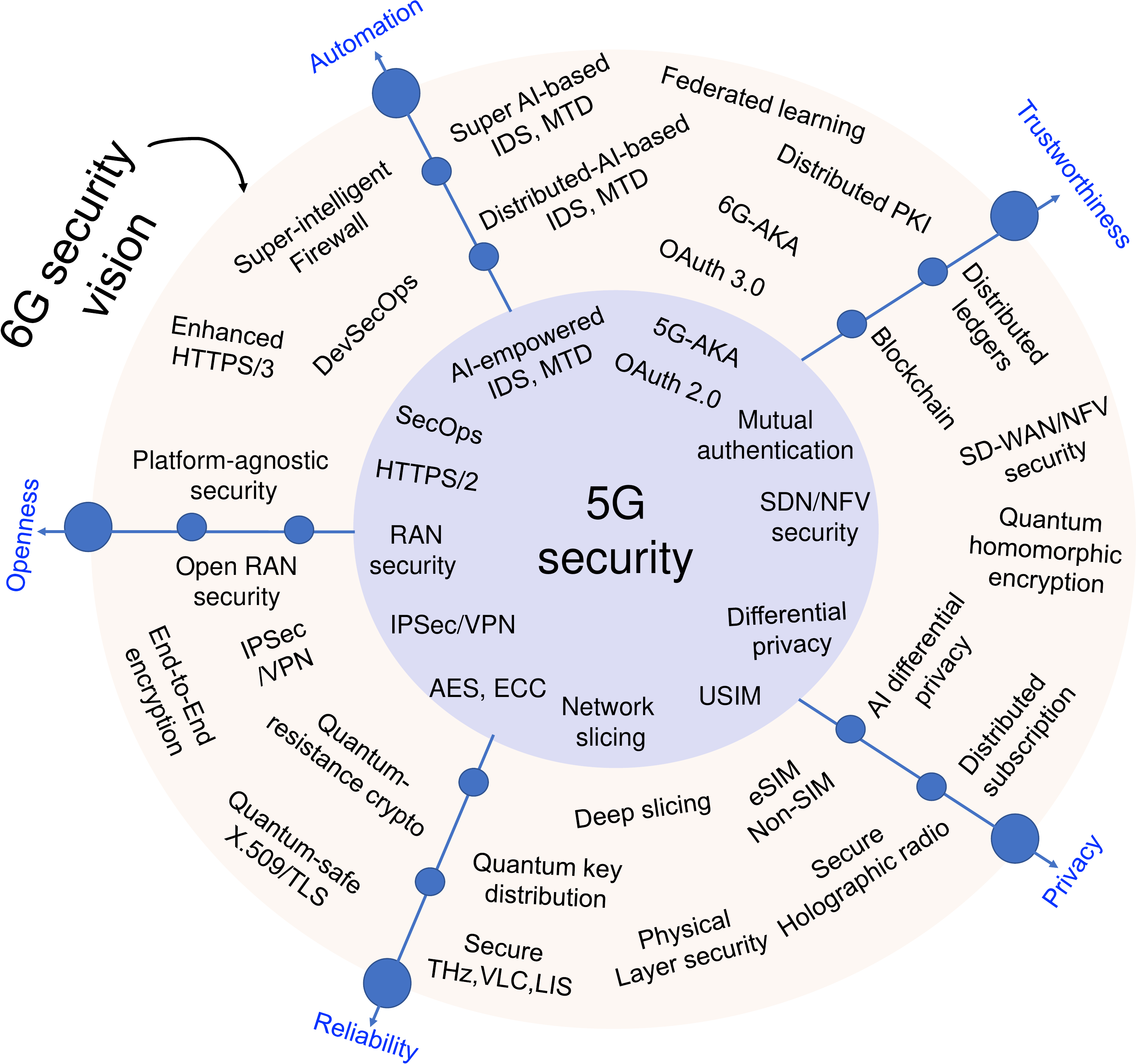}
	\end{center}
   \caption{The illustration of our vision on the evolution path of security technologies in 6G, from 5G. 6G will continue the trajectory of many 5G security technologies with significant upgrades of performance capabilities and models to satisfy new applications.}
   \label{fig:6G-security-probabilities}
\end{figure}

6G is a unique opportunity for significantly improving security and privacy. There are two main approaches to advance this goal. One is to secure high-impact enabling technologies for 6G, and the other is to improve 5G technologies which will likely be carried over to 6G. The summary of some prospective technologies for enhancing 6G security and privacy are illustrated in Fig.~\ref{fig:6G-security-probabilities}. These assessments are classified by our vision on the criteria of automation, trustworthiness, privacy, reliability, and openness. For example, before the full quantum-safe TLS protocols operate, a period transition with temporary quantum-resistance ciphersuites can be reasonable. In the other example, the distributed subscription may come after a transition of using nuSIM or non-ID for subscriber identity management. The positions of the technologies can change periodically, depending on the standardization and the market demands.

\subsection{Future research directions for 6G security and privacy}

\subsubsection{Securing high-impact enabling technologies for 6G: AI, THz, quantum computing, ultra-massive MIMO, and more}

The biggest upgrade on 6G networks will likely occur in the physical layer. To satisfy requirements such as network speeds greater that 1Tbps, many bet on THz communications and ultra-massive MIMO antenna technologies. Despite impressive initial results, these technologies have proved to be vulnerable to physical layer attacks, such as jamming and pilot contamination attacks \cite{Xu21}. Therefore, enhancing security of THz communications and relevant technologies such as NOMA/LIS/Holographic Radio is an important issue. An emerging solution is to build AI/ML models to improve secrecy rates and minimize the secrecy outage probabilities in these technologies, particularly with the existence of fading influence of partial or imperfect CSI. Such integration can be applied to improve key physical generation performance, physical layer authentication, radio slicing, and anti-jamming. However, 6G physical layer security will need a breakthrough in implementation; otherwise, the vision of commercial deployment will be unlikely. In the connection layer, quantum-safe encryption and key distribution are expected to be the next-generation standards for communication security, likely to be first applied to 6G. While the winner of the standardization process race is unclear, related studies such as \cite{Carames20} have revealed technical challenges of efficient key exchange that may take years to solve, at least with the current infrastructure. At the service layer, microservices and serverless security will likely dominate the protection models in 6G, in which security will be integrated into functions instead of using firewalls and secure web gateways as is currently done.   

\subsubsection{Enhancing 5G technologies for 6G: SDN, network slicing, vRAN/Open RAN, Edge, and their successors}

6G continues to perfect many of the technologies and features of 5G security. MmWave bands above 300 GHz are likely important forces for 6G-compatible communications (with 5G/5.5G network infrastructure) in the physical layer. Therefore, fixing vulnerabilities of mmWave and massive MIMO technologies is still an important task for maintaining 6G security, given that a complete transition to THz technology may take years. Security challenges in vRAN and cloud paradigms (C-RAN) are further concerns. Given the complexity of vRAN management, more testing on vRAN API control vulnerabilities is required. Any bug or design flaw in the shared infrastructure of vRAN/C-RAN can also compromise entire security isolation functions and result in networks in chaos. It is certainly critical to address these vulnerabilities since vRAN/C-RAN will shortly be deployed. 

Although it is only at an early stage of development, there is a high expectation of a breakthrough in SD-WAN security, which will be crucial because 6G, or at least autonomous systems in 6G, will likely be managed by this technology. SD-WAN security will be an important step toward the vision of intent-based networking and network-as-a-service models, the key goal of 6G. In the service layer, enhanced versions of biometric authentication, AI-empowered firewalls/IDS, and open-source security will be the pillars of 6G service security, given the reputation of efficiency of its predecessors, proved by commercial deployment.  

\subsubsection{Enhancing privacy technologies that satisfy GDPR}

The complexity of networks and the diversity of applications in 6G will likely complicate data privacy preservation more than ever. As we noted in Section~\ref{sec:privacy-in-6G}, privacy concerns are not unique to 6G, but will probably become worse. Privacy challenges rooted in many factors (e.g., tricks of data-driven business platforms or lack of advanced security protection in low-cost devices) are not easy to fix quickly. However, since big data is critical for AI and knowledge mining, balancing data collection needs for learning in beneficial applications and business is extremely hard. Blockchain, distributed learning, federated learning, homomorphic encryption, and differential privacy are potential technologies that support protecting personal data, complying with GDPR standards, and satisfying the needs for information mining models. With many remaining high complexity and computation challenges, more enhancements and stress tests to verify their feasibility for real network environments will then be needed. 

\subsection{Open challenges and issues}

\subsubsection{Real-time adaptive security -- challenge for a breakthrough}

6G will deliver ultra-low latency (less than 0.1 milliseconds), which will enable many latency-sensitive applications, such as autonomous vehicles, industrial automation, and telesurgery. These applications must meet their deadlines. Conventional security solutions are designed to defend both IT and corporate networks, but they will likely fail to satisfy such 6G-enabled real-time applications if there are no further breakthroughs. A real-time security system should ensure that these time-critical applications meet their deadlines regardless of malicious activities. There are three design challenges for real-time security described briefly below.
\begin{enumerate}
    \item Seeking no interference influence in the system to preserve the time constraints of applications. Such interference influences may introduce infinite delays which will impair the predictability and determinism of real-time applications.
    \item Employing real-time adaptive responses to security incidents: Adaptive security software must be able to learn network behavior dynamically and respond to security alerts.
    \item Implementing rapid and real-time context-aware security for target applications: A security solution is expected to demonstrate its efficiency in intrusion detection in two ways, by performing real-time anomaly detection compatible with 6G ``wire'' speed or by supporting context-aware security protection with diversified data generated from target applications.
\end{enumerate}

\begin{figure}
    \centering
   \begin{center}
			\includegraphics[width=1\linewidth]{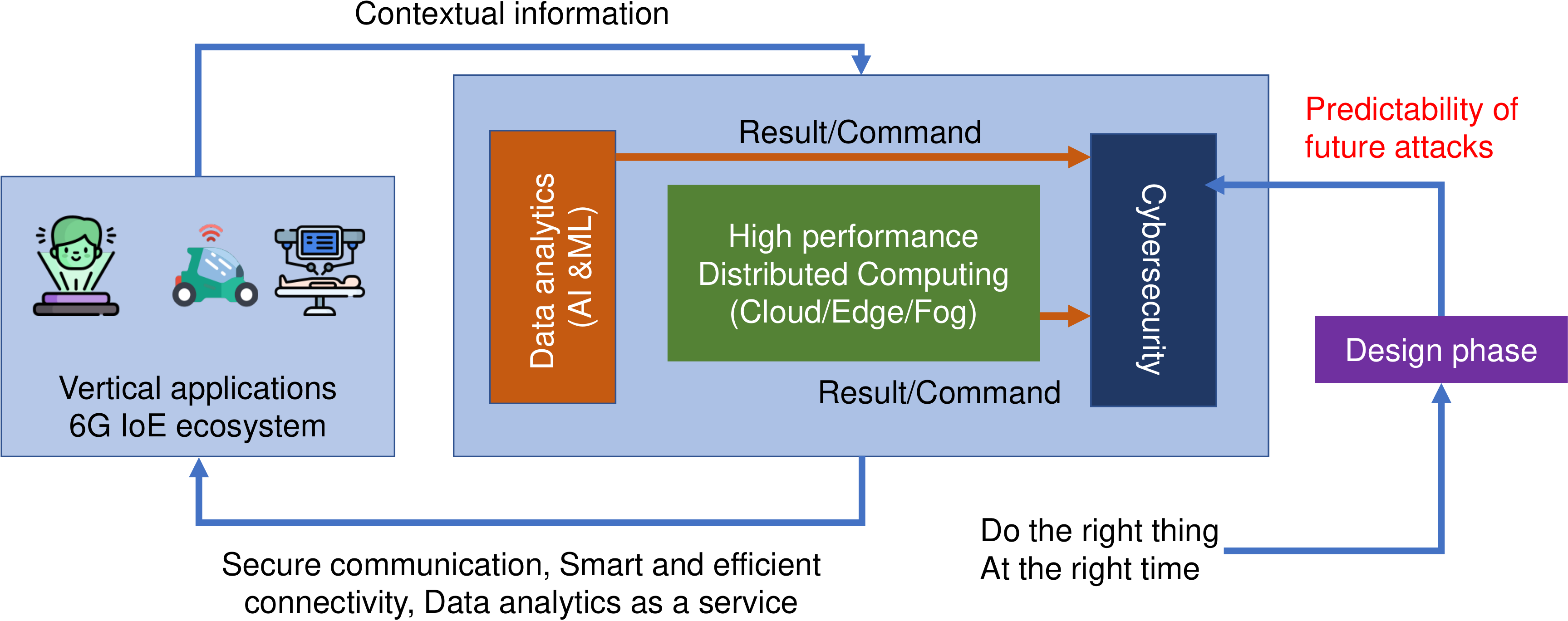}
	\end{center}
   \caption{An illustration of the key role of timely processing and predictability in 6G security to support two other pillars of 6G networks (updated from the ``Smart Networks'' concept of \cite{5GIA2020}).}
   	\label{fig:6G-smart-security}
\end{figure}

We believe that implementing an adaptive security defense system to respond to attacks in real-time is a challenge, given the difficulty of processing massive data transmission without delay. Advancements of optical networks, superluminal communication, high-performance computing, and intelligent edge technologies may in the future partially address this problem by enabling super-fast processing and one-hop communication. Also, apply AI to support predictability in protection systems is another promising approach. Fig.~\ref{fig:6G-smart-security} illustrates the key role of timely processing and predictability in 6G security, besides availability-integrity-confidentiality principles.

\subsubsection{Simplifying the security architecture and relevant technologies -- a not-easy-at-all target}

Simplicity and transparency are the keys to keeping a network system more secure and introducing fewer vulnerabilities because of more oversight of the community. Many hope that 6G security architecture will be simpler or even enable better oversight rights for users on how operators protect their data, which 5G is not supposed to have achieved. Supposing 6G network coverage will be expanded substantially, as well as the diversity of applications and services with different protection requirement goals, achieving these goals will not be a trivial task. Several initiatives such as slicing, open encryption standards, open security orchestration, and open authentication protocols may boost this target. 

\subsubsection{Maintaining backward compatibility is a complicated issue but likely a must-have feature in 6G security}

As we noted in Section~\ref{sub:5G-sa-nsa-strategy}, selecting a non-stand-alone deployment strategy, which most operators favor, requires the capability of backward compatibility maintenance in network access authentication and mutual authentication between the serving network and the home network. However, the backward compatibility feature can open the door to exposing old vulnerabilities when 6G must request 5G security architecture to authenticate deployed devices. The details of a backward compatibility feature and how to implement it in 6G security architecture may need to be given more thought in the future.  

\subsubsection{Supply chain security -- an emerging issue for the development of 6G security}

Supply chain digitalization has become unavoidable in modern business processes by making them more flexible and accurate. The shortage of supply chains recently highlighted the vital role of supply assurance. A timeline for enabling many 6G key technologies, such as AI and high-performance computing, will likely significantly impact a supply chain if it is cut off or disrupted on a large scale. Security and privacy issues become an even bigger challenge to overcome when supply-chain management becomes a potential attack target because of the wide attack surfaces and consequences from system breaches. Also, software and hardware from untrusted sources can threaten the trust of networks and be a real danger to business assets. 

A recent hearing called ``5G Supply Chain Security: Threats and Solutions'' held on March 2020 in the US Senate examined the security and integrity of telecommunications supply chains and protecting the network transition to 5G. Further, the ATIS Supply Chain Working Group is working on the development of supply chain standards for public and private sectors. We can expect that stakeholders in 6G technologies and infrastructure will be concerned about supply chain security in terms of hardware, software and communications aspects, and make efforts to develop mechanisms to detect corrupted components before use. For example, hardware supply chain authentication and security are intended as countermeasures to various threats and attacks on networks, and to validate 6G hardware/equipment’s authenticity. Three key technologies to ensure supply chain security are based on, (1) blockchain, (2) AI, and (3) physically unclonable functions \cite{Hassija_2020}. In the future, AI can be the key technology for reinforcing security protection for supply chain management platforms and enhance accuracy performance for key processes, such as demand forecasting and secure shipping. From a telecom operator’s point of view, attention to security validations and measurements of hardware (e.g., chipsets, semiconductors, and equipment) and software (e.g., inhouse built software, commercial software, and open-source software) could significantly contribute to for supply chain security and effectiveness. 

\subsubsection{Transforming the hardware-driven security platforms into the software-driven security platforms can create new risks}

Top network operators are major forces who lobby strongly for the idea of transforming network control functions from a hardware-based model into a software-based platform or platform-agnostic system. The goal is to avoid locking in a specific vendor, increase modularity and diversify supply chains. At first glance, this transition seems a good strategy and will benefit operators with a long-term vision. However, we argue that open software-driven platforms may still have certain drawbacks in security aspects. First, software-based platforms are susceptible to more attack surfaces, such as from conventional techniques like DDoS. Although issuing software patches is undoubtedly faster than going with hardware, these platforms are also at risk of more defections and human errors when managing millions of line codes.